\documentclass[12pt,preprint]{aastex}
\usepackage{natbib}
\newcommand{\be}{\begin{equation}}
\newcommand{\ee}{\end{equation}}
\newcommand{\nass}{\mbox{$N_{\rm ass}$}}
\newcommand{\nfalse}{\mbox{$N_{\rm false}$}}
\newcommand{\nfalsemc}{\mbox{$\langle \hat N_{\rm false} \rangle$}}

\newcommand{\id}{\mbox{A}}
\newcommand{\cc}{\mbox{$\overline{\rm A}$}}
\newcommand{\data}{\mbox{$r,\phi$}}
\newcommand{\post}{\mbox{$P_{ik}(\id|\data)$}}
\newcommand{\prior}{\mbox{$P_{i}(\id)$}}
\newcommand{\idp}{\mbox{H$_k$}}
\newcommand{\ccp}{\mbox{H$_{-}$}}
\newcommand{\nlat}{\mbox{$N_{\rm LAT}$}}

\newcommand{\pthres}{\mbox{$P_{\rm thr}$}}

\shorttitle{Fermi LAT First Catalog}
\shortauthors{Abdo et al.}

\begin{document}


\title{Fermi Large Area Telescope First Source Catalog}

\author{
A.~A.~Abdo\altaffilmark{2,3}, 
M.~Ackermann\altaffilmark{4}, 
M.~Ajello\altaffilmark{4}, 
A.~Allafort\altaffilmark{4}, 
E.~Antolini\altaffilmark{5,6}, 
W.~B.~Atwood\altaffilmark{7}, 
M.~Axelsson\altaffilmark{8,9,10}, 
L.~Baldini\altaffilmark{11}, 
J.~Ballet\altaffilmark{12,1}, 
G.~Barbiellini\altaffilmark{13,14}
% ...
}
\altaffiltext{1}{Corresponding authors: J.~Ballet, jean.ballet@cea.fr; S.~W.~Digel, digel@stanford.edu; J.~Kn\"odlseder, knodlseder@cesr.fr.}
\altaffiltext{2}{Space Science Division, Naval Research Laboratory, Washington, DC 20375, USA}
\altaffiltext{3}{National Research Council Research Associate, National Academy of Sciences, Washington, DC 20001, USA}
\altaffiltext{4}{W. W. Hansen Experimental Physics Laboratory, Kavli Institute for Particle Astrophysics and Cosmology, Department of Physics and SLAC National Accelerator Laboratory, Stanford University, Stanford, CA 94305, USA}
\altaffiltext{5}{Istituto Nazionale di Fisica Nucleare, Sezione di Perugia, I-06123 Perugia, Italy}
\altaffiltext{6}{Dipartimento di Fisica, Universit\`a degli Studi di Perugia, I-06123 Perugia, Italy}
\altaffiltext{7}{Santa Cruz Institute for Particle Physics, Department of Physics and Department of Astronomy and Astrophysics, University of California at Santa Cruz, Santa Cruz, CA 95064, USA}
\altaffiltext{8}{Department of Astronomy, Stockholm University, SE-106 91 Stockholm, Sweden}
\altaffiltext{9}{Lund Observatory, SE-221 00 Lund, Sweden}
\altaffiltext{10}{The Oskar Klein Centre for Cosmoparticle Physics, AlbaNova, SE-106 91 Stockholm, Sweden}
\altaffiltext{11}{Istituto Nazionale di Fisica Nucleare, Sezione di Pisa, I-56127 Pisa, Italy}
\altaffiltext{12}{Laboratoire AIM, CEA-IRFU/CNRS/Universit\'e Paris Diderot, Service d'Astrophysique, CEA Saclay, 91191 Gif sur Yvette, France}
\altaffiltext{13}{Istituto Nazionale di Fisica Nucleare, Sezione di Trieste, I-34127 Trieste, Italy}
\altaffiltext{14}{Dipartimento di Fisica, Universit\`a di Trieste, I-34127 Trieste, Italy}
% ...

\begin{abstract}
We present a catalog of high-energy gamma-ray sources detected by the Large Area Telescope (LAT), the primary science instrument on the {\it Fermi Gamma-ray Space Telescope (Fermi)}, during the first 11 months of the science phase of the mission, which began on 2008 August 4.  The First $Fermi$-LAT catalog (1FGL) contains 1451 sources detected and characterized in the 100~MeV to 100~GeV range.  Source detection was based on the average flux over the 11-month period, and the threshold likelihood Test Statistic is 25, corresponding to a significance of just over 4$\sigma$.  The 1FGL catalog includes source location regions, defined in terms of elliptical fits to the 95\% confidence regions and power-law spectral fits as well as flux measurements in 5 energy bands for each source.  In addition, monthly light curves are provided.  Using a protocol defined before launch we have tested for several populations of gamma-ray sources among the sources in the catalog.  For individual LAT-detected sources we provide firm identifications or plausible associations with sources in other astronomical catalogs.  Identifications are based on correlated variability with counterparts at other wavelengths, or on spin or orbital periodicity.  For the catalogs and association criteria that we have selected, 630 of the sources are unassociated.  Care was taken to characterize the sensitivity of the results to the model of interstellar diffuse gamma-ray emission used to model the bright foreground, with the result that 161 sources at low Galactic latitudes and toward bright local interstellar clouds are flagged as having properties that are strongly dependent on the model or as potentially being due to incorrectly modeled structure in the Galactic diffuse emission.

\end{abstract}



\section{Introduction}

The {\it Fermi Gamma-Ray Space Telescope} has been routinely surveying the sky with the Large Area Telescope (LAT) since the science phase of the mission began in 2008 August.  The combination of  deep and fairly uniform exposure, good per-photon angular resolution, and stable response of the LAT have made for the most sensitive, best-resolved survey of the sky to date in the 100~MeV to 100~GeV energy range.  

Observations at these high energies reveal non-thermal sources and a wide range of processes by which Nature accelerates particles.  The utility of a uniformly-analyzed catalog such as this is both for identifying special sources of interest for further study and for characterizing populations of $\gamma$-ray emitters.  The LAT survey data analyzed here allow much more detailed characterizations of variability and spectral shapes than has been possible before.

Here we expand on the Bright Source List \citep[][BSL]{LATBSL}, which was an early release of 205 high-significance (likelihood Test Statistic $TS >$100; see \S~\ref{run:TS}) sources detected with the first 3 months of science data.  The expansion is in terms of time interval considered (11 months vs. 3 months), energy range (100~MeV -- 100~GeV vs. 200~MeV -- 100~GeV), significance threshold ($TS >$ 25 vs. $TS >$ 100), and detail provided for each source.  Regarding the latter, we provide elliptical fits to the confidence regions for source location (vs. radii of circular approximations), fluxes in 5 bands (vs. 2 for the BSL) for the range 100~MeV -- 100~GeV, and monthly light curves for the integral flux over that range.

We also provide associations with previous $\gamma$-ray catalogs, for EGRET \citep{3EGcatalog,EGRcatalog} and AGILE \citep{AGILEcatalog}, and with likely counterpart sources from known or suspected source classes.  The number of sources for which no plausible associations are found is 630, at the specified confidence level for source association (80\%).  The First LAT AGN Catalog \citep[1LAC, ][]{LATAGNCatalog} is based on the 1FGL sources, and applies the same association methods, but provides associations for AGNs down to the 50\% confidence level.

As with the BSL, the First $Fermi$-LAT catalog of $\gamma$-ray sources (1FGL, for first $Fermi$ Gamma-ray LAT) is not flux limited and hence not uniform.  As described in \S~\ref{run:construction}, the sensitivity limit depends on the region of the sky and on the hardness of the spectrum.  Only sources with $TS >$ 25 (corresponding to just over 4~$\sigma$ statistical significance) are included, as described below.

\section{Gamma-ray Detection with the Large Area Telescope}
\label{run:LAT}

The LAT is a pair-production telescope \citep{LATinstrument}. The tracking section has 36 layers of silicon strip detectors to record the tracks of charged particles, interleaved with 16 layers of tungsten foil (12 thin layers, 0.03 radiation length, at the top or {\it Front} of the instrument, followed by 4 thick layers, 0.18 radiation length, in the {\it Back} section) to promote $\gamma$-ray pair conversion.
Beneath the tracker is a calorimeter comprised of an 8-layer array of CsI crystals (1.08 radiation length per layer) to determine the 
$\gamma$-ray energy. The tracker is surrounded by segmented charged-particle anticoincidence detectors (plastic scintillators 
with photomultiplier tubes) to reject cosmic-ray background events. The LAT's improved sensitivity 
compared to EGRET stems from a large peak effective area ($\sim$8000~cm$^2$, or $\sim$6 times greater than 
EGRET's), large field of view ($\sim$2.4~sr, or nearly 5 times greater than EGRET's), good background rejection, superior angular resolution (68\% containment angle $\sim$0.6$^\circ$ at 1~GeV for the {\it Front} section and about a factor of 2 larger for the {\it Back} section, vs. $\sim$1.7$^\circ$ at 1~GeV for EGRET; \citealt{EGRETcalib}), and improved observing efficiency (keeping the sky in the field of view with scanning observations, vs. inertial pointing for EGRET). Pre-launch predictions of the instrument performance are described in \cite{LATinstrument}. 

The data analyzed for the 1FGL catalog were obtained during 2008 August 4 --
2009 July 4 (LAT runs 239557414 through 268411953, where the numbers refer
to the Mission Elapsed Time (MET) in seconds since 00:00 UTC on 1 January 2001, at the start of the data acquisition runs). During most of this
time $Fermi$ was operated in sky-scanning survey mode (viewing direction rocking
35$\degr$ north and south of the zenith on alternate orbits).  During May 7--20 the rocking angle was increased to 39$\degr$ for operational reasons.  In addition, a few
hours of special calibration observations during which the
rocking angle was much larger than nominal for survey mode or the
configuration of the LAT was different from normal for science operations were obtained during the period analyzed. 
Time intervals when the rocking angle was larger than 43$\degr$ have been
excluded from the analysis, because the bright limb of the
Earth enters the field of view (see below).

In addition, two short time intervals associated with $\gamma$-ray bursts (GRB)
that were detected in the LAT have been excluded. These intervals
correspond to GRB 080916C \citep[MET 243216749--243217979, ][]{GRB080916C}
and GRB 090510 \citep[MET 263607771--263625987, ][]{GRB090510}. 

Observations were nearly continuous during the survey interval, although a few data gaps are present due to operational issues, special calibration runs, or in rare cases, data loss in transmission.  Table~\ref{tab:gaps} lists all data gaps longer than 1~h.   The longest gap by far is 3.9~d starting early on March 16; together the gaps longer than 1~h amount to $\sim$7.9~d or 2.4\% of the interval analyzed for the 1FGL Catalog.  

\begin{deluxetable}{lr}
\setlength{\tabcolsep}{0.04in}
\tabletypesize{\scriptsize}
\tablewidth{0pt}
\tablecaption{Gaps Longer Than One Hour in Data\label{tab:gaps}}
\tablehead{
\colhead{Start of Gap (UTC)} &
\colhead{Duration (h)}
}
\startdata
2008-09-30 14:27 &   1.16\\
2008-10-11 03:14 &   1.59\\
2008-10-11 11:04 &   3.82\\
2008-10-14 12:23 &   3.83\\
2008-10-14 17:11 &   3.49\\
2008-10-14 20:22 &   1.59\\
2008-10-15 17:03 &   3.47\\
2008-10-16 15:18 &   1.83\\
2008-10-22 19:20 &   1.59\\
2008-10-22 23:43 &   2.06\\
2008-10-23 11:16 &   1.91\\
2008-10-30 16:43 &   1.59\\
2008-12-11 17:41 &   6.37\\
2009-01-01 00:35 &   1.72\\
2009-01-06 20:43 &   6.98\\
2009-01-13 13:26 &   2.10\\
2009-01-17 12:58 &   2.05\\
2009-01-28 19:28 &   4.78\\
2009-02-01 15:46 &   1.59\\
2009-02-15 10:15 &   1.05\\
2009-03-16 00:27 & 116.78\\
2009-05-02 19:04 &   8.94\\
2009-05-07 15:21 &   5.46\\
2009-06-26 12:59 &   3.19\\
\enddata
\end{deluxetable}

The total live time included is 245.6~days (21.22 Ms).  This corresponds to an absolute efficiency of 73.5\%.  Most of the inefficiency is due to time lost during passages through the South Atlantic Anomaly ($\sim$13\%) and to readout dead time (9.2\%).

The standard onboard filtering, event reconstruction, and classification were applied to the data \citep{LATinstrument}, and for this analysis the `Diffuse' event class\footnote{See http://fermi.gsfc.nasa.gov/ssc/data/analysis/documentation/Cicerone/Cicerone\_Data/LAT\_DP.html.} is used.  This is the class with the least residual contamination from charged-particle background events, released to the public.  The tradeoff for using this event class relative to the `looser' Source class is primarily reduced effective area, especially below 500~MeV.

The instrument response functions (IRFs) -- effective area, energy redistribution, and point-spread function (PSF) -- used in the likelihood analyses described below were derived from GEANT4-based Monte Carlo simulations of the LAT using the event-selection criteria corresponding to the Diffuse event class.  The Monte Carlo simulations themselves were calibrated prior to launch using accelerator tests of flight-spare `towers' of the LAT \citep{LATinstrument} and have since been updated based on observation of pile-up effects on the reconstruction efficiency in flight data \citep{rando2009}.  The effect introduces an inefficiency that is proportional to the trigger rate and dependent on energy.  The likelihood analysis for characterizing the sources uses the P6\_V3 IRFs (see \S~\ref{run:TS}), which have the effective areas corrected for the inefficiency corresponding to the overall average trigger rate seen by the LAT.  The use of the P6\_V3 IRFs allows the energy range of the analysis for the catalog to be extended down to 100~MeV (vs. 200~MeV for the BSL analysis, which used P6\_V1).  Below 100~MeV the  effective area is relatively small and strongly dependent on energy.  These considerations, together with the increasing breadth of the PSF at low energies (scaling approximately as $0.8^\circ (E/{\rm 1 GeV})^{-0.8}$), motivated the selection of 100~MeV as the lower limit for this analysis.

The alignment of the $Fermi$ observatory viewing direction with the z-axis of the LAT was found to be stable during survey-mode observations \citep{LATcalib}.  Analyses of flight data suggest that the PSF is somewhat broader than the calculated Diffuse class PSF at energies greater than $\sim$10~GeV; the primary effect on the current analysis is to decrease the localization capability somewhat.  As discussed below, this is taken into account in the catalog by increasing the derived sizes of source location regions by 10\%.

For the analysis, a cut on zenith angle (angle between the boresight of the LAT and the local zenith)
was applied to the Diffuse class events to limit the contamination from
albedo $\gamma$-rays from interactions of cosmic rays with the upper
atmosphere of the Earth.  These interactions make the limb of the Earth
(zenith angle $\sim$113$^\circ$  at the 565~km, nearly-circular orbit of
{\it Fermi}) an intensely-bright $\gamma$-ray source
\citep{Thompson1981}.  The limb is very far off axis in survey-mode observations, at least 70$^\circ$ for the data set considered here because of the rocking angle requirement described above.  
Removing events at zenith angles greater than 105$^\circ$ affects the exposure calculation
negligibly but  reduces the overall background rate.  After these cuts, the
data set contains $1.1 \times 10^7$ Diffuse-class events with energies $>$100~MeV.

\begin{figure}

\epsscale{.80}
\plotone{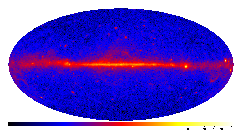}
\caption{Sky map of the LAT data for the time range analyzed in this paper, Aitoff projection in Galactic coordinates. The image shows $\gamma$-ray intensity for energies $>$300~MeV, in units of photons m$^{-2}$ s$^{-1}$ sr$^{-1}$. \label{fig:intensity_map}}
\end{figure}

The intensity map of Figure~\ref{fig:intensity_map} summarizes the data set used for this analysis and shows the dramatic increase of the brightness of the $\gamma$-ray sky at low Galactic latitudes.  The corresponding exposure is relatively flat and featureless as was the case for the shorter time interval analyzed for the BSL.  The degree of exposure nonuniformity is relatively small (about 30\% difference between minimum and maximum), with the deficit around the south celestial pole due to loss of exposure during passages of $Fermi$ through the South Atlantic Anomaly \citep{LATinstrument}.

\section{Diffuse emission model}
\label{run:diffusemodel}
An essential input to the analyses for detecting and characterizing $\gamma$-ray sources in the LAT data is a model of the diffuse $\gamma$-ray intensity of the sky.  Interactions between cosmic rays and interstellar gas and photons make the Milky Way a bright, structured celestial foreground.  Unresolved emission from extragalactic sources contributes an isotropic component as well.  In addition, residual charged-particle background, i.e., cosmic rays that trigger the LAT and are misclassified as $\gamma$-rays, provides another approximately isotropic background.  For the analyses described in this paper we used models for the Galactic diffuse emission (\texttt{gll\_iem\_v02.fit})  and isotropic backgrounds that were developed by the LAT team and made publicly available as models recommended for high-level analyses.  The models, along with descriptions of their derivation, are available from the $Fermi$ Science Support Center\footnote{http://fermi.gsfc.nasa.gov/ssc/data/access/lat/BackgroundModels.html}.    

Briefly, the model for the Galactic diffuse emission was developed using spectral line surveys of H~{\sc I} and CO (as a tracer of H$_2$) to derive the distribution of interstellar gas in Galactocentric rings.  Infrared tracers of dust column density were used to correct column densities as needed, e.g., in directions where the optical depth of H~{\sc I} was either over or under-estimated. The model of the diffuse $\gamma$-ray emission was then constructed by fitting the $\gamma$-ray emissivities of the rings in several energy bands to the LAT observations.  The fitting also required a model of the inverse Compton emission that was calculated using GALPROP \citep{strong04b,strong07} and a model for the isotropic diffuse emission.  

The isotropic component was derived as the residual of a fit of the Galactic diffuse emission model to the LAT data at Galactic latitudes above $|b|$ = 30$^\circ$ and so by construction includes the contribution of residual (misclassified) cosmic rays for the event analysis class used (Pass 6 Diffuse; see \S~\ref{run:LAT}).  Treating the residual charged particles as effectively an isotropic component of the $\gamma$-ray sky brightness rests on the assumption that the acceptance for residual cosmic rays is the same as for $\gamma$-rays.  This approximation has been found to be acceptable; the numbers of residual cosmic-ray background events scale as the overall livetime and any acceptance differences from $\gamma$-rays would not introduce small-scale structure in the models for likelihood analysis.

\section{Construction of the Catalog}
\label{run:construction}
The procedure used to build the 1FGL catalog follows the same steps
described in \citet{LATBSL} for the BSL,
with a number of improvements.
We review those steps in this section, highlighting what was done
differently for 1FGL.

Three steps were applied in sequence:
detection, localization, significance estimation.
In this scheme the threshold for inclusion in 1FGL is defined at the last step,
but the completeness is controlled by the first one.
After the list was defined we determined the source characteristics
(flux in 5 energy bands, time variability). The 1FGL catalog includes much more information for each source
than the BSL. In what follows, flux $F$ means photon flux and spectral index $\Gamma$
is for photons (i.e., $F \propto E^{-\Gamma}$).

In constructing the catalog the source detection step was applied only to the data from the full 11-month period as a whole.  That is to say, we did not search for potentially flaring sources which might only be detectable on shorter timescales.  Independently of this work, the LAT Automated Science Processing \citep{LATinstrument} and Flare Advocate activity provide a framework through which such flaring sources are detected in a timely manner and reported as Astronomer's Telegrams (ATels).  However, since all bright flaring sources that were reported as ATels were also bright enough to be detected over 11 months based on their average fluxes, they are included
in the 1FGL catalog anyway.   No GRB is detected over the full interva; the time ranges of the two brightest GRBs were excluded from the analysis (see \S~\ref{run:LAT}).

The pulsars \citep{LATPulsars} and X-ray binaries
\citep{LATLSI+61,LATLS5039,LATCygX3} which are identified via their
rotation or orbital period, were detected and localized as ordinary
sources. But they were entered explicitly at their true positions
in the main maximum likelihood analysis (\S~\ref{run:TS}), in order
not to bias their characteristics and those of their surroundings
if the Galactic diffuse model is imperfect (\S~\ref{run:diffusemodel}).
For the LAT-detected pulsars, we used the radio or $\gamma$-ray timing localization
\citep{LATPulsars}
which is always more precise than that based on the spatial distribution
of the events.
We have checked that the positions of the brightest pulsars found by
the localization algorithm (\S~\ref{run:localization}) were consistent
with their true positions at the 95\% level (using only the statistical error,
without any systematic correction).

\subsection{Detection}
\label{run:detection}

The detection step used the same ideas that were detailed in \citet{LATBSL}.
It was based on the same three energy bands, combining $Front$ and $Back$
events to preserve spatial resolution.
The detection does not use events below 200~MeV, which have poor
angular resolution. It uses events up to 100~GeV.
The full band ($6.7 \times 10^6$ counts) starts at 200~MeV for $Front$
and 400~MeV for $Back$ events.
The medium band ($12.0 \times 10^5$ counts) starts at 1~GeV for $Front$
and 2~GeV for $Back$ events.
The hard band ($10.7 \times 10^4$ counts) starts at 5~GeV for $Front$
and 10~GeV for $Back$ events.

We used the same partitioning of the sky into 24 planar projections
as in the BSL, and the same two wavelet-based detection methods:
{\it mr\_filter} \citep{sp98} and {\it PGWave} \citep{dmm97,PGWave}.
The methods looked for sources on top of the diffuse emission model
described in \S~\ref{run:diffusemodel}.
For {\it mr\_filter} the threshold was set in each image using the
False Discovery Rate procedure \citep{bh95} at 5\% of false detections.
For {\it PGWave} we used a flat threshold at 4 $\sigma$.
For comparison with the BSL, the number of `seed' sources from {\it mr\_filter}
was 857 in the full band, 932 in the medium band and 331 in the hard band.
Contrary to the BSL procedure,
we combined the results of those two methods (eliminating duplicates)
rather than choosing a baseline
method and using the other for comparison.
The rationale was to limit
the number of missed sources to a minimum, since the later steps
do not introduce any additional sources.
Duplicates were defined after the first localization
({\it pointfit} in \S~\ref{run:localization}, run separately on each list
of seeds). If two resulting
positions were consistent within the quadratic sum of 95\% error radii
only one source was kept (that with highest significance estimate).
Where {\it pointfit} did not converge, the 95\% error radius was set to
$0.3\degr$, typical for faint sources (\S~\ref{run:localization}).

To that same end we also introduced for 1FGL two other detection methods:
\begin{itemize}
\item $pointfind$, a tool that searches for candidate point sources by maximizing the likelihood function for trial point sources at each direction in a HEALPix \citep{Gorski2005} order 9 (pixel size $\sim$0.1~deg$^2$) tessellation of the sky.  The algorithm for evaluating the likelihood is optimized for speed by using energy-dependent binning of the photon data, choosing 4 energy bands per decade starting at 700~MeV, and a HEALPix order commensurate with the PSF width in each band.  A first pass examines the significance of a trial point source at the center of each pixel, on the assumption that the diffuse background is adequately described by the model for Galactic diffuse emission and ignoring any nearby point sources. The likelihood is optimized with respect to the signal fraction (i.e., the source and diffuse intensities are not fit separately) in each energy band, with the total likelihood being the product over all the bands. This makes the result independent of the spectrum of the point source or of the diffuse background.  While the search is quite efficient, it produces many false signals, so a second pass is used to optimize a more detailed likelihood function which includes nearby detected sources and fits the test source flux and diffuse background normalization independently.  The result of the second pass is a map of Test Statistic  from which the coordinates of candidate point sources can be derived.

\item the minimum spanning tree \citep{cmg08} looks for clusters of
high-energy events ($>$ 4~GeV outside the Galactic plane and $>$ 10~GeV
at $|b| < 15\degr$). It is restricted to high energies because it does not account for structured background, but can efficiently detect very hard sources.
\end{itemize}
We combined the `seed' positions from those two methods with
those from the wavelet-based methods, using the same procedure
for removing duplicates as above.

Finally, we introduced external seeds from the BZCAT \citep{BZcatalog} and WMAP \citep{WMAP5catalog} catalogs.  The BZCAT catalog is not homogenous but includes the great majority of known, well-characterized blazars.  It is a superset of the CGRaBS \citep{CGRaBS} catalog and has broader sky coverage.  The WMAP catalog includes mainly bright FSRQ blazars, and was used primarily to try to recover soft-spectrum sources that might have been missed by the source-detection algorithms.

In order to not bias the 1FGL catalog toward those external sources, we used
them as seeds only when there was no seed from the detection methods
within its 95\% error radius.
Of the 335 BZCAT seeds introduced, 24 survived as LAT $\gamma$-ray sources in this catalog.  Of the 7 WMAP seeds, 3 remain in the catalog.

The variety of seeds that we used means that the catalog is not
homogeneous.
Because of the strong underlying diffuse emission, achieving a truly homogeneous catalog was not possible in any case.
Our aim was to provide enough seeds to 
allow the main maximum likelihood analysis (\S~\ref{run:TS}) to be the defining step of the catalog construction.
The total number of seeds was 2433.

\subsection{Localization}
\label{run:localization}

The localization of faint or soft sources is more sensitive to the diffuse emission
and to nearby sources than for brighter sources, so we proceeded in three steps 
instead of just one for the bright sources considered in the BSL:
\begin{enumerate}
\item The first step consisted of localizing the sources before
the main maximum likelihood analysis (\S~\ref{run:TS})
as we did for the BSL (using {\it pointfit}),
treating each source independently but in descending order of significance
and incorporating the bright sources into
the background for the fainter ones.
This is fast and provides a good enough starting point for step 2.
\item The second step consisted of improving the localization within the
main maximum likelihood analysis (\S~\ref{run:TS})
using the {\it gtfindsrc} utility
in the Science Tools\footnote{Available from the $Fermi$ Science Support Center, http://fermi.gsfc.nasa.gov/ssc.}. Again sources are considered in descending order
of significance. When localizing one source, the others are fixed in position,
but the fluxes and spectral indicies of sources within $2\degr$ are left free to accommodate the 
loss of low energy photons in the model if the source that is being localized moves away.
At the end of that step we have a good
representation of the location, flux and spectral shape of the sources
over the entire sky, but a single error radius to describe the error box.
\item The third step is new and described in more detail below.
It uses a similar framework as the first step, but incorporates
the results of the main maximum likelihood analysis for all sources
other than the one being considered, so it has a good representation
of the source's surroundings.
It is faster than {\it gtfindsrc} and {\it gttsmap}
and returns a full Test Statistic map around each source
and an elliptical representation as well as an indicator of
the quality of the elliptical fit.
\end{enumerate}

The first and third steps used a likelihood analysis tool ({\it pointfit}) that provides speed at little sacrifice of precision by maximizing a specially-constructed binned likelihood function.  Photons are assigned  to twelve energy bands (four per decade from 100~MeV to 100~GeV) and HEALpix-based spatial bins for which the size is selected to be small compared with the scale set by the PSF.  Since the PSF for $Front$-converting photons is significantly smaller than that for $Back$ conversions, there are separate spatial bins for $Front$ and $Back$. Note that the width of the PSF at a given energy is only a weak function of incidence angle.  For {\it pointfit} the likelihood function is evaluated using the PSF averaged over the full field of view for each energy band. For each band, we define the likelihood as a function of the position and flux of the assumed point source, and adopt as the background the sum of Galactic diffuse, isotropic diffuse (see \S~\ref{run:diffusemodel}) and any nearby (i.e., within 5$\degr$), other point sources in the catalog. The flux for each band is then evaluated by maximizing the likelihood of the data given the model using the coordinates defined by {\it gtfindsrc}. The overall likelihood function, as a function of the source position, is then the product of the band likelihoods. We define a function of the position $\mathbf{p}$, as  $2 (\log(L_{\rm max}) - \log(L(\mathbf{p}))$, where $L$ is the likelihood function described above. This function, according to Wilks' theorem \citep{wilks1938}, is the probability distribution for the coordinates of the point source consistent with the observed data.  Note that the width of this distribution is a measure of the uncertainty, and that it scales directly with the width of the PSF.
 
We then fit the distribution to a 2-dimensional quadratic form with 5 parameters describing the expected elliptical shape: the coordinates (R.A. and Dec.) of the center of the ellipse, semi-major and -minor axis extents ($\alpha$ and $\beta$), and the position angle $\phi$ of the ellipse\footnote{In the FITS version of the 1FGL catalog, $\alpha$ is {\tt CONF\_95\_SEMIMAJOR}, $\beta$ is {\tt CONF\_95\_SEMIMINOR}, and $\phi$ is {\tt CONF\_95\_PosAng}; see Appendix D}. A `quality' factor is evaluated to represent the goodness of the fit: it is the square root of the sum of the squares of the deviations for 8 points sampled along the contour where the value is expected to be 4.0, that is, 2 $\sigma$ from the maximum likelihood coordinates of the source. 
 
We quote the parameters of the ellipse that would contain 95\% of the probability for the location of the source; for Gaussian errors this would be a radius of 2.45 $\sigma$. An analysis of the deviations of 396 AGNs at high latitudes from the positions of the nearest LAT point sources indicated that the PSF width is underestimated, on average, by a factor of $1.10\pm 0.05$. Thus the final uncertainties reported by {\it pointfit} were scaled up by a factor of 1.1. To visually assess the fits, a Test Statistic map was made for each source, and these were considered in evaluating the analysis flags that are discussed in \S~\ref{run:flags}.

Twelve sources did not converge at the third step, converged to a point far away
($> 1\degr$) or were in crowded regions where the procedure (which does
not have free parameters for the fluxes of nearby sources) may not be reliable.
Those 12 were left at their {\it gtfindsrc} positions. They can be easily
identified in the 1FGL catalog because they have identical semimajor and semiminor axes for the source location uncertainty, and position angle 0.
The LAT-detected pulsars and X-ray binaries, which were placed at the high-precision positions of these identified sources,
have null values in the localization parameters.

\begin{figure}
\epsscale{.80}
\plotone{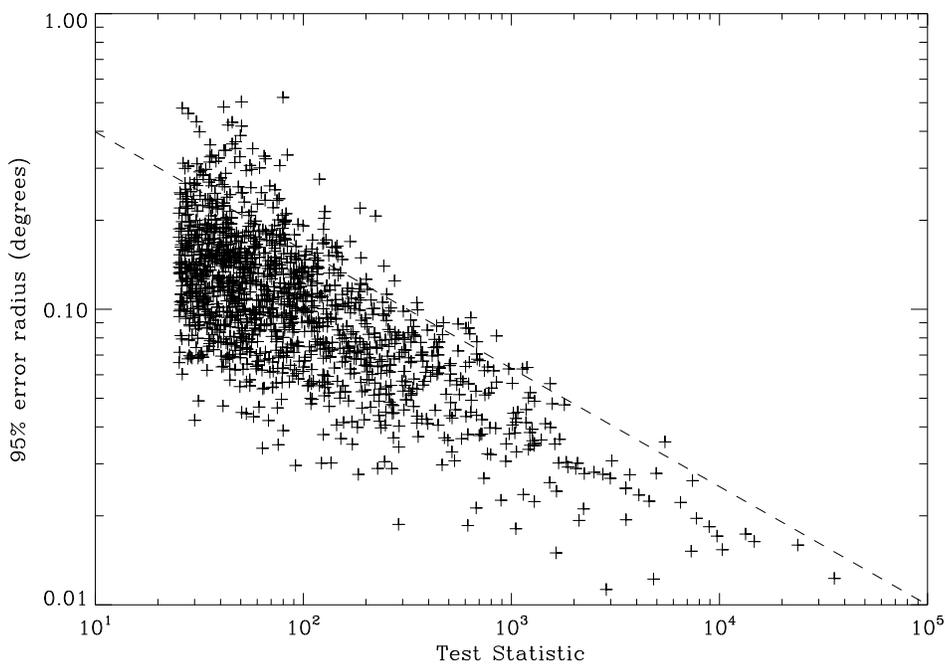}
\caption{95\% source location error
(geometric mean of the two axes of the ellipse)
as a function of Test Statistic (\S~\ref{run:TS}).
The dashed line is a $(TS)^{-0.4}$ trend for reference
(not adjusted vertically).}
\label{fig:loc}
\end{figure}

\begin{figure}
\epsscale{.80}
\plotone{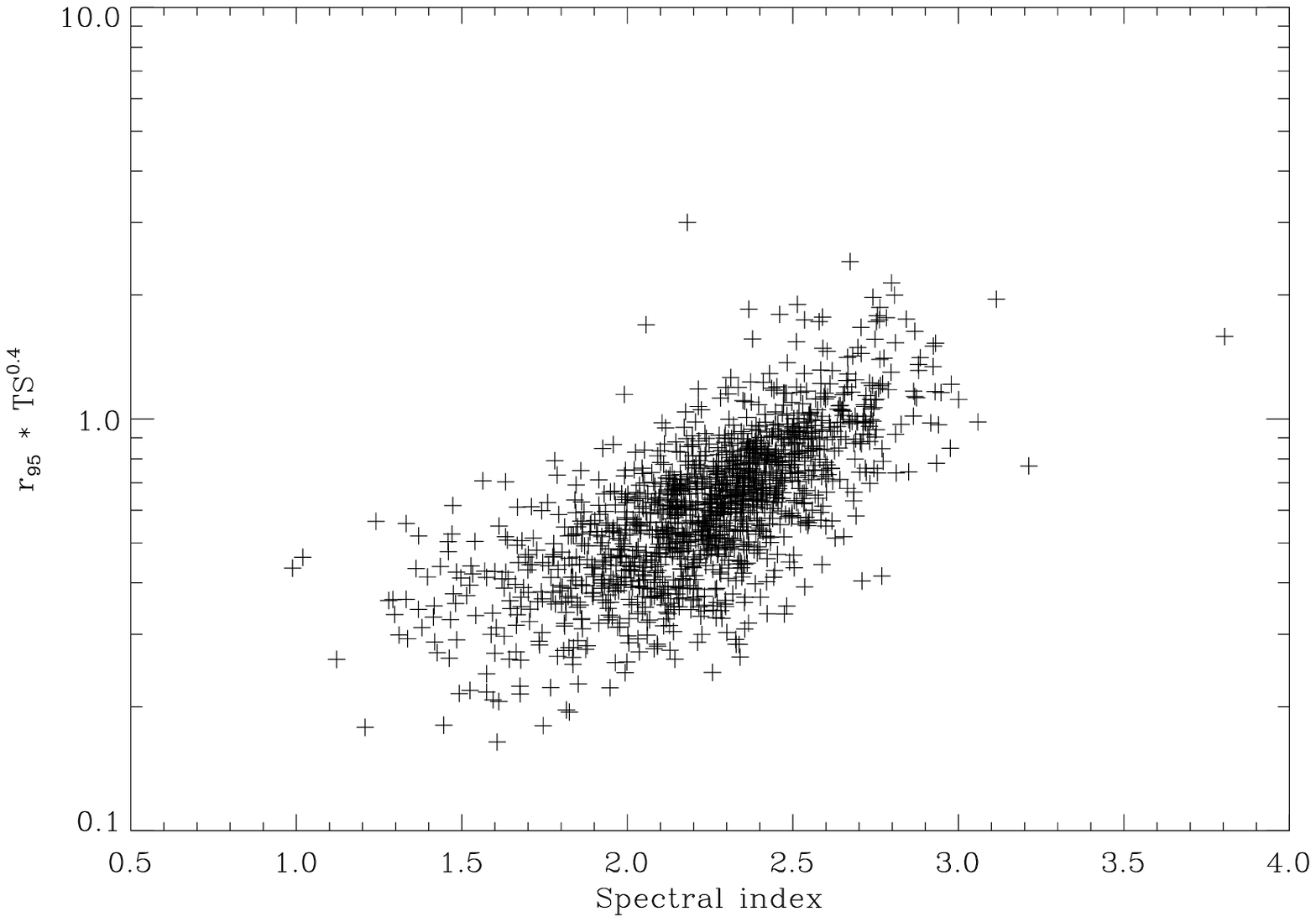}
\caption{95\% source location error
multiplied by $(TS)^{0.4}$ to remove the global trend (Fig.~\ref{fig:loc})
as a function of the photon spectral index from \S~\ref{run:TS}.}
\label{fig:locspec}
\end{figure}

Figure~\ref{fig:loc} illustrates the resulting position errors 
as a function of the Test Statistic $(TS)$ values obtained in
\S~\ref{run:TS}.
The relatively large dispersion that is seen at a given $TS$
is in part due to the local conditions (level of diffuse $\gamma$-ray emission) but 
primarily depends upon the source spectrum. Hard-spectrum sources are better localized
than soft ones for the same $TS$ (Fig.~\ref{fig:locspec})
because the PSF is so much
narrower at high energy.
At our threshold of $TS$ = 25 the typical 95\% position error is
about 10$\arcmin$, and most 95\% errors are below 20$\arcmin$.

\subsection{Significance and thresholding}
\label{run:TS}

The detection and localization steps provide estimates of source significances.  However, since the detection step does not use the energy information and the localization step fits only one source at a time, these estimates are not sufficiently accurate for use in the catalog.
To better estimate the source significances we use a 3-dimensional
maximum likelihood algorithm ({\it gtlike}) in unbinned mode
(i.e., the position and energy of
each event is considered individually)
applied on the full energy range from 100~MeV to 100~GeV using the P6\_V3 IRFs (see \S~\ref{run:LAT}). 
This is part of the standard Science Tools software package,
currently at version 9r15p5.
The tool does not vary the
source position, but does adjust the source spectrum.  The underlying
optimization engine is Minuit\footnote{http://lcgapp.cern.ch/project/cls/work-packages/mathlibs/minuit/doc/doc.html}.
The code works well with up to $\sim$30 free parameters,
an important consideration for regions where sources are close enough
together to partially overlap. The {\it gtlike} tool provides 
the best-fit parameters for each source
and the Test Statistic $TS=2\Delta$log(likelihood) between models
with and without the source.  The $TS$ associated with each source is a measure of the source significance.
Error estimates (and a full covariance matrix) are obtained from Minuit
in the quadratic approximation around the best fit.
For this stage we modeled the sources with simple power-law spectra.  It should be noted
that
{\bf \it gtlike}
does not include the energy dispersion in the $TS$ calculation
(i.e., it assumes that the measured energy is the true energy).
Given the 8 to 10\% energy resolution of the LAT over the wide energy bands
used in the present analyses, this approximation is justified.  

Because the fitted fluxes and spectra of the sources
can be very sensitive to even slight errors in the spectral shape of the
diffuse emission we allow the Galactic diffuse model (\S~\ref{run:diffusemodel})
to be corrected (i.e., multiplied) locally
by a power law in energy with free normalization and spectral slope.
The slope varies between 0 and 0.07 (making it harder) in the Galactic plane
and the normalization by $\pm$ 10\% (down from 0.15 and 20\% for the BSL).
The smaller excursions of that corrective slope when compared to the BSL
reflect the better fit of the current diffuse model to the data.
The normalization of the isotropic component of the diffuse emission
(which represents the extragalactic and residual backgrounds) was left free.
The three free parameters were separately adjusted in each
Region of Interest (RoI).

We split the sky into overlapping circular RoIs.
The parameters are free for sources
in the central part of each RoI
(RoI radius minus $7\degr$), 
such that all free sources are well within
the RoI even at low energy ($7\degr$ is larger than $r_{68}$ at 100~MeV).
It is advantageous (for the global convergence
over the entire sky) to use large RoIs, but at the same time smaller RoIs allow spectral variations of the diffuse emission relative to the model to be corrected in more detail.
We set the RoI sizes so that not more than
8 sources are free at a time. Adding 3 parameters for the diffuse
model, the total number of free parameters in each RoI is normally 19 at most.
We needed 445 RoIs to cover the 2433 seed positions.
The RoI radii range between $9\degr$ and $15\degr$.

We proceed iteratively.
All RoIs are processed in parallel and a global current model
is assembled after each step in which the best-fit parameters
for each source are taken from the RoI whose center is closest to the source.
The local model for each RoI includes sources up to $7\degr$ outside the RoI
(which can contribute at low energy due to the broad PSF).
Their parameters are fixed to their values in the global model
at the previous step.
The parameters of the sources inside the RoI but within $7\degr$
of the border are also fixed except in two cases
(not considerered for the BSL analysis):
\begin{itemize}
\item Sources within $2\degr$
of any source inside the central part, because they can influence the
inner source. $2\degr$ is chosen to be larger than twice the containment
radius at 1~GeV ($2 \times 0.8\degr$) where the LAT sensitivity peaks
(Fig.~ \ref{fig:tsweight}).
We leave both flux and spectral index free for these.
\item Very bright sources contributing more than 5\% of the total counts
in the RoI because they can influence the diffuse emission parameters.
We leave only the flux free for these.
\end{itemize}
All seed sources start at 0
flux at the first step; the starting point for the slope is 2.
We iterate over 5 steps; the fits change very little after the fourth.
To facilitate the convergence the seed sources are not entered all at once.
The brightest 10\% of the sources are entered at the first step, 30\% at the second step,
and finally all at the third step.
At each step we remove seed sources with low $TS$, raising the threshold
for inclusion into the global model
from 10 at the third step to 15 at the fourth and finally 25 at the last step.
All seeds are reentered at the fourth step to avoid losing faint sources
before the global model has fully converged.
We have checked via simulations that removing the faint sources
has little impact on the bright ones, much less than changing
the diffuse model (\S~\ref{run:systematics}).
This procedure left 1451 sources above threshold.  The variation of the detection threshold across the sky and the dependence of the threshold on source spectrum are discussed in Appendix A.

The $TS$ of each source can be related to the probability that such an excess can be obtained
from background fluctuations alone.
The probability distribution in such a situation (source over background)
is not known precisely \citep{pdc02}.
However since we consider only positive fluctuations, and each fit
involves four degrees of freedom
(two for position, plus flux and spectral index), 
the probability to get at least $TS$ at a given position in the sky
is close to 1/2 of the $\chi^2$ distribution with four degrees of freedom
\citep{mattox96}, so that $TS$ = 25
corresponds to a false detection probability of $2.5 \times 10^{-5}$ or
4.1 $\sigma$ (one sided).
For the BSL we considered only two degrees of freedom because the localization
was based on a simpler algorithm which did not involve explicit minimization
of the same likelihood function.

The sources that we see are best (most strongly) detected around 1~GeV. This is approximately the median of the Pivot\_Energy
quantity in the catalog, i.e., the energy at which the uncertainties in normalization and spectral index for the power-law fit are uncorrelated.  At 1~GeV the 68\% containment radius is
approximately $r_{68} = 0.8\degr$. The number of independent elements
in the sky (trials factor) is about $4 \pi / (\pi r_{68}^2)$ in which $r_{68}$
is converted to radians. This is about $2 \times 10^4$ so at a threshold
of $TS$ = 25 we expect less than 1 spurious source by chance only.
If any, there might be a few very hard spurious sources in the catalog
because hard sources have a smaller effective PSF so that the trials factor
is larger. The main reason for potentially spurious sources, though,
is our imperfect knowledge of the underlying diffuse emission
(\S~\ref{run:systematics}).

\subsection{Flux Determination}
\label{run:flux}

The maximum likelihood method described in \S~\ref{run:TS}
provides good estimates of the source significances
and the overall spectral slope, but not very
accurate estimates of the fluxes. This is because the spectra of most sources
do not follow a single power law over that broad an energy range
(three decades). Within the two most populous categories,
the AGN often have broken power-law spectra and the pulsars have power-law spectra with an exponential cutoff.
In both cases fitting a single power law over the entire range over-predicts the flux in the low-energy region of the spectrum, which contains the majority of the photons from the source, biasing the fluxes high.
On the other hand the effect on the significance is low
due to the broad PSF and high background at low energies.

In addition, the significance is mostly obtained from GeV photons
(Fig.~\ref{fig:tsweight}) whereas the photon flux in the full range
(above 100~MeV) is dominated by lower energy events so that the uncertainty
on that flux can be quite large even for highly significant sources.
For example, the typical relative uncertainty on the photon flux above 100~MeV
is 23\% for a $TS = 100$ source with spectral index 2.2.

To provide better estimates of the source fluxes, we decided
to split the range into five energy bands
from 100 to 300~MeV, 300~MeV to 1~GeV, 1 to 3~GeV, 3 to 10~GeV
and 10 to 100~GeV (the number of counts does not justify dividing the last decade into two bands).
The list of sources remained the same in all bands.
It is generally not possible to fit the spectral index
in each of those relatively narrow energy bands
(and the flux estimate does not depend very much on the index),
so we simply froze the spectral
index of each source to the best fit over the full interval.
The spectral bias to the Galactic diffuse emission
(\S~\ref{run:TS}) was also frozen.

The estimate from the sum of the five bands is on average within 30\%
of the flux obtained from the global power-law fit (as described in \S~\ref{run:LAT}, with excursions
up to a factor of 2.
We have also compared those estimates with a more precise
spectral model for the three bright pulsars
(Vela, Geminga and the Crab). The sum of the five fluxes is
within 5\% of the more precise flux estimate, whereas the power-law
estimate is 25\% too high for Vela and Geminga.
However because it is not based on extrapolating a relatively well defined
power-law fit the relative uncertainty on that flux is even larger
than that on the power-law fit, typically 50\% for a $TS = 100$ source
with spectral index 2.2.
For that reason we do not show this very poorly measured quantity
in Table~\ref{tab:sources}.
We provide instead the photon flux between 1 and 100~GeV (the sum of
the three high energy bands), which is much better defined.
The relative uncertainty on this flux is typically 18\% for a $TS = 100$ source
with spectral index 2.2.

\begin{deluxetable}{lrrrrrrrrrrrrrrccclcclc}
\setlength{\tabcolsep}{0.02in}
\tabletypesize{\scriptsize}
\rotate
\tablewidth{0pt}
\tablecaption{LAT 1FGL Catalog\label{tab:sources}}
\tablehead{
\colhead{Name 1FGL} &
\colhead{R.A.} &
\colhead{Decl.} &
\colhead{$l$} &
\colhead{$b$} &
\colhead{$\theta_{\rm 1}$} &
\colhead{$\theta_{\rm 2}$} &
\colhead{$\phi$} &
\colhead{$\sigma$} &
\colhead{$F_{35}$} &
\colhead{$\Delta F_{35}$} &
\colhead{$S_{25}$} &
\colhead{$\Delta S_{25}$} &
\colhead{$\Gamma_{25}$} &
\colhead{$\Delta \Gamma_{25}$} &
\colhead{Curv.} &
\colhead{Var.} &
\colhead{Flags} &
\colhead{$\gamma$-ray Assoc.} &
\colhead{TeV} &
\colhead{Class} &
\colhead{ID or Assoc.} &
\colhead{Ref.}
}
\startdata
 J0000.8+6600c &   0.209 & 66.002 & 117.812 & 3.635 & 0.112 & 0.092 & $-$73 &      9.8 &     2.9 &   0.6 &   35.2 &    5.7 & 2.60 & 0.09 & \nodata & \nodata & 6 & \nodata & \nodata & \nodata & \nodata & \nodata \\
 J0000.9$-$0745 &   0.236 & $-$7.763 &  88.903 & $-$67.237 & 0.179 & 0.130 & 16 &      5.6 &     1.0 &   0.0 &    9.2 &    3.0 & 2.41 & 0.20 & \nodata & \nodata & \nodata & \nodata & \nodata & bzb & CRATES J0001$-$0746 & \nodata \\
 J0001.9$-$4158 &   0.482 & $-$41.982 & 334.023 & $-$72.028 & 0.121 & 0.116 & 53 &      5.5 &     0.5 &   0.2 &   14.4 &    0.0 & 1.92 & 0.25 & \nodata & \nodata & \nodata & \nodata & \nodata & \nodata & \nodata & \nodata \\
 J0003.1+6227 &   0.798 & 62.459 & 117.388 & 0.108 & 0.119 & 0.112 & $-$19 &      7.8 &     2.1 &   0.5 &   19.9 &    4.9 & 2.53 & 0.10 & T & \nodata & 3 & \nodata & \nodata & \nodata & \nodata & \nodata \\
 J0004.3+2207 &   1.081 & 22.123 & 108.757 & $-$39.448 & 0.183 & 0.157 & 58 &      4.7 &     0.6 &   0.2 &    5.3 &    2.5 & 2.35 & 0.21 & \nodata & \nodata & \nodata & \nodata & \nodata & \nodata & \nodata & \nodata \\
 J0004.7$-$4737 &   1.187 & $-$47.625 & 323.864 & $-$67.562 & 0.158 & 0.148 & $-$5 &      6.6 &     0.8 &   0.3 &   10.9 &    3.3 & 2.56 & 0.17 & \nodata & \nodata & \nodata & \nodata & \nodata & bzq & PKS 0002$-$478 & \nodata \\
 J0005.1+6829 &   1.283 & 68.488 & 118.689 & 5.999 & 0.443 & 0.307 & $-$4 &      6.1 &     1.4 &   0.5 &   17.0 &    4.8 & 2.58 & 0.12 & \nodata & \nodata & 1,4 & \nodata & \nodata & \nodata & \nodata & \nodata \\
 J0005.7+3815 &   1.436 & 38.259 & 113.151 & $-$23.743 & 0.216 & 0.186 & 32 &      8.4 &     0.6 &   0.3 &   13.6 &    3.1 & 2.86 & 0.13 & \nodata & \nodata & \nodata & \nodata & \nodata & bzq & B2 0003+38A & \nodata \\
 J0006.9+4652 &   1.746 & 46.882 & 115.082 & $-$15.311 & 0.194 & 0.124 & 32 &     10.2 &     1.1 &   0.3 &   18.3 &    3.4 & 2.55 & 0.11 & \nodata & \nodata & \nodata & \nodata & \nodata & \nodata & \nodata & \nodata \\
 J0007.0+7303 &   1.757 & 73.052 & 119.660 & 10.463 & \nodata & \nodata & \nodata &    119.7 &    63.4 &   1.5 &  432.5 &   10.1 & 1.97 & 0.01 & T & \nodata & \nodata & 0FGL J0007.4+7303     & \nodata & PSR & LAT PSR J0007+7303 & 1,2,3 \\
  &  &  &  &  &  &  &  &  &  &  &  &  &  &  &  &  &  & EGR J0008+7308        &  &  & \\
  &  &  &  &  &  &  &  &  &  &  &  &  &  &  &  &  &  & 1AGL J0006+7311       &  &  & \\
 J0008.3+1452 &   2.084 & 14.882 & 107.655 & $-$46.708 & 0.144 & 0.142 & $-$42 &      4.7 &     0.8 &   0.2 &    9.6 &    0.0 & 2.00 & 0.21 & \nodata & \nodata & \nodata & \nodata & \nodata & \nodata & \nodata & \nodata \\
 J0008.9+0635 &   2.233 & 6.587 & 104.426 & $-$54.751 & 0.120 & 0.114 & 65 &      5.0 &     0.8 &   0.0 &    6.1 &    3.0 & 2.28 & 0.22 & \nodata & \nodata & \nodata & \nodata & \nodata & bzb & CRATES J0009+0628 & \nodata \\
 J0009.1+5031 &   2.289 & 50.520 & 116.089 & $-$11.789 & 0.119 & 0.108 & 72 &      8.5 &     1.3 &   0.3 &   15.6 &    3.4 & 2.41 & 0.13 & \nodata & \nodata & \nodata & \nodata & \nodata & \nodata & \nodata & \nodata \\
\enddata
\tablerefs{1 \citet{LATCTA1}, 2 \citet{LATPulsars}, 3 \citet{LATBSPs}}
\tablecomments{Photon flux units for $F_{35}$ are $10^{-9}$ cm$^{-2}$
 s$^{-1}$; energy flux units for $S_{25}$ are $10^{-12}$ erg cm$^{-2}$ s$^{-1}$.
The prefix ``FRBA'' in the column of source associations refers to sources observed at 8.4 GHz as part of VLA program AH996 (``Finding and Rejecting Associations for {\it Fermi}-LAT $\gamma$-ray sources'').
This table is published in its entirety in the electronic edition of the Astrophysical Journal Supplements.  A portion is shown here for guidance regarding its form and content.}
\end{deluxetable}

In contrast, the energy flux over the full band is better defined
than the photon flux
because it does not depend as much on the poorly-measured low-energy fluxes.
So we provide this quantity in Table~\ref{tab:sources}.
Here again the sum of the energy fluxes in the five bands provides
a more reliable estimate of the overall flux than the power-law fit.
The relative uncertainty on the energy flux between 100~MeV and 100~GeV
is typically 26\% for a $TS = 100$ source
with spectral index 2.2.

An additional difficulty that does not exist when considering the full data
is that, because we wish to provide the fluxes in all bands for all sources,
we must handle the case of sources that are not significant in one
of the bands. Many sources have $TS < 10$ in one or several bands:
1135 in the 100 to 300~MeV band, 630 in the 300~MeV to 1~GeV band, 
359 in the 1 to 3~GeV band, 503 in the 3 to 10~GeV band and 800
in the 10 to 100~GeV band.
There are even a number of sources which have upper limits in all bands,
even though they are formally significant (as defined in \S~\ref{run:TS})
over the full energy range.
It is particularly difficult to measure fluxes below 300~MeV because
of the large source confusion and the modest effective area of the LAT
at those energies with the current event cuts (\S~\ref{run:LAT}).
For the sources with poorly-measured fluxes
(where $TS < 10$ or the nominal uncertainty of the flux is larger than half the flux itself),
we replace the flux value from the likelihood analysis
by a 2 $\sigma$ upper limit,
indicating the upper limit by a 0 in the flux uncertainty column
of Table~\ref{tab:bandfluxes}; the corresponding columns of the FITS version of the 1FGL catalog are described in Appendix D.
The upper limit is obtained by looking for 2$\Delta$log(likelihood) = 4
when increasing the flux from the maximum-likelihood value.
When the maximum-likelihood value is very close to 0 (i.e., the flux that maximizes the likelihood would be negative), solving 2$\Delta$log(likelihood) = 4
tends to underestimate the upper limit.
Whenever $TS < 1$ we switch to the Bayesian method proposed
by \citet{helene83}. We do not use that method throughout because it is
about five times slower to compute.

\begin{deluxetable}{lrrcrrrrrrrrrrrrrrr}
\setlength{\tabcolsep}{0.03in}
\tabletypesize{\scriptsize}
\tablecaption{First LAT Catalog:  Spectral Information\label{tab:bandfluxes}}
\tablewidth{0pt}
\tablehead{
\colhead{} & \colhead{}  & \colhead{}  & \colhead{} & \multicolumn{3}{c}{100~MeV -- 300~MeV} & \multicolumn{3}{c}{300~MeV -- 1 GeV} & \multicolumn{3}{c}{1 GeV -- 3 GeV} & \multicolumn{3}{c}{3 GeV -- 10 GeV} & \multicolumn{3}{c}{10 GeV -- 100 GeV} \\ \cline{5-7} \cline{8-10} \cline{11-13} \cline{14-16} \cline{17-19} \\
\colhead{Name 1FGL} &
\colhead{$\Gamma$} &
\colhead{$\Delta \Gamma$} &
\colhead{Curv.} &
\colhead{$F_{\rm 1}$\tablenotemark{a}} &
\colhead{$\Delta F_{\rm 1}$\tablenotemark{a}} &
\colhead{$\sqrt{TS_{\rm 1}}$} &
\colhead{$F_{\rm 2}$\tablenotemark{a}} &
\colhead{$\Delta F_{\rm 2}$\tablenotemark{a}} &
\colhead{$\sqrt{TS_{\rm 2}}$} &
\colhead{$F_{\rm 3}$\tablenotemark{b}} &
\colhead{$\Delta F_{\rm 3}$\tablenotemark{b}} &
\colhead{$\sqrt{TS_{\rm 3}}$} &
\colhead{$F_{\rm 4}$\tablenotemark{c}} &
\colhead{$\Delta F_{\rm 4}$\tablenotemark{c}} &
\colhead{$\sqrt{TS_{\rm 4}}$} &
\colhead{$F_{\rm 5}$\tablenotemark{c}} &
\colhead{$\Delta F_{\rm 5}$\tablenotemark{c}} &
\colhead{$\sqrt{TS_{\rm 5}}$} 
}
\startdata
 J0000.8+6600c & 2.60 & 0.09 & \nodata &    8.3 &    0.0 &    2.6 &    1.9 &    0.3 &    6.4 &    2.6 &    0.6 &    5.2 &    3.7 &    1.5 &    4.1 &    0.8 &    0.0 &    0.0 \\
 J0000.9$-$0745  & 2.41 & 0.20 & \nodata &    2.9 &    0.0 &    2.5 &    0.5 &    0.0 &    3.0 &    0.7 &    0.0 &    2.2 &    3.8 &    0.0 &    3.7 &    1.5 &    0.0 &    2.4 \\
 J0001.9$-$4158  & 1.92 & 0.25 & \nodata &    2.1 &    0.0 &    1.8 &    0.2 &    0.0 &    1.1 &    0.6 &    0.0 &    2.2 &    2.9 &    1.1 &    6.0 &    1.7 &    0.0 &    0.0 \\
 J0003.1+6227  & 2.53 & 0.10 & T &    3.0 &    0.0 &    0.0 &    1.7 &    0.3 &    6.8 &    2.0 &    0.5 &    5.0 &    4.3 &    0.0 &    1.4 &    1.4 &    0.0 &    1.9 \\
 J0004.3+2207  & 2.35 & 0.21 & \nodata &    1.8 &    0.0 &    0.9 &    0.4 &    0.0 &    1.8 &    0.4 &    0.2 &    3.2 &    1.9 &    0.9 &    3.8 &    0.9 &    0.0 &    0.0 \\
 J0004.7$-$4737  & 2.56 & 0.17 & \nodata &    2.4 &    0.8 &    3.3 &    0.3 &    0.1 &    3.6 &    0.8 &    0.2 &    5.0 &    2.0 &    0.0 &    2.0 &    1.5 &    0.0 &    0.2 \\
 J0005.1+6829  & 2.58 & 0.12 & \nodata &    3.9 &    0.0 &    0.6 &    1.3 &    0.3 &    5.3 &    2.2 &    0.0 &    3.1 &    4.9 &    0.0 &    2.0 &    1.0 &    0.0 &    0.0 \\
 J0005.7+3815  & 2.86 & 0.13 & \nodata &    3.3 &    0.9 &    3.7 &    0.5 &    0.1 &    4.4 &    1.1 &    0.0 &    3.1 &    2.5 &    0.0 &    1.6 &    1.1 &    0.0 &    0.0 \\
 J0006.9+4652  & 2.55 & 0.11 & \nodata &    2.9 &    0.9 &    3.3 &    0.7 &    0.1 &    6.4 &    0.8 &    0.3 &    3.7 &    3.0 &    1.3 &    4.3 &    1.5 &    0.0 &    2.1 \\
 J0007.0+7303  & 1.97 & 0.01 & T &   20.9 &    1.2 &   19.8 &   12.0 &    0.3 &   60.5 &   49.0 &    1.3 &   82.3 &  135.6 &    6.5 &   57.6 &    8.5 &    1.6 &   15.3 \\
 J0008.3+1452  & 2.00 & 0.21 & \nodata &    0.9 &    0.0 &    0.0 &    0.2 &    0.0 &    0.5 &    0.6 &    0.2 &    3.8 &    2.0 &    0.9 &    4.5 &    1.2 &    0.0 &    0.0 \\
 J0008.9+0635  & 2.28 & 0.22 & \nodata &    1.6 &    0.0 &    0.3 &    0.5 &    0.0 &    3.1 &    0.5 &    0.0 &    1.0 &    2.2 &    1.0 &    4.0 &    1.5 &    0.0 &    2.9 \\
 J0009.1+5031  & 2.41 & 0.13 & \nodata &    4.0 &    0.0 &    2.3 &    0.5 &    0.1 &    4.6 &    0.9 &    0.3 &    4.6 &    3.5 &    1.2 &    5.1 &    1.4 &    0.0 &    2.0 \\
 J0011.1+0050  & 2.51 & 0.15 & \nodata &    1.2 &    0.0 &    0.0 &    0.4 &    0.1 &    4.8 &    0.5 &    0.2 &    4.1 &    1.7 &    0.0 &    0.3 &    0.9 &    0.0 &    0.0 \\
 J0013.1$-$3952  & 2.09 & 0.22 & \nodata &    1.2 &    0.0 &    0.1 &    0.3 &    0.0 &    1.4 &    0.9 &    0.0 &    2.9 &    2.3 &    0.0 &    1.4 &    2.1 &    0.0 &    4.3 \\
\enddata
\tablecomments{This table is published in its entirety in the electronic edition of the Astrophysical Journal Supplements.  A portion is shown here for guidance regarding its form and content.}
\tablenotetext{a}{In units of $10^{-8}$ photons cm$^{-2}$ s$^{-1}$}
\tablenotetext{b}{In units of $10^{-9}$ photons cm$^{-2}$ s$^{-1}$}
\tablenotetext{c}{In units of $10^{-10}$ photons cm$^{-2}$ s$^{-1}$}
\end{deluxetable}

\begin{figure}
\epsscale{.80}
\plotone{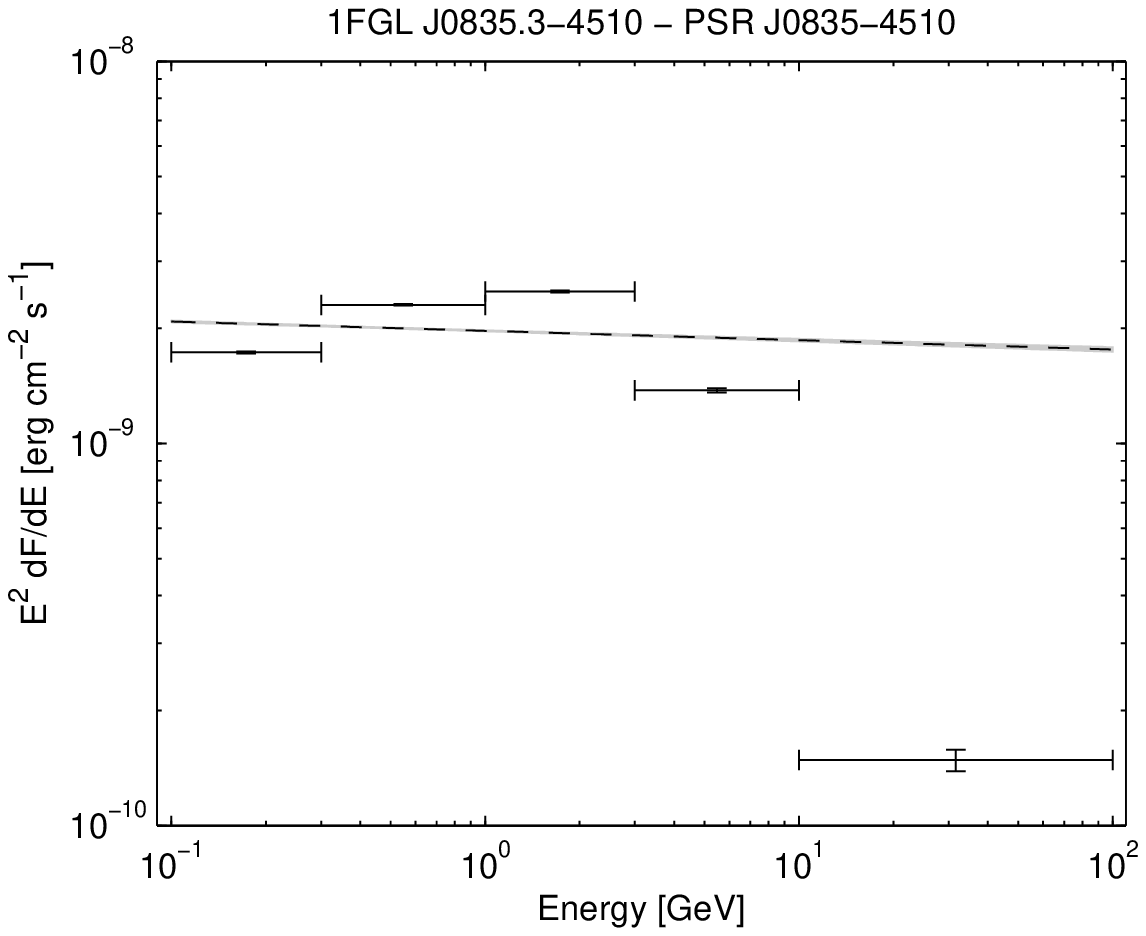}
\caption{Sample spectrum of Vela (1FGL J0835.3$-$4510) generated from the five energy-band flux measurements in the catalog and plotted as $E_i^2\Delta F_i/\Delta E_i$, with $E_i$ chosen to be the center of the energy bin in log space. The energy range of the integration is indicated by a horizontal bar. The vertical bar indicates the statistical error on the flux. The point at which these bars cross is not the same as the differential power per unit log bandwidth, $E^2dF/dE$ at $E_i$. The dashed lines (nearly coincident for this very bright source) reflect the uncertainties on the flux and index of the power-law fit to the full energy range in \S~\ref{run:TS}.}
\label{fig:spec_vela}
\end{figure}

\begin{figure}
\epsscale{.80}
\plotone{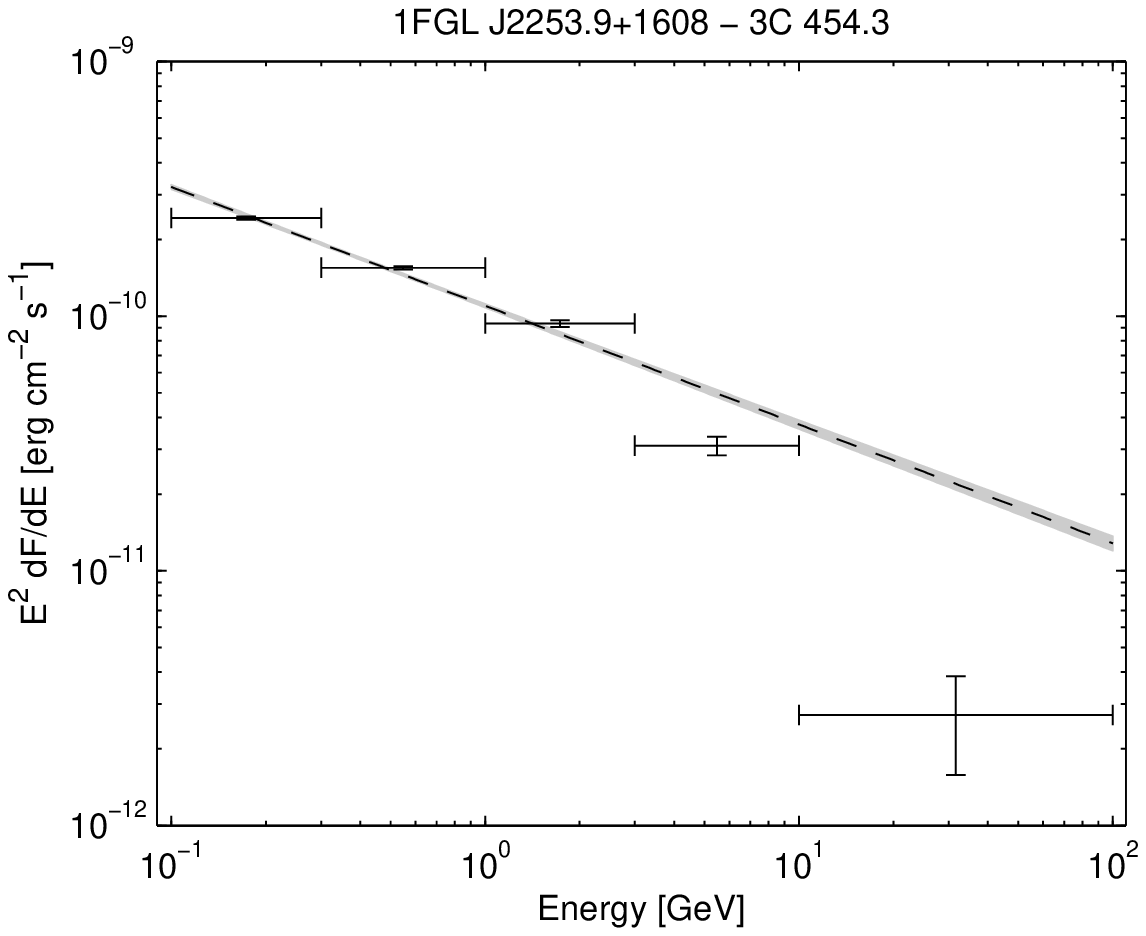}
\caption{Spectrum of the bright blazar 3C~454.3 (1FGL J2253.9+1608).}
\label{fig:spec_3C454}
\end{figure}

The five fluxes provide a rough spectrum, allowing
departures from a power law to be judged.
This is the main advantage over the BSL scheme
which involved only two bands.
Examples of those rough spectra are given in Figures~\ref{fig:spec_vela} and
\ref{fig:spec_3C454} for a bright pulsar (Vela) and
a bright blazar (3C~454.3).
In order to quantify departures from a power-law shape, we introduce
a Curvature\_Index 
\begin{equation}
\label{eq:curindex}
C = \sum_i \frac{(F_i - F_i^{\rm PL})^2}{\sigma_i^2 + (f_i^{\rm rel} F_i)^2}
\end{equation}
where $i$ runs over all bands and $F_i^{\rm PL}$ is the flux predicted
in that band from the global power-law fit.
$f_i^{\rm rel}$ reflects the relative systematic uncertainty on effective area
described in \S~\ref{run:systematics}. It is set to 10, 5, 10, 15 and 20\%
in the bands [0.1,0.3], [0.3,1], [1,3], [3,10] and [10,100]~GeV respectively.
Note that this systematic uncertainty on the effective area is not included
in the uncertainties reported in Table~\ref{tab:bandfluxes} (or in the FITS file), because
this systematic factor cancels when comparing each of the band fluxes between different sources.
We use for $F_i$ and $\sigma_i$
the best-fit and 1 $\sigma$ estimates even when the values are reported
as upper limits in the table, both for computing the Curvature\_Index
and the sums (photon flux and energy flux). 

Since the power-law fit involves two parameters
(normalization and spectral index), $C$ would be expected to follow
a $\chi^2$ distribution with $5 - 2 = 3$ degrees of freedom if the
power-law hypothesis was true. At the 1\% confidence level, the spectral
shape is significantly different from a power law if $C > 11.34$.
That condition is met by 225 sources
(at 1\% confidence, we expect 15 false positives).
The curvature index is by no means an estimate of curvature itself,
just a statistical indicator. A faint source
with a strongly curved spectrum can have the same curvature index
as a bright source with a slightly curved spectrum.
Since the relative uncertainties on the fluxes in each band are quite
different and depend on the spectral index itself, it is difficult
to build a curvature indicator similar to the fractional variability
for the light curves.
The curvature index is also not exclusively an indicator of curvature.  Any kind of deviation from the best fit
power-law can trigger that index, although curvature is by far the most common.

\subsection{Variability}
\label{run:variability}

Variability is very common at $\gamma$-ray energies (particularly among
accreting sources) and it is useful to estimate it.
To that end we derive a variability index for each source by splitting the LAT data into a number of time intervals and deriving a flux for each source in each interval, using the same energy range as in \S~\ref{run:TS}
(100~MeV to 100~GeV).  We split the full 11-month interval into $N_{\rm int}$ = 11 intervals
of about one month each (2624~ks or 30.37~days).
This is much more than the week used in the BSL,
in order to preserve some statistical precision for the
majority of faint sources we are dealing with here.
It is also far enough from half the precession period of the orbit ($\approx0.5\times53.4$ = 26.7 days)
that we do not expect possible systematic effects
as a function of off-axis angle
to be coherent with those intervals.

To avoid ending up with too large error bars
in relatively short time intervals, we froze the spectral
index of each source to the best fit over the full interval.
Sources do vary in spectral shape as well as in flux, of course, but
we do not aim at characterizing source variability here, just detecting it.
It is very unlikely that a true variability in shape will be such
that it will not show up in flux at all.
In addition, little spectral variability was found in bright AGN
where it would be detectable if present \citep{LATAGNSpectra}.
Because we do not expect the diffuse emission to vary, we freeze
the spectral adjustment of the Galactic diffuse component to the local
(in the same RoI) best-fit index from the full interval.
So the fitting procedure is the same as in \S~\ref{run:flux}
with all spectral shape parameters frozen.

The variability index is defined as a simple $\chi^2$ criterion:
\begin{eqnarray}
w_i = \frac{1}{\sigma_i^2 + (f_{\rm rel} F_i)^2} \\
F_{\rm wt} = \frac{\sum_i w_i F_i}{\sum_i w_i} \\
V = \sum_i w_i (F_i - F_{\rm wt})^2
\label{eq:varindex}
\end{eqnarray}
where $i$ runs over the 11 intervals and
$\sigma_i$ is the statistical uncertainty in $F_i$.
As for the BSL we have added in quadrature
a fraction $f_{\rm rel}$ = 3\% of the flux for each interval  $F_{\rm i}$
to the statistical error estimates $\sigma_i$ (for each 1-month time interval)
used to compute the variability index\footnote{In the FITS version of the 1FGL catalog, the {\tt Flux\_History} and {\tt Unc\_Flux\_History} columns contain $F_i$ and $\sqrt{\sigma_i^2 + (f_{\rm rel} F_i)^2}$, respectively; see Table~\ref{tab:columns}}.
Since the weighted average flux $F_{\rm wt}$ is not known a priori, $V$ is expected,
in the absence of variability, to follow a $\chi^2$ distribution
with 10 (= $N_{\rm int} -$ 1) degrees of freedom.
At the 1\% confidence level, the light curve
is significantly different from a flat one if $V > 23.21$.
That condition is met by 241 sources
(at 1\% confidence, we expect 15 false positives).
For those sources we provide directly in the FITS version of the table the maximum monthly flux
(Peak\_Flux) and its uncertainty, as well as the time when it occurred
(Time\_Peak); see Table~\ref{tab:columns} for the column specifications.

As in \S~\ref{run:flux} it often happens that a source is not
significant in all intervals. To preserve the variability index
(Eq.~\ref{eq:varindex}) we keep the best-fit value and its estimated
error even when the source is not significant. This does not work,
however, when the best fit is close to zero
because in that case the log(likelihood) as a function of flux
is very asymmetric.
Whenever $TS < 10$ or
the nominal flux uncertainty is larger than half the flux itself
we compute the 2 $\sigma$ upper limit and replace the
error estimate for that interval ($\sigma_i$) with
half the difference between that upper limit and the best fit.
This is an estimate of the error on the positive side only.
Because the parabolic extrapolation often exceeds the log(likelihood) profile
at 2 $\sigma$ this is more conservative than computing
the 1 $\sigma$ upper limit directly.
The best fit itself is retained.
Note that this error estimate can be a large overestimate of the error
on the negative side, particularly in the deep Poisson regime at high energy.
This explains why $\sigma_i/F_i$ can be as high as 1
even when $TS$ is 4 or so in that interval.
As in \S~\ref{run:flux} we switch to the Bayesian method whenever $TS < 1$.

\begin{figure}
\plotone{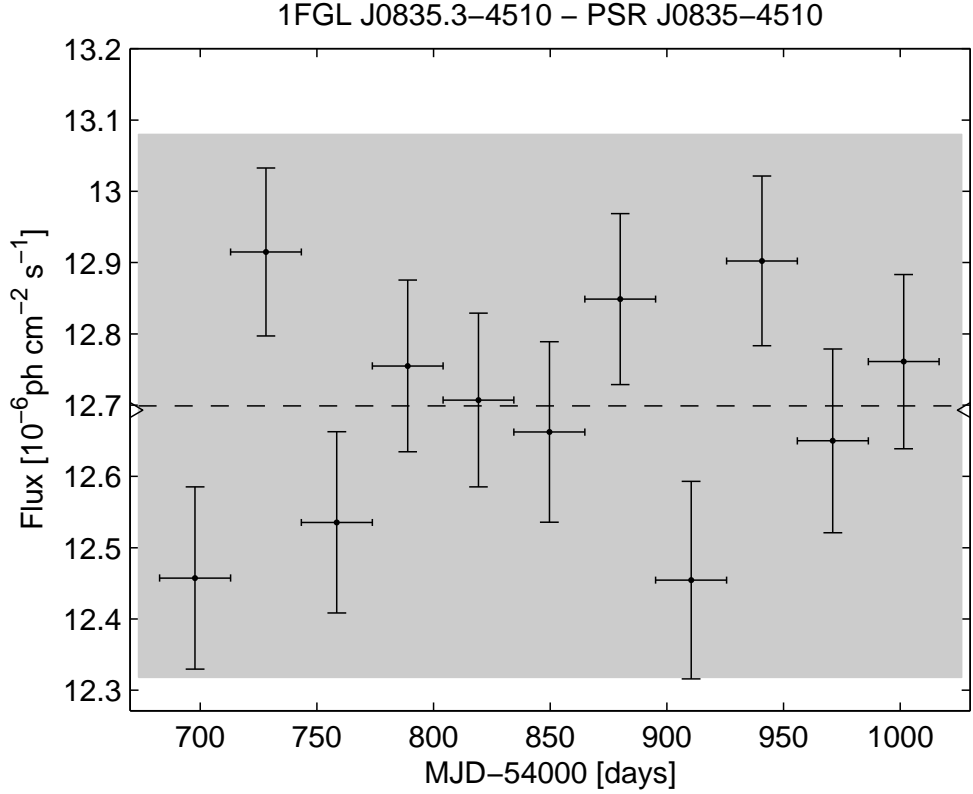}
\caption{Light curve of Vela (1FGL J0835.3$-$4510) for the 11-month interval analyzed for the 1FGL catalog.  The fluxes are integrated from 100~MeV to 100~GeV using single power-law fits and the error bars indicate the 1 $\sigma$ statistical errors.  The grey band shows the time-averaged flux with the conservative 3\% systematic error that we have adopted for evaluating the variability index.  Vela is not seen to be variable even at the level of the statistical uncertainty.  The spectrum of Vela is not well described by a power law and the fluxes shown here overestimate the true flux, but the overestimate does not depend on time.}
\label{fig:lc_vela}
\end{figure}

\begin{figure}
\plotone{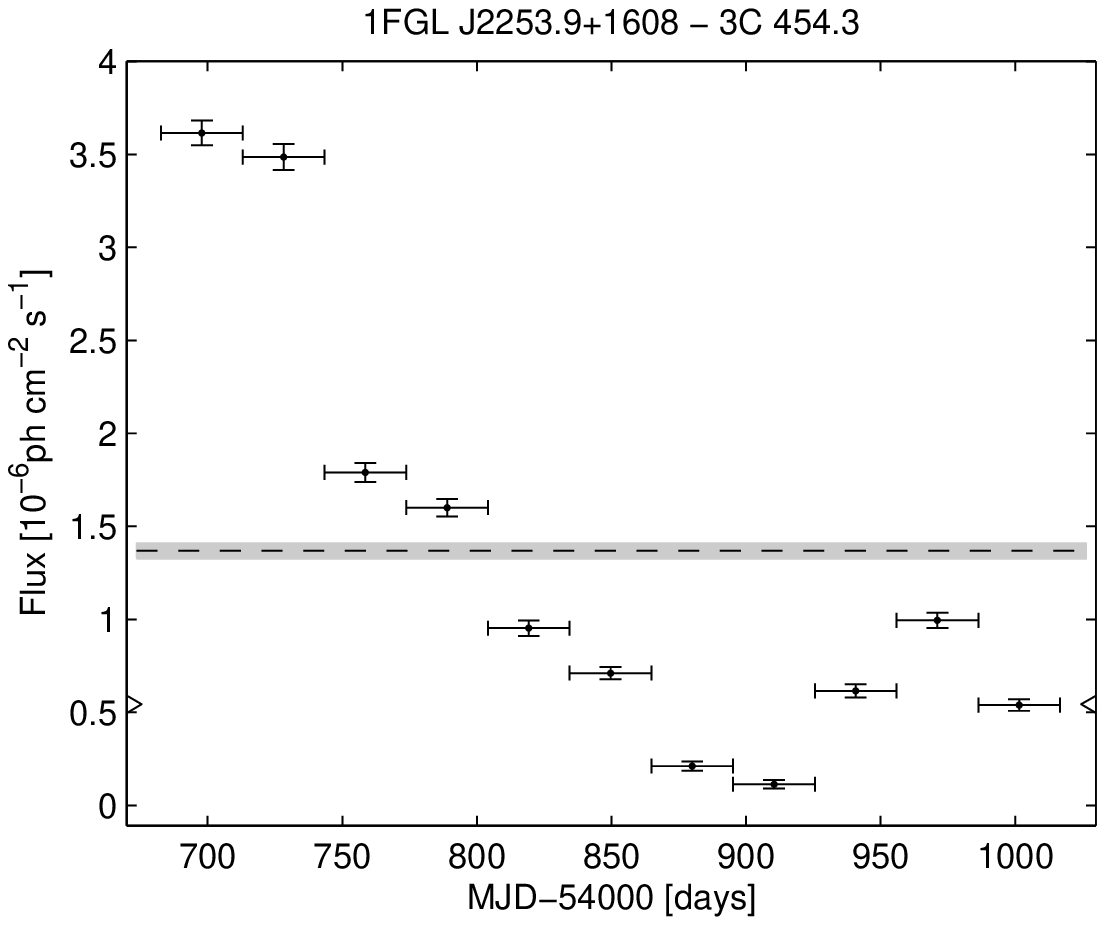}
\caption{Light curve of 3C~454.3 (1FGL J2253.9+1608), which exhibits extreme variability.  The grey band is the same 3\% systematic uncertainty that we have adopted for evaluating the variability index.  The triangles on the left and right indicate the value of the weighted average flux $F_{\rm wt}$ that minimizes $V$ in Eq.~\ref{eq:varindex}.  Owing to the systematic uncertainty term, for bright, highly-variable sources $F_{\rm wt}$ can differ from the time-averaged flux (which we derive from a power-law fit to the integrated data set).  }
\label{fig:lc_3C454}
\end{figure}

Examples of light curves are given in Figures~\ref{fig:lc_vela} and
\ref{fig:lc_3C454} for a bright constant source (the Vela pulsar) and
a bright variable source (the blazar 3C~454.3).  With the 3\% systematic relative uncertainty no pulsar is found to be variable.
The very brightest pulsars (Vela and Geminga) appear to have observed
variability below 3\%,
so this may be overly conservative.
It is not a critical parameter though,
as it affects only the very brightest sources.

\begin{figure}
\epsscale{1.1}
\plotone{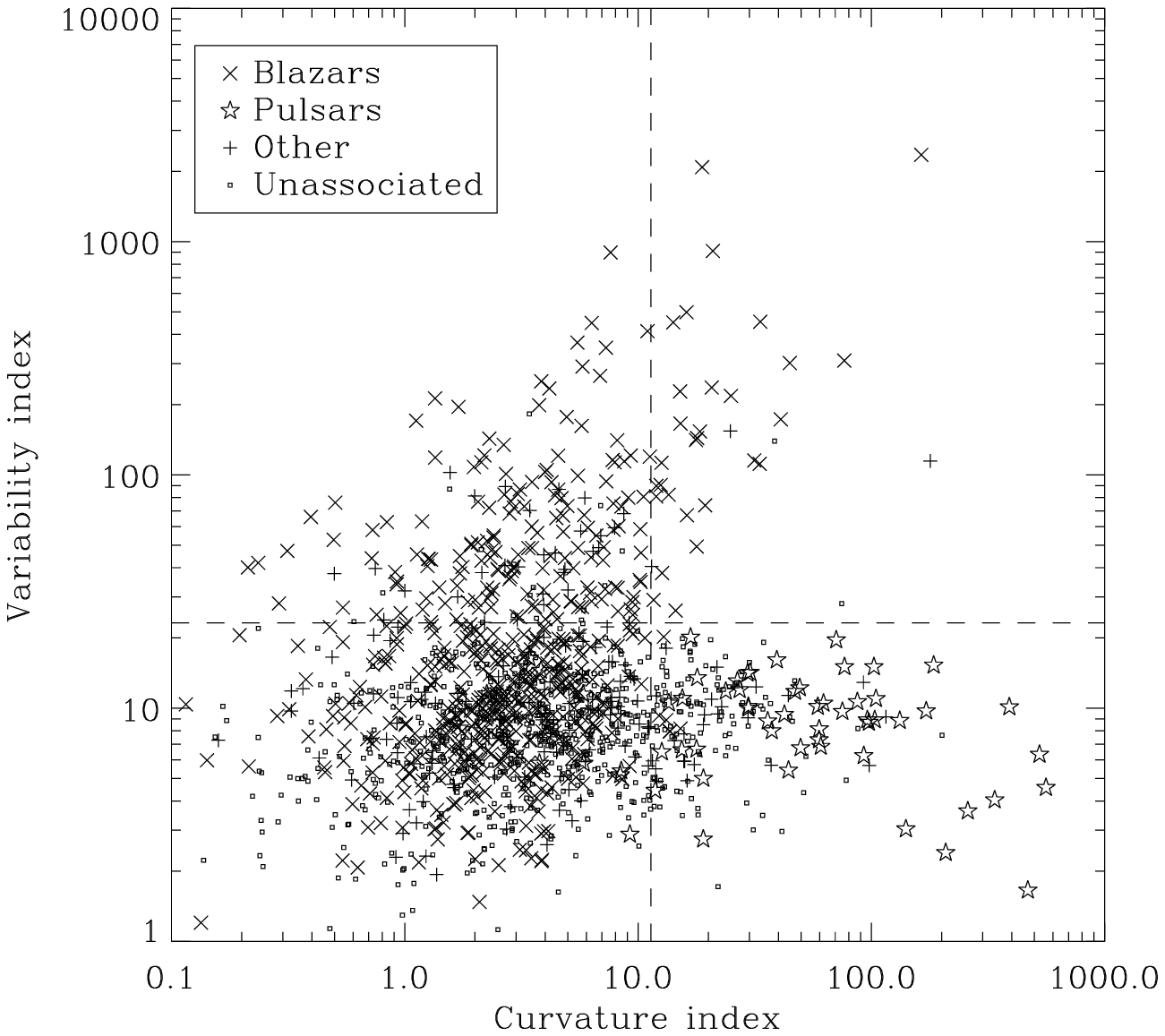}
\caption{Variability index plotted as a function of curvature index
(\S~\ref{run:flux}).
The horizontal dashed line shows where we set
the variable source limit, at $V > 23.21$. The vertical dashed line
shows where the spectra start deviating from a power-law, at $C > 11.34$.
The cross standing out as very significantly curved and variable is the source associated with LS~I~+61~303 \citep{LATLSI+61}.}
\label{fig:curvar}
\end{figure}

The fractional variability of the sources
is defined from the excess variance on top of the statistical
and systematic fluctuations:
\begin{equation}
\label{eq:relvar}
\delta F/F = \sqrt{\frac{\sum_i (F_i - F_{\rm av})^2}
                        {(N_{\rm int}-1) F_{\rm av}^2} -
                   \frac{\sum_i \sigma_i^2}{N_{\rm int} F_{\rm av}^2} -
                   f_{\rm rel}^2}
\end{equation}
The typical fractional variability is 50\%, with only a few strongly variable
sources beyond $\delta F/F$ = 1.
This is qualitatively similar to what was reported on
Figure~8 of \citet{LATBSL}.
The criterion we use is not sensitive to relative variations smaller
than 60\% at $TS$ = 100.
That limit goes down to 20\% as $TS$ increases to 1000.
We are certainly missing many variable AGN below $TS$ = 100
and up to $TS$ = 1000.
There is no indication that fainter sources are less variable than brighter
ones; we simply cannot measure their variability.

Both the curvature index and the variability index highlight certain
types of sources. This is best illustrated on Figure~\ref{fig:curvar}
in which one is plotted against the other for the main types of identified or associated sources
(from the association procedure described in \S~\ref{run:assoc}).
One can clearly separate the pulsar branch at large curvature and small
variability from the blazar branch at large variability and smaller curvature.

\subsection{Limitations and Systematic Uncertainties}
\label{run:systematics}

In this work we
did not test for or account for source extension. All sources are
assumed to be point-like. This is true for the major source populations
in the GeV range (blazars, pulsars).
On the other hand the TeV instruments
have detected many extended sources in the Galactic plane, mostly pulsar
wind nebulae  and supernova remnants (SNRs), 
\citep[e.g.][]{2005HESSPlane} and the LAT has already started detecting
extended sources \citep[e.g.][]{LATW51C}.
Because measuring extension over a PSF which varies so much with energy
is delicate, we are not yet ready to address this matter systematically across all the sources in a large catalog such as this.

\begin{figure}
\epsscale{.80}
\plotone{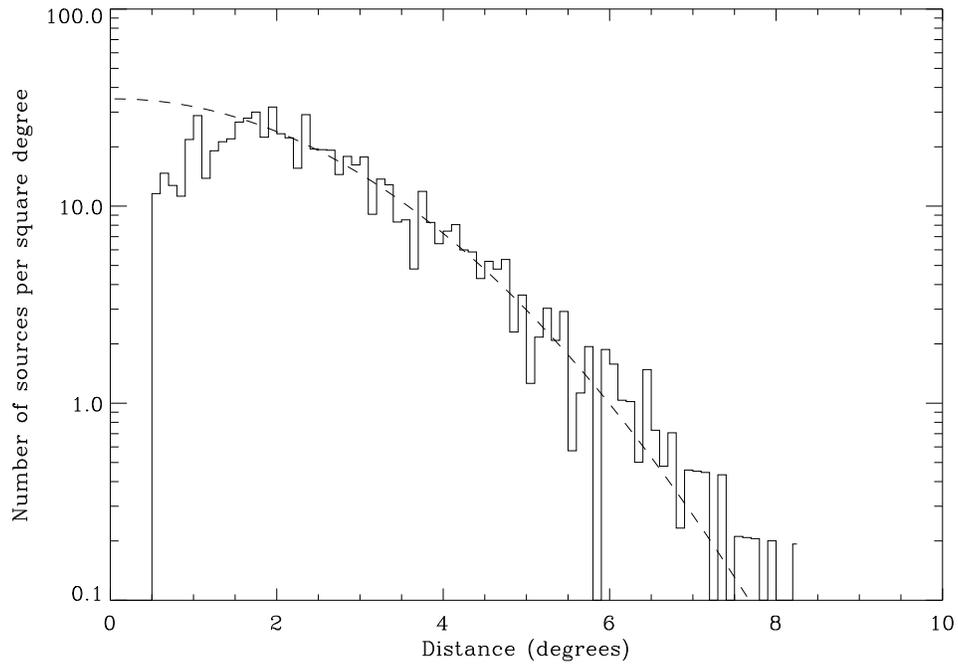}
\caption{Distribution of the distances $D_n$ to the nearest neighbors
of all detected sources at $|b| > 10\degr$. The number of entries is divided
by $2 \pi D_n \, \Delta D_n$ in which $\Delta D_n$ is the distance bin,
in order to eliminate the 2-dimensional geometry.
The overlaid curve is the expected Gaussian distribution
for a uniform distribution of sources with no confusion
(Eq.~\ref{eq:dnear} normalized using Eq.~\ref{eq:ntrue}).}
\label{fig:confusion}
\end{figure}

We have addressed the issue of systematics for localization
in \S~\ref{run:localization}. Another related limitation is that of source confusion.
This is of course strong in the inner Galaxy (\S~\ref{run:ridge})
but it is also a significant issue elsewhere.
The average distance between sources outside the Galactic plane
is $3\degr$ in 1FGL, to be compared with a per photon containment radius
$r_{68} = 0.8\degr$ at 1~GeV where the sensitivity is best.
The ratio between both numbers is not large enough
that confusion can be neglected.
The simplest way to quantify this is to look at the distribution
of distances between each source and its nearest neighbor ($D_n$)
in the area of the sky where the source density is approximately uniform, i.e.,
outside the Galactic plane. 
This is shown in Figure~\ref{fig:confusion}.
The source concentration in the Galactic
plane is very narrow (less than $1\degr$) but we need to make sure
that those sources do not get chosen as nearest neighbors
so we select $|b| > 10\degr$.
The histogram of $D_n$ (after taking out the geometric factor as
in Figure~\ref{fig:confusion}) should follow
\begin{equation}
\label{eq:dnear}
H(D_n) = N_{\rm true} \, \rho_{\rm src} \, \exp \left(- \pi D_n^2 \rho_{\rm src} \right)
\end{equation}
where $\rho_{\rm src}$ is the source density
(number of sources per square degree) and $N_{\rm true}$ is the true number
of sources (after correcting for missed sources due to confusion).
The exponential term is the probability that no nearest source exists.
It is apparent that, contrary to expectations, the histogram falls off toward $D_n=0$.
This indicates that confusion is important, even in the extragalactic sky.
The effect disappears only at distances larger than $1.5\degr$.
To get $N_{\rm true}$, one may solve for the number of observed sources
at distances beyond $1.5\degr$.
Since $ \rho_{\rm src} = N_{\rm true}/A_{\rm tot}$
in which $A_{\rm tot}$ is the sky area at $|b| > 10\degr$, 
this amounts to solving
\begin{equation}
\label{eq:ntrue}
N_{\rm obs}(>1.5\degr) = N_{\rm true} \, \exp \left(- N_{\rm true} A_0 / A_{\rm tot} \right)
\end{equation}
in which $A_0$ is the area up to $1.5\degr$.
This results in $N_{\rm true} - N_{\rm obs}$ = 80 missed sources
on top of the $N_{\rm obs}$ = 1043 sources observed at $|b| > 10\degr$.
Those missed sources are probably the reason for some of the
asymmetries in the $TS$ maps discussed
in \S~\ref{run:localization}.
The conclusion is that globally we missed nearly 10\% of the extragalactic
sources. But because of the worse PSF at low
energy soft sources are comparatively more affected than hard sources.
This is approximately indicated by the difference between the full and
the dashed lines on Figure~\ref{fig:sensspec}.

Another important issue is the systematic
uncertainties on the effective area of the instrument.
At the time of the BSL we used pre-launch calibration and cautioned
that there were indications that our effective area was reduced in flight
due to pile-up. Since then, the pile-up effect has been integrated
in the simulation of the instrument \citep{rando2009}
and many tests have shown that the resulting calibration (P6\_V3)
is consistent with the data. The estimate of the remaining systematic
uncertainty is 10\% at 100~MeV, 5\% at 500~MeV rising to 20\%
at 10~GeV and above. This
uncertainty applies uniformly to all sources.
Our relative errors (comparing one source to another
or the same source as a function of time) are much smaller,
as indicated in \S~\ref{run:variability}.
The fluxes resulting from this new calibration are systematically higher
than the BSL fluxes. For example, the fluxes of the three brightest pulsars
(Vela, Geminga and Crab) are about 30\% larger in 1FGL than in the BSL.
The differences are more pronounced for soft sources than hard ones.
This implies also that the 1FGL fluxes are significantly larger than the EGRET
fluxes in the 3EG catalog \citep{3EGcatalog} which happened to be close
to the BSL fluxes.  As shown by diffuse \citep{LATGeVexcess} and point source \citep{LATVela,LATCrab} observations, the LAT data produce spectra systematically steeper than those reported in EGRET analysis.  LAT fluxes are greater at energies below 200~MeV and less at energies above a few GeV.

The model of diffuse emission is the other important source
of uncertainties. Contrary to the effective area, it does not affect all sources
equally: its effects are smaller outside the Galactic plane ($|b| > 10\degr$)
where the diffuse emission is faint and varying on large angular scales.
It is also less of a problem in the high energy bands ($>$ 3~GeV) where
the PSF is sharp enough that the sources dominate the background
under the PSF.
But it is a serious issue inside the Galactic plane
($|b| < 10\degr$)
in the low energy bands ($<$ 1~GeV) and particularly
inside the Galactic ridge ($|l| < 60\degr$) where the diffuse emission
is strongest and very structured, following the molecular cloud distribution.
It is not easy to assess precisely how large the uncertainty is, for lack
of a proper reference model.
We discuss the Galactic ridge more specifically in \S~\ref{run:ridge}.
For an automatic assessment we have tried re-extracting the source fluxes
assuming a different diffuse model, derived from GALPROP
(as we did for the BSL) but with protons and electrons adjusted to the data
(globally). The model reference is 54\_87Xexph7S.
The results show that
the systematic uncertainty more or less follows the statistical one
(i.e., it is larger for fainter sources in relative terms) and is of the same order.
More precisely, the dispersion is 0.7 $\sigma$ on flux and 0.5 $\sigma$
on spectral index at $|b| > 10\degr$, and
1.8 $\sigma$ on flux and 1.2 $\sigma$
on spectral index at $|b| < 10\degr$.
We have not increased the errors accordingly, though, because this
alternative model does not fit the data as well as the reference model.
From that point of view we may expect this estimate to be an upper limit.
On the other hand both models rely on nearly the same set of H~{\sc I} and CO maps of the gas in the interstellar medium, which we know are an imperfect representation of the mass.
That is, potentially large systematic uncertainties are not accounted for by the comparison.
So we present the figures as qualitative estimates.

\subsection{Sources Toward Local Interstellar Clouds and the Galactic Ridge}
\label{run:ridge}
\begin{figure}
\epsscale{.80}
\plotone{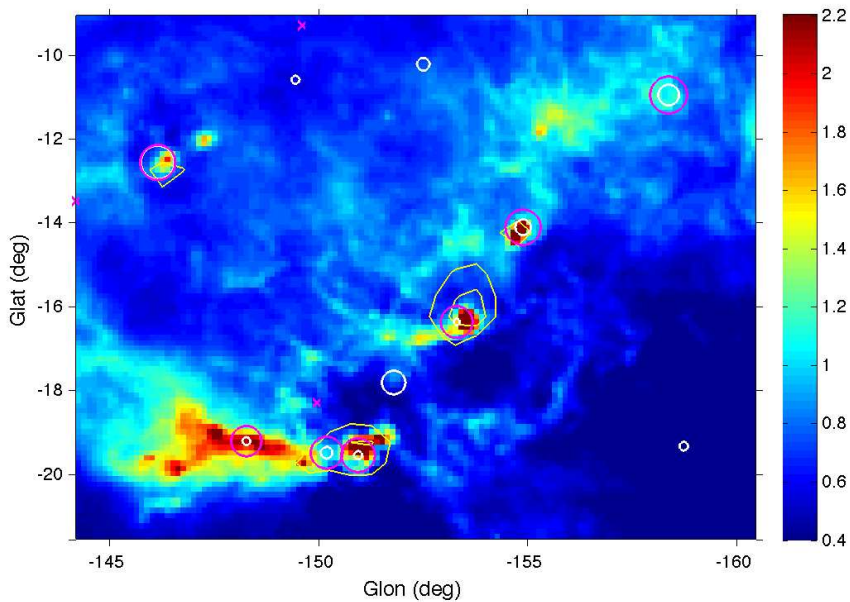}
\caption{Overlay of 1FGL sources on a square root color scale representation of E(B-V) reddening in Orion \citep{schlegel98}. The units for the color bar are magnitudes of reddening.  The white circles indicate the positions and 95\% confidence regions for the 1FGL sources in the field. The magenta circles indicate the effective (spectrally-weighted) 68\% containments for photons $>$500 MeV associated with each source that is positionally correlated with the clouds; these circles can be considered to represent the region of the sky most relevant for the definition of each source. The yellow contours around $(l, b)$ = ($-$151\degr, $-$19.2\degr), ($-$153.5\degr, $-$16.2\degr), and ($-$146.3\degr, $-$12.8\degr) outline the regions with negative reddening residuals caused by errors in the dust infrared color corrections near young clusters of IR sources.  The Orion Nebula is near $l, b \sim 209$\degr, $-$19.5$\degr$.
}
\label{fig:orion}
\end{figure}

Figure~\ref{fig:orion} shows an example of the striking, and physically unlikely, correspondence between the 1FGL sources and tracers of the column density of interstellar gas, in this case E(B--V) reddening.
The sources in Orion appear to be tightly associated with the regions with greatest column densities. Yet no particular classes of $\gamma$-ray emitters are known to be associated with interstellar cloud complexes.  Young SNRs would be resolved in the radio and in $\gamma$-rays in the nearby clouds outside the Galactic plane.  Even if radio-quiet pulsars were the sources, they would not be expected to be aligned so closely with the regions of highest column densities.  The implication is that peak column densities are being systematically underestimated in the model for the Galactic diffuse emission used in the analysis.  E(B--V) is not directly used in the model; as described in \S~\ref{run:diffusemodel} the column densities are derived from surveys of H~{\sc I} and CO line emission, the latter as a tracer of molecular hydrogen.  An E(B--V) `residual' map E(B--V)$_{\rm res}$, representing interstellar reddening that is not correlated with $N$(H~{\sc I}) or $W$(CO), is included in the model.  So the peak column densities would need to be underestimated both in CO and E(B--V).  We are studying the effect and strategies for validating the model for Galactic diffuse emission at high column densities.  

In addition to the concerns about the accuracy of column densities toward the peaks of interstellar clouds, self absorption of H~{\sc I} at low latitudes can introduce small angular-scale underestimates of the column densities, and intensities of the diffuse emission.  The current, half-degree binned, model of the interstellar emission used for the source analysis also cannot capture structure on smaller angular scales.  Bright structure on finer scales could be detected as unresolved point sources.

\begin{figure}
\epsscale{.80}
\plotone{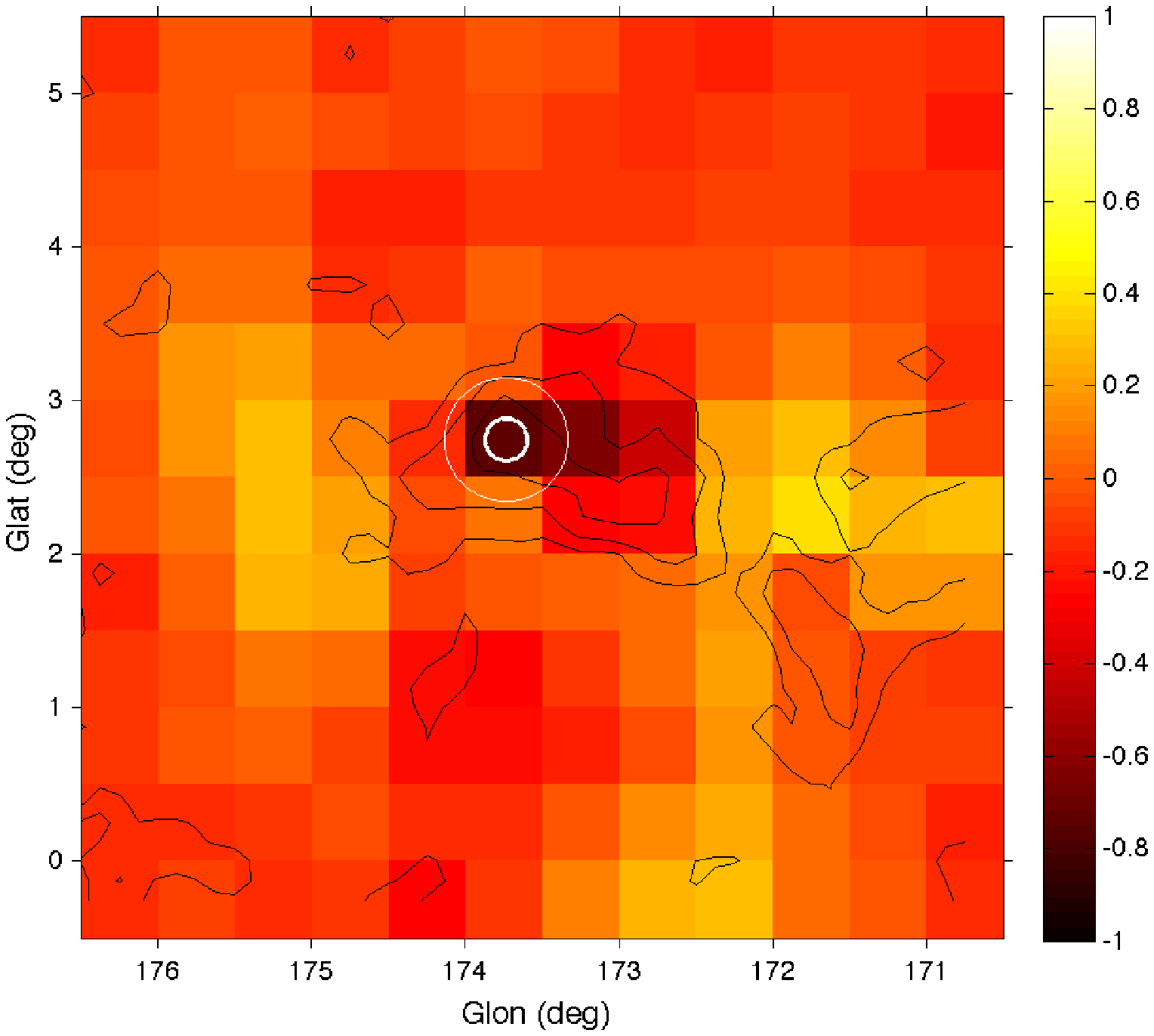}
\caption{Overlay of 1FGL J0541.1+3542 sources on a square root color scale representation of E(B--V)$_{\rm res}$ residual reddening in the S235 H~{\sc II} region. The units for the color bar are magnitudes of reddening.  The small circle indicates the position and 95\% confidence region for the 1FGL source. The large circle indicates the effective (spectrally-weighted) 68\% containment for photons with energies $>$500 MeV. The $W$(CO) intensity contours \citep[from ][]{Dame2001} at 6, 15, and 30 K km s$^{-1}$ trace the column density of molecular gas.
}
\label{fig:s225}
\end{figure}

Figure~\ref{fig:orion} illustrates another, better understood issue with the model for Galactic diffuse emission.  In the Orion nebula (near $l, b \sim 209$\degr, $-$19.5$\degr$), a massive star-forming region that is extremely bright in the infrared, the infrared color corrections used to evaluate E(B--V) are inaccurate and the column densities inferred from E(B--V) are underestimated; the depression in E(B--V) in the nebula does not correspond to a decreased column density of gas.  The model of Galactic diffuse emission was constructed with E(B--V)$_{\rm res}$ allowed to be a signed correction for the column densities inferred from H~{\sc I} and CO lines.  
This made essential improvements in large regions,  
but in discrete directions toward massive star-forming regions, negative E(B--V)$_{\rm res}$ can introduce deep depressions in the predicted diffuse emission.  Figure~\ref{fig:s225} illustrates the depression around the S225 star-forming region, and its close correspondence with a 1FGL source.

\begin{figure}
\epsscale{.80}
\plotone{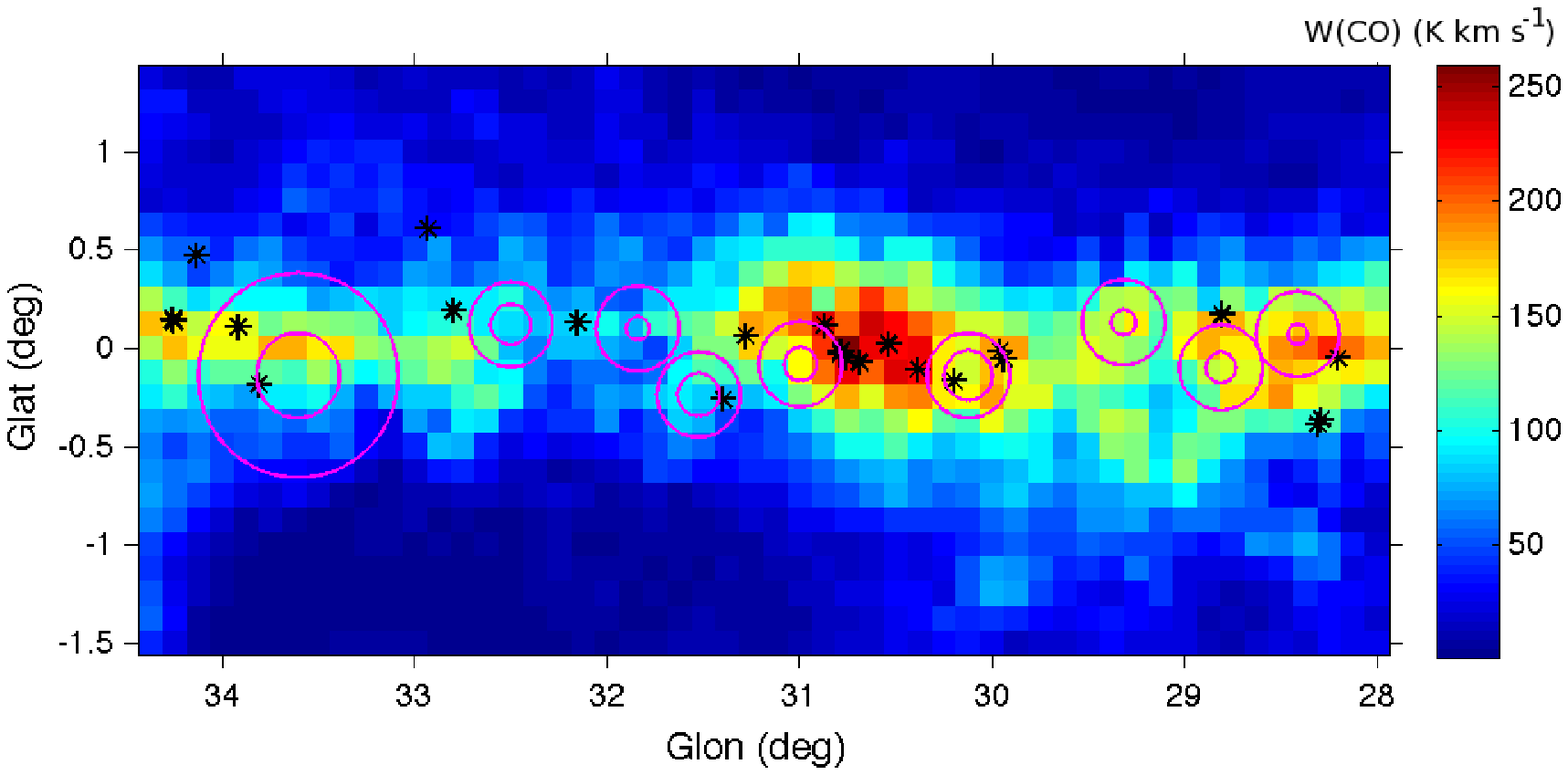}
\caption{Overlay of 1FGL sources (magenta circles) on an image of the intensity $W$(CO) of the 2.6~mm line of CO \citep{Dame2001}, in a segment of the Galactic ridge in the first quadrant.  The black asterisks indicate the positions of ultracompact H~{\sc II} regions \citep{Giveon2005}, which are similarly narrowly distributed about the Galactic equator.  For the 1FGL sources, the inner circles indicate the 95\% confidence regions for the locations and the outer circles the approximate extents of the spectrally-weighted PSF for energies $>$1~GeV.  The latter give an approximate sense of the `regions of influence' for the 1FGL sources.
}
\label{fig:ridge}
\end{figure}

\begin{figure}
\epsscale{.80}
\plotone{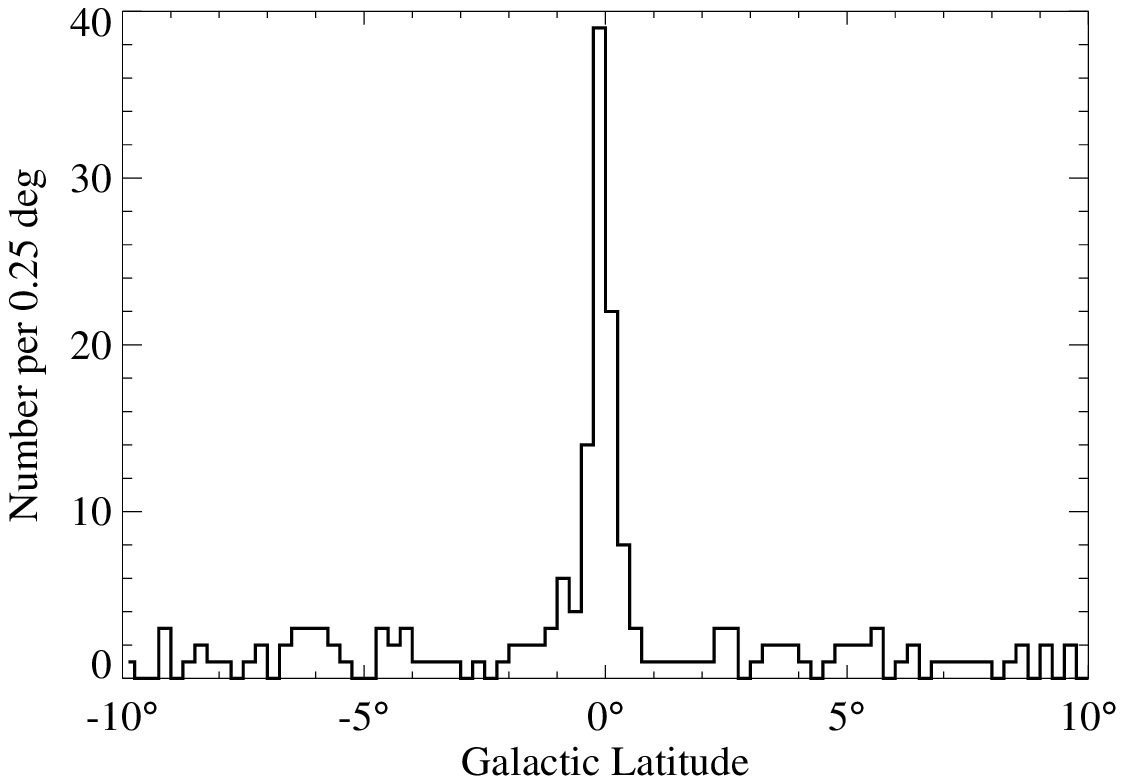}
\caption{Latitude distribution of unassociated/unidentified 1FGL sources in the Galactic ridge ($300\degr < l < 60\degr$). \label{fig:latdist} }

\end{figure}

Similar considerations relate to the sources at low latitudes in the inner Galaxy.  The density of unassociated sources in the Galactic ridge ($300\degr < l < 60\degr, |b| < 1\degr$) is very high (Fig.~\ref{fig:ridge}), and their latitude distribution is exceedingly narrow (Fig.~\ref{fig:latdist}).  If these 1FGL sources are true $\gamma$-ray emitters they must have a very small scale height in the Milky Way, like that of the youngest massive star-forming regions, traced by ultracompact H~{\sc II} regions \citep[$\sim$25$^\prime$ FWHM, e.g., ][]{Giveon2005}, or be quite distant and hence very luminous.  The 1FGL sources do not have an obvious correspondence with the ultracompact H~{\sc II} regions, and the latter are not plausible $\gamma$-ray sources, but owing to the effects described above, embedded star-forming regions can influence tracers of gas and dust and thereby potentially introduce small-scale errors in the model of Galactic diffuse  emission.   The inferred luminosities of the 1FGL sources in the Galactic ridge would be quite high if the scale height of their distribution is characteristic of most tracers of Population~I objects.  For a relatively narrow dispersion of 40~pc, the characteristic distances of these sources are $\sim$11~kpc (i.e., more distant than the Galactic center) and the $\gamma$-ray luminosities exceed 10$^{36}$ erg s$^{-1}$ \citep[i.e., more than an order of magnitude more luminous than the Vela pulsar, ][]{LATVela}.  For broader dispersions about the plane, the distances and luminosities would increase correspondingly.

The 1FGL sources toward the peaks of local interstellar clouds and the Galactic ridge all have analysis flags set (\S~\ref{run:flags}) in the catalog.  {\bf We have also added a designator `c' to their names to indicate that they are to be considered as potentially confused with interstellar diffuse emission or perhaps spurious.}  In addition, the `c' designator is used for unidentified 1FGL sources in crowded regions of high source density outside the Galactic ridge, as a caution about the complications due to PSF overlaps {\bf The `c' designator, thus applied to 161 of the 1FGL sources, is a warning that the existence of the source or its measured properties (location, flux, spectrum) may not be reliable.}

\subsection{Analysis Flags}
\label{run:flags}

We have identified a number of conditions
that can shed doubt on a source. 
They are described in Table~\ref{tab:flags}.
As noted, setting of flags 4 and 5 depends on the energy band in which a source is detected.
The high energy bands are always more favorable for source detection and characterization, as a result of the narrower PSF at high energy,
so the flags are set on the basis of the highest band in which a source
is significant. Flag 5 signals confusion and depends on a reference distance
$\theta_{\rm ref}$.
Because statistics are better at low energy (enough events
to sample the core of the PSF), $\theta_{\rm ref}$ is set
to the Full Width at Half Maximum (FWHM) there
(minimum distance to have two peaks with a local minimum in between
in the counts map).
At high energy there are fewer events so $\theta_{\rm ref}$ is set
to the larger value $2 r_{68}$. In the intermediate bands
we interpolate between FWHM and $2 r_{68}$.
In the FITS version of the catalog, these flags are summarized in a single integer column (Flags; see Appendix D).  Each condition is indicated by one bit among the 16 bits forming Flags. The bit is raised (set to 1)
in the dubious case, so that good sources have Flags = 0.

\begin{deluxetable}{cl}

\tablecaption{Definitions of the Analysis Flags \label{tab:flags}}
\tablewidth{0pt}
\tablehead{
\colhead{Flag\tablenotemark{a}} &  
\colhead{Meaning}
}

\startdata
  1  & Source with $TS > 35$ which went to $TS < 25$ when changing\\
        &  the diffuse model (\S~\ref{run:systematics}). Note that sources with $TS < 35$\\
       &    are not flagged with this bit because normal statistical \\
       &  fluctuations can push them to $TS < 25$. \\
  2  & Moved beyond its 95\% error ellipse when changing the diffuse \\
        & model. \\
  3  & Flux or spectral index changed by more than 3 $\sigma$ when \\
        &  changing the diffuse model. Requires also that the flux change\\ 
        & by more than 35\% (to not flag strong sources). \\
  4  & Source-to-background ratio less than 30\% in highest band in \\
        & which $TS > 25$. Background is integrated over $\pi r_{68}^2$ \\
        & or 1 square degree, whichever is smaller.\\
  5  & Closer than $\theta_{\rm ref}$ from a brighter neighbor.  $\theta_{\rm ref}$ is defined in the \\
       &  highest-energy band in which source $TS > 25$.  $\theta_{\rm ref}$ is set to \\
        & $2.6\degr$ (FWHM) below 300~MeV, $1.52\degr$ between 300~MeV and\\
       &   1~GeV, $0.84\degr$ between 1~GeV and 3~GeV, and $2 \, r_{68}$ above 3~GeV.\\
  6  & On top of an interstellar gas clump or small-scale defect in the \\
       &   model of diffuse emission. \\
  7  & Unstable position determination; result from {\it gtfindsrc} outside\\
       & the 95\% ellipse from {\it pointlike} (see \S~\ref{run:localization}). \\
  8  & {\it pointlike} did not converge. Position from {\it gtfindsrc}. \\
  9  & Elliptical quality $>$ 10 in {\it pointlike} (i.e., $TS$ contour does not\\
       &    look elliptical). \\
 \enddata
 
 \tablenotetext{a}{In the FITS version the values are encoded in a single column, with Flag $n$ having value $2^{(n-1)}$.  For information about the FITS version of the table see Appendix D and  \S~\ref{run:sourcelist}.}
\end{deluxetable}

\section{The 1FGL Catalog}
\label{run:sourcelist}

In this section we tabulate the quantities listed in Table~\ref{tab:sources} for each source; see Table~\ref{tab:desc} for descriptions of the columns.  The source designation is \texttt{1FGL JHHMM.m+DDMM} where the \texttt{1} refers to this being the first LAT catalog, \texttt{FGL} represents {\it Fermi} Gamma-ray LAT.  Sources close to the Galactic ridge and some nearby interstellar cloud complexes are assigned names of the form \texttt{1FGL JHHMM.m+DDMMc}, where the \texttt{c} indicates that caution should be used in interpreting or analyzing these sources.  Errors in the model of interstellar diffuse emission, or an unusually high density of sources, are suspected to affect the measured properties or even existence of these sources (see \S~\ref{run:ridge}).  

The designations of the classes that we use to categorize the 1FGL sources are listed in Table~\ref{tab:classes} along with the numbers of sources assigned to each class.  We distinguish between associated and identified sources, with associations depending primarily on close positional correspondence (see \S~\ref{run:autoassoc}) and identifications requiring measurement of correlated variability at other wavelengths or characterization of the 1FGL source by its angular extent (see \S~\ref{run:ids}).  Sources associated with SNRs are often also associated with PWNs and pulsars, and the SNRs themselves are often not point-like.  We do not attempt to distinguish among the possible classifications and instead in Table~\ref{tab:snr_over} list plausible associations of each class for unidentified 1FGL sources that are found to be associated with SNRs.

\begin{figure}
\epsscale{.80}
\plotone{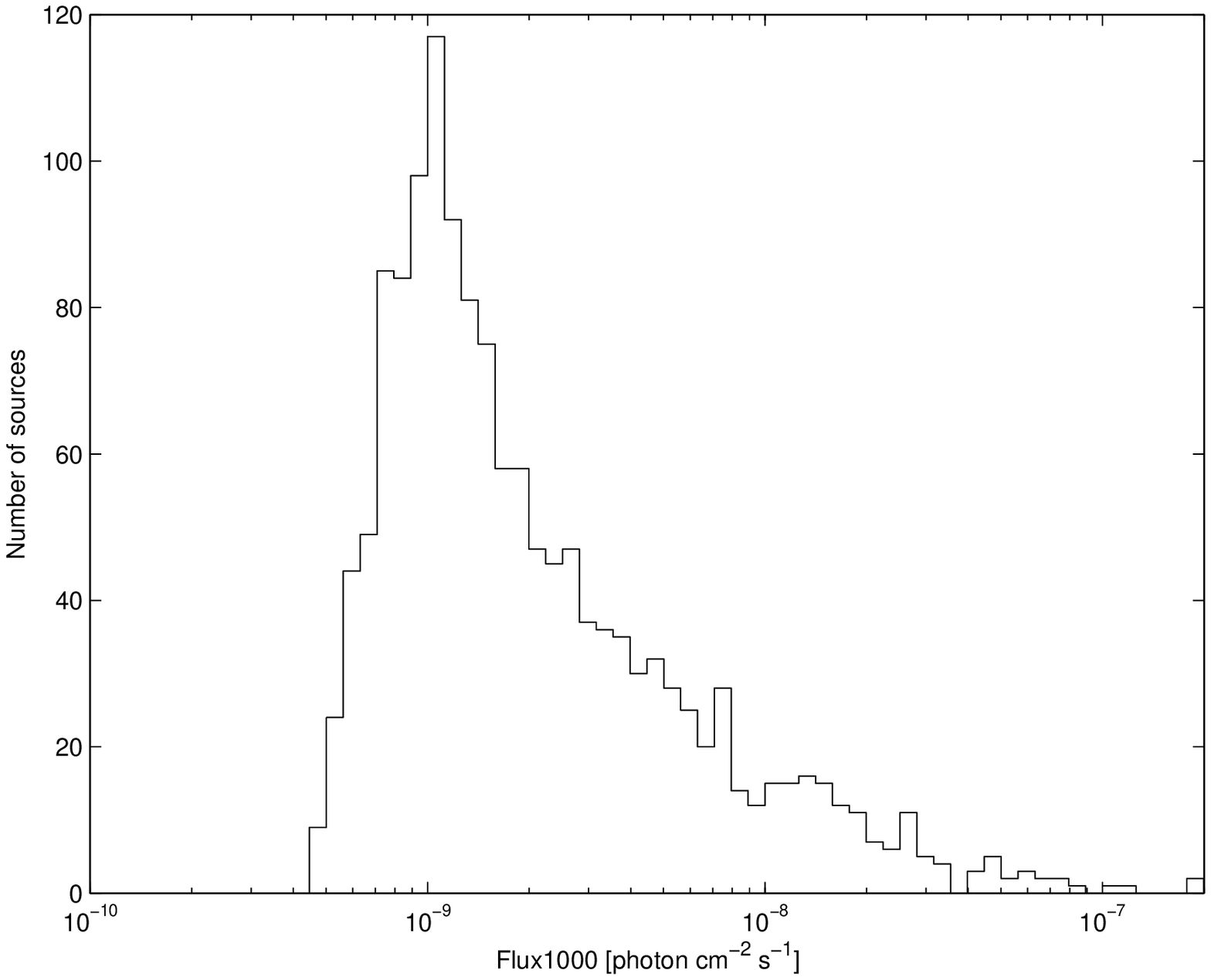}
\caption{Distribution of the fluxes of the 1FGL sources in the energy range 1~GeV -- 100~GeV. \label{fig:1fgl_fluxes}}
\end{figure}

The summed photon flux for 1--100~GeV ($F_{35}$; the subscript $ij$ indicates the energy range as 10$^i$ -- 10$^j$ MeV) and the energy flux for 100~MeV to 100~GeV in Table~\ref{tab:sources} are evaluated from the fluxes in bands presented in Table~\ref{tab:bandfluxes}.  The energy fluxes in each band are derived on the assumption that the spectral shape is the same as for the overall power-law fit ($\Gamma_{25}$).  This is an approximation but the bands are narrow enough that the energy fluxes are not strongly dependent on the spectral index.  We do not present the integrated photon flux for 100~MeV to 100~GeV because the relative uncertainties tend to be very large, because of the limited effective area in the lower energy bands.  Figure~\ref{fig:1fgl_fluxes} shows the distribution of integrated fluxes (1~GeV -- 100~GeV) for all of the sources in the catalog.  The dynamic range is approximately 2.5 orders of magnitude, owing both to the intrinsic range of fluxes of the $\gamma$-ray sources as well as their spectral hardnesses.

\begin{deluxetable}{ll}
\setlength{\tabcolsep}{0.04in}
\tabletypesize{\scriptsize}
\tablecaption{LAT First Catalog Description\label{tab:desc}}
\tablehead{
\colhead{Column} &
\colhead{Description} 
}
\startdata
Name &  1FGL JHHMM.m+DDMM[c], constructed according to IAU Specifications for Nomenclature; m is decimal\\
  &  minutes of R.A.; in the name R.A. and Decl. are truncated at 0.1 decimal minutes and 1\arcmin, respectively; \\
  &  c indicates that based on the region of the sky the source is considered to be potentially confused \\
  & with Galactic diffuse emission \\
R.A. & Right Ascension, J2000, deg, 3 decimal places  \\
Decl. & Declination, J2000, deg, 3 decimal places \\
 $l$  & Galactic Longitude, deg, 3 decimal places \\
 $b$  & Galactic Latitude, deg, 3 decimal places \\
 $\alpha$ & Semimajor radius of 95\% confidence region, deg, 3 decimal places\\
 $\beta$ & Semiminor radius of 95\% confidence region, deg, 3 decimal places\\
 $\phi$ & Position angle of 95\% confidence region, deg. East of North, 0 decimal places\\
$\sigma$ & Significance derived from likelihood Test Statistic for 100~MeV--100~GeV analysis, 1 decimal place \\
 $F_{35}$ &  Photon flux for 1~GeV--100~GeV, 10$^{-9}$ ph cm$^{-2}$ s$^{-1}$, summed over 3 bands, 1 decimal place \\
$\Delta F_{35}$ & 1-$\sigma$ uncertainty on  $F_{35}$ , same units and precision \\
$S_{25}$  &  Energy flux for 100~MeV--100~GeV, 10$^{-12}$ erg cm$^{-2}$ s$^{-1}$, from power-law fit, 1 decimal place \\
 $\Delta S_{25}$ & 1-$\sigma$ uncertainty on $S_{25}$, same units and precision \\
 $\Gamma$  & Photon number power-law index, 100~MeV--100~GeV, 2 decimal places \\
 $\Delta \Gamma$ & 1 $\sigma$ uncertainty of photon number power-law index, 100~MeV--100~GeV, 2 decimal places \\
Curv. & T indicates $<$~1\% chance that the power-law spectrum is a good fit to the 5-band fluxes; see note in text  \\
 Var. & T indicates $<$~1\% chance of being a steady source; see note in text  \\
Flag & See Table 1 for definitions of the flag numbers  \\
$\gamma$-ray Assoc.  & Positional associations with 0FGL, 3EG, EGR, or AGILE sources \\
TeV & Positional association with a TeVCat source, P for angular size $<$20$^\prime$, E for extended \\
 Class & Like `ID' in 3EG catalog, but with more detail (see Table \ref{tab:classes}).  Capital letters indicate firm identifications;\\
  &  lower-case letters indicate associations. \\
 ID or Assoc.  & Designator of identified or associated source \\
 Ref. & Reference to associated paper(s) \\  
\enddata
\end{deluxetable}

\begin{deluxetable}{llr}
\setlength{\tabcolsep}{0.04in}
\tablewidth{0pt}
\tabletypesize{\scriptsize}
\tablecaption{LAT 1FGL Source Classes\label{tab:classes}}
\tablehead{
\colhead{Description} & 
\colhead{Designator} &
\colhead{Number Assoc. (ID)}
}
\startdata
 Pulsar, X-ray or radio, identified by pulsations & psr (PSR) & 7 (56)\\
 Pulsar, radio quiet (LAT PSR, {\it subset of above}) & PSR & 24\\
 Pulsar wind nebula & pwn (PWN) & 2 (3)\\
 Supernova remnant & $\dagger$ (SNR) & 41 (3) \\
 Globular Cluster & glc (GLC) & 8 (0) \\
 Micro-quasar object:  X-ray binary (black hole & mqo (MQO) & 0 (1) \\
  or neutron star) with radio jet & &  \\
 Other X-ray binary & hxb (HXB) & 0 (2)\\
 BL Lac type of blazar & bzb (BZB) & 295 (0)\\
 FSRQ type of blazar & bzq (BZQ) & 274 (4) \\
 Non-blazar active galaxy & agn (AGN) & 28 (0)\\
 Active galaxy of uncertain type & agu (AGU) & 92 (0)\\
 Normal galaxy & gal (GAL) & 6 (0)\\
 Starburst galaxy & sbg (SBG) & 2 (0) \\
 Unassociated & & 630 \\
\enddata
\tablecomments{ The designation `$\dagger$' indicates potential association with SNR or PWN (see Table~\ref{tab:snr_over}).  Designations shown in capital letters are firm identifications; lower case letters indicate associations. In the case of AGN, many of the associations have high confidence \citep{LATAGNCatalog}. Among the pulsars, those with names beginning with LAT were discovered with the LAT.  For the normal galaxy class, 5 of the associations are with the Large Magellanic Cloud.  In the FITS version of the 1FGL catalog, the $\dagger$ designator is replaced with `spp'; see Appendix D.}
\end{deluxetable}

\begin{deluxetable}{lrrl}
\setlength{\tabcolsep}{0.04in}
\tablewidth{0pt}
\tabletypesize{\scriptsize}
\tablecaption{Potential Associations for Sources Near SNRs\label{tab:snr_over}}
\tablehead{
\colhead{Name 1FGL} &
\colhead{$l$} &
\colhead{$b$} &
\colhead{Assoc.} 
}
\startdata
 J0218.8+6158c & 133.01 & 0.82 & SNR G132.7+01.3 \\
 J0220.0+6257 & 132.80 & 1.80 & SNR G132.7+01.3 \\
 J0500.1+5237 & 155.63 & 6.34 & SNR G156.2+05.7 \\
 J0503.2+4526 & 161.64 & 2.35 & SNR G160.9+02.6 \\
 J0538.6+2717 & 180.59 & $-$2.16 & SNR G180.0$-$01.7 \\
 J0553.9+3105 & 179.08 & 2.65 & SNR G179.0+02.6 \\
 J0636.0+0458c & 206.74 & $-$1.15 & SNR G205.5+00.5 \\
 J0823.3$-$4248 & 260.37 & $-$3.15 & SNR G260.4$-$03.4 \\
 J0841.9$-$4620 & 265.18 & $-$2.58 & SNR G263.9$-$03.3 \\
 J0854.0$-$4632 & 266.64 & $-$1.09 & SNR G266.2$-$01.2 (Vela Junior)\\
 J1018.6$-$5856 & 284.32 & $-$1.70 & SNR G284.3$-$01.8 (MSH 10$-$53) \\
 J1119.4$-$6127c & 292.17 & $-$0.52 & SNR G292.2$-$00.5, PWN G292.2$-$0.5, PSR J1119$-$6127 \\
 J1134.8$-$6055 & 293.77 & 0.57 & SNR G293.8+00.6, PWN G293.8+0.6 \\
 J1213.7$-$6240c & 298.63 & $-$0.12 & SNR G298.6$-$00.0 \\
 J1343.7$-$6239c & 308.89 & $-$0.39 & SNR G308.8$-$00.1 \\
 J1350.8$-$6212c & 309.80 & $-$0.12 & SNR G309.8+00.0 \\
 J1410.3$-$6128c & 312.21 & $-$0.03 & SNR G312.4$-$00.4, PSR J1410$-$6132\ \\
 J1442.0$-$6000c & 316.34 & $-$0.05 & SNR G316.3$-$00.0 \\
 J1501.6$-$4204 & 327.30 & 14.55 & SNR G327.6+14.6 \\
 J1514.7$-$5917 & 320.33 & $-$1.35 & SNR G320.4$-$01.2 (MSH 15$-$52) \\
 J1521.8$-$5734c & 322.03 & $-$0.38 & SNR G321.9$-$00.3 \\
 J1552.4$-$5609 & 326.25 & $-$1.71 & SNR G326.3$-$01.8, PWN G326.3$-$1.8 \\
 J1617.5$-$5105c & 332.39 & $-$0.40 & SNR G332.4$-$00.4 \\
 J1640.8$-$4634c & 338.29 & $-$0.06 & SNR G338.3$-$00.0, PWN G338.3$-$0.0 \\
 J1711.7$-$3944c & 347.15 & $-$0.19 & SNR G347.3$-$00.5 \\
 J1725.5$-$2832 & 357.96 & 3.91 & SNR G358.0+03.8 \\
 J1745.6$-$2900c & 359.94 & $-$0.05 & SNR G000.0+00.0, PWN G359.95$-$0.04 \\
 J1801.3$-$2322c &   6.57 & $-$0.22 & SNR G006.4$-$00.1 (W28) \\
 J1805.2$-$2137c &   8.55 & $-$0.14 & SNR G008.7$-$00.1 (W30)\\
 J1806.8$-$2109c &   9.13 & $-$0.24 & SNR G008.7$-$00.1 (W30)\\
 J1834.3$-$0842c &  23.24 & $-$0.20 & SNR G023.3$-$00.3 (W41) \\
 J1834.7$-$0709c &  24.67 & 0.43 & SNR G024.7+00.6 \\
 J1916.0+1110c &  45.67 & $-$0.31 & SNR G045.7$-$00.4 \\
 J2020.0+4049 &  78.37 & 2.53 & SNR G078.2+02.1 \\
 J2042.3+5041 &  88.66 & 5.19 & SNR G089.0+04.7 \\
 J2046.0+4954 &  88.42 & 4.24 & SNR G089.0+04.7 \\
 J2046.4+3041 &  73.40 & $-$7.79 & SNR G074.0$-$08.5 \\
 J2049.1+3142 &  74.56 & $-$7.60 & SNR G074.0$-$08.5 \\
 J2055.2+3144 &  75.40 & $-$8.59 & SNR G074.0$-$08.5 \\
 J2057.4+3057 &  75.11 & $-$9.46 & SNR G074.0$-$08.5 \\
 J2323.4+5849 & 111.74 & $-$2.12 & SNR G111.7$-$02.1 \\
\enddata
\tablecomments{See text.  These sources are marked with a $\dagger$ in Table 6.  They may be pulsars or PWN rather than the SNR named.}
\end{deluxetable}

\begin{figure}
\epsscale{.80}
\plotone{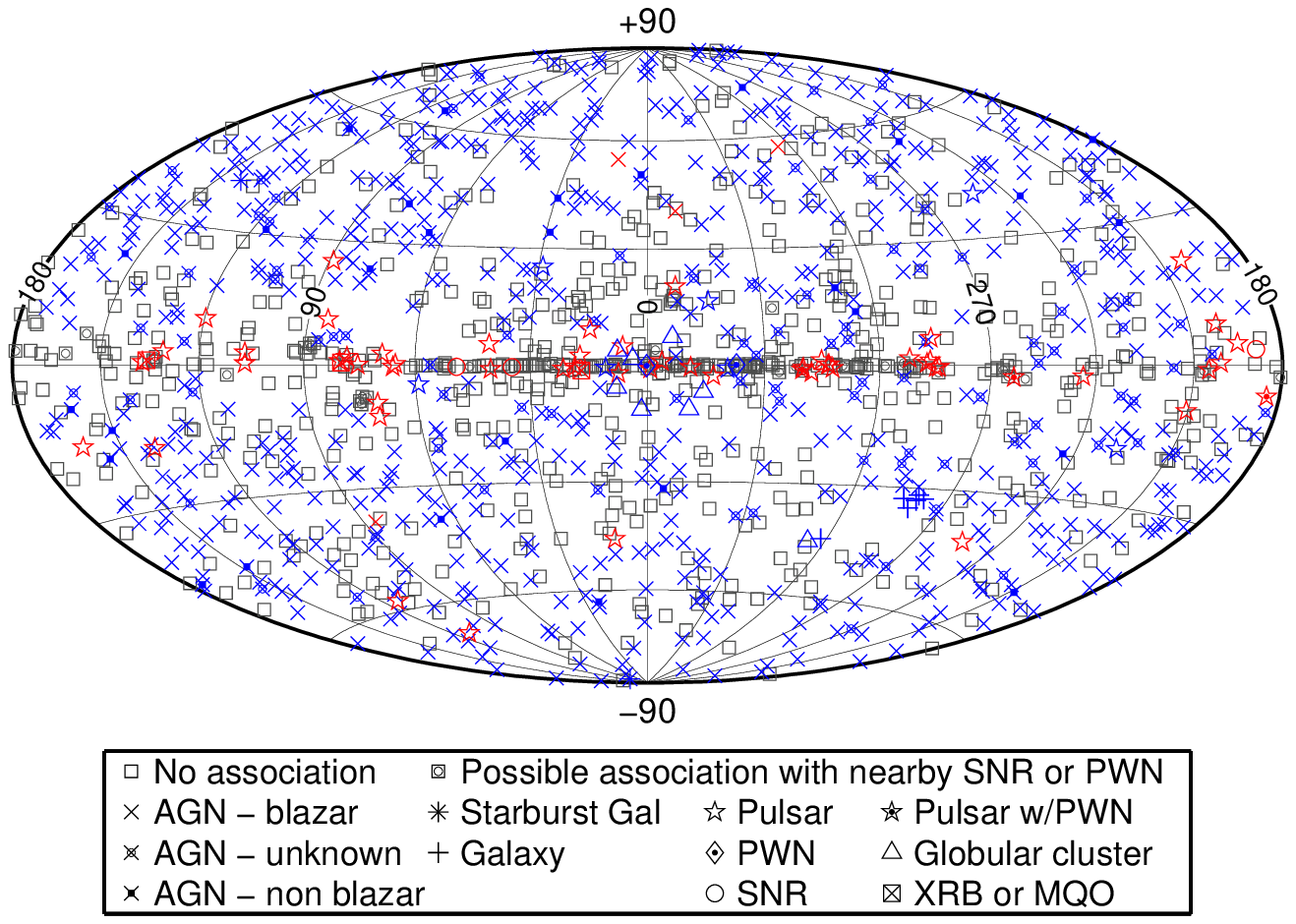}
\caption{The 1451 1FGL catalog sources, showing locations on the sky (in Galactic coordinates with Aitoff projection) and associated source class, coded according to the legend.  Gray symbols indicate unassociated sources, blue indicate associated sources, and red symbols firmly identified sources (primarily pulsars). For this plot the bzb and bzq  designators have been combined (``AGN-blazar''), as have hxb and mqo (``XRB or MQO''). The sources possibly associated with SNR, PSR or PWN (those indicated by a dagger in Table~\ref{tab:sources}) are listed as ``Potential SNR''.\label{fig:1fgl_sky}}
\end{figure}

\begin{figure}
\epsscale{.80}
\plotone{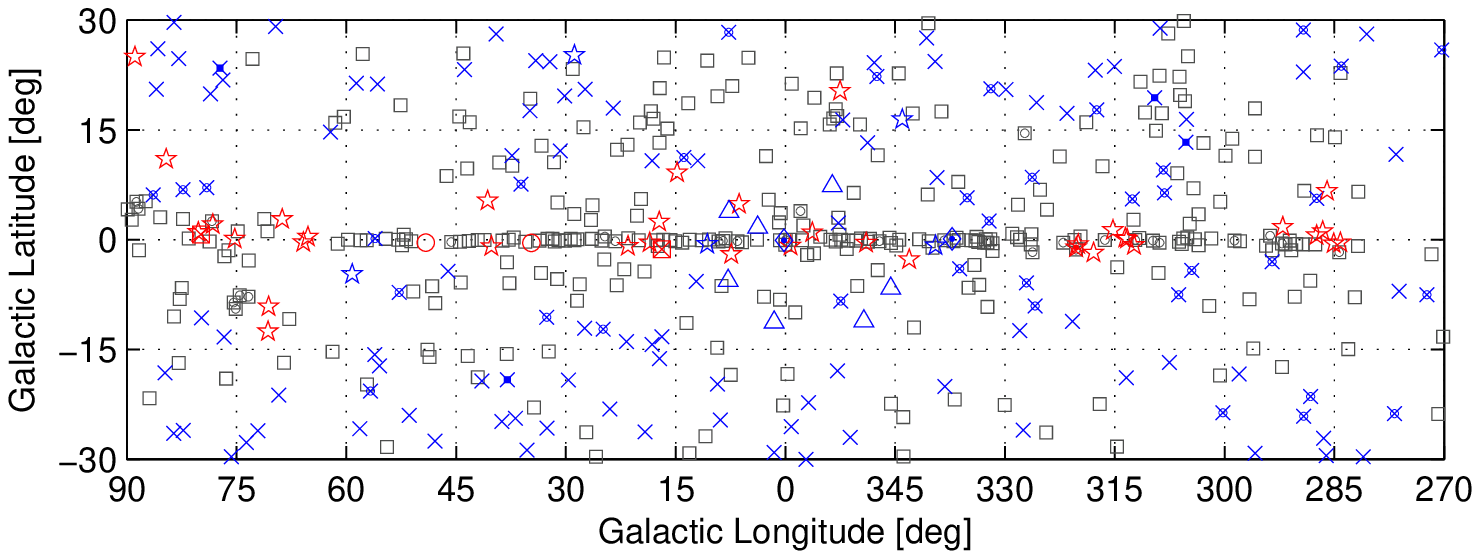}
\caption{The 1FGL catalog sources in the inner Galactic region. Sources are indicated in the same manner as Figure~\ref{fig:1fgl_sky}.\label{fig:1fgl_inner}}
\end{figure}

The procedure used to designate associated sources and class designations is described in \S~\ref{run:autoassoc}.  Figure~\ref{fig:1fgl_sky} shows the distribution of the 1FGL sources on the sky by source class, and Figure~\ref{fig:1fgl_inner} shows just the inner Galaxy.  

\begin{figure}
\epsscale{1.2}
\plottwo{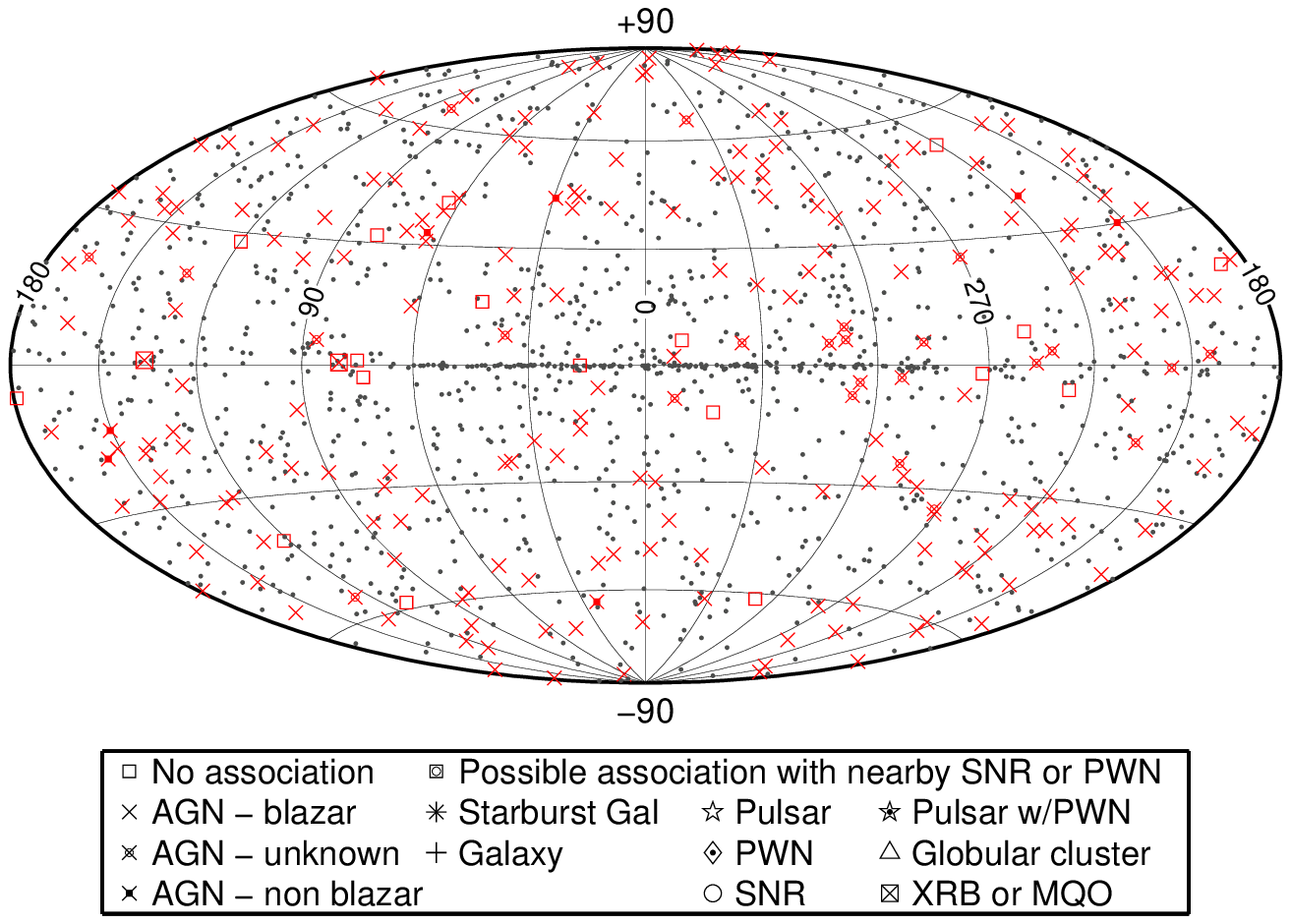}{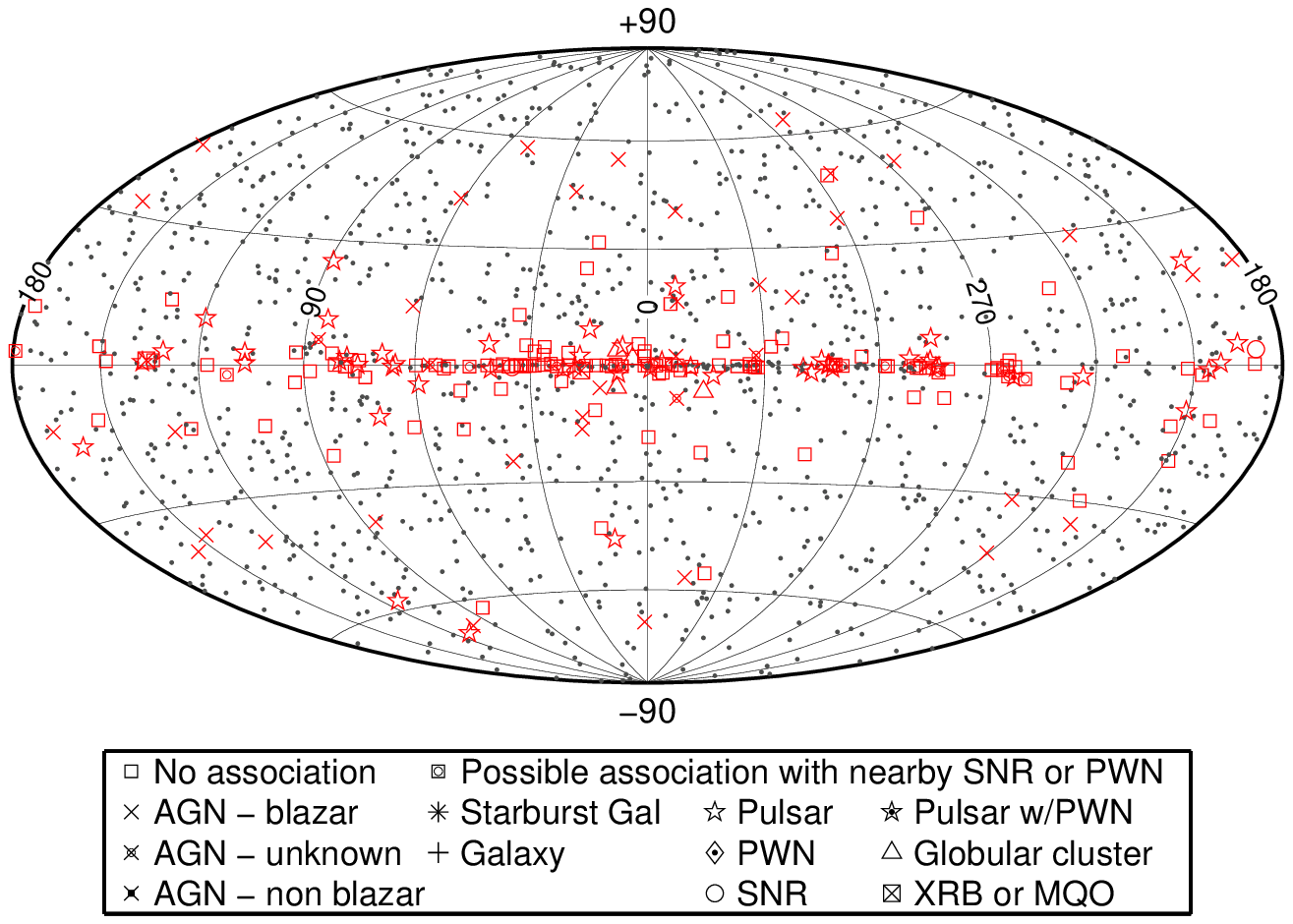}
\caption{The 1FGL catalog sources with variable ($left$) and spectrally-curved ($right$) sources highlighted. Sources with the variability flag or spectral curvature flags set are shown according to their associated class, as red symbols, coded as marked as in the legend while sources showing no evidence of variability or with spectra compatible with power laws are shown as black dots.\label{fig:1fgl_varcurve}}
\end{figure}

Figure~\ref{fig:1fgl_varcurve} selects just the variable sources, i.e., those with $<$1\% chance of being steady sources in the monthly light curves and those with spectra incompatible with power laws.  The variable sources are seen to be predominantly outside the Galactic plane, and many are associated with blazars.  The spectrally-curved sources have a distribution much more confined to the Galactic equator.

The electronic version of the 1FGL catalog, available in FITS format from the $Fermi$ Science Support Center, includes all of the information in these tables plus the monthly light curves from which the variability index values were derived and pivot energies for the overall power-law fits; see Table~\ref{tab:columns}.

\section{Source Association and Identification}
\label{run:assoc}
Even with the good angular resolution of LAT, source location accuracy is typically not  precise enough to make a firm identification based on positional coincidence alone.  A typical LAT error region contains numerous stars, galaxies, X-ray sources, infrared sources, and radio sources. Determination of the nature of a given LAT source must therefore rely on more information than simply location, including time variability, spectral information, and availability of sufficient energy and a plausible physical process at the source to produce $\gamma$-rays.  

In this analysis, we make a clear distinction between a source {\it identification} and an {\it association} with an object at another wavelength.  A firm identification of a source is based on a timing characteristic such as a periodicity for a pulsar or binary or a variability correlated with observations at another wavelength in the case of a blazar, or on measurement of finite angular extent, which is the case for some Galactic sources, e.g., SNRs. An association is defined as a positional coincidence that is statistically unlikely to have occurred by chance between a plausible $\gamma$-ray-producing object and a LAT source.  

For the 1FGL catalog, the approach to designating associations and identifications involves three steps:

\begin{enumerate}
\item  A test for statistically-significant detections of classes of sources, based on a defined protocol as described by \citet{torres2005} and \citet{reimer07} has been carried out.  Based on these results, some potential source classes are deemed unlikely even if individual sources in such classes might be positionally coincident with 1FGL sources. 
\item A general automated source association analysis, enhanced from the version used for the BSL \citep{LATBSL}, has been applied to the sources in the catalog.  This method relies principally, although not exclusively, on comparing the local space density of plausible source classes with the number of positional associations found for a class with 1FGL sources. 
\item For pulsars and binary systems with clearly-identified periodic emission, a firm identification can be established. 
\end{enumerate}

Each of these methods is described below. 

\subsection{Protocol for population identification}
\label{run:pops}

The idea for a protocol for population testing in $\gamma$-ray source data such as represented by the {\it Fermi} 1st-year catalog was introduced by \citet{torres2005}, a paper to which we refer for general background.  The test was devised to provide high levels of confidence 
in population classification with small number statistics. Essential to this protocol is the use of \citet{feldman1998} 
confidence level intervals in a priori, physically selected samples of plausible $\gamma$-ray emitters. 

Thus, we consider spatial correlations between the 1st-year catalog sources (1451 detections) and a priori selected (on physical grounds or earlier hints in $\gamma$-ray data of previous missions) sets of astrophysical sources, details on which are given below. 
To test for spatial correlations, each of the {\it Fermi} sources is described by a centroid position and an uncertainty. The latter is described by an ellipse, with semi-minor and semi-major axis and a position angle; with all these values being taken from the catalog at 95\% CL. Those sources that are identified beyond doubt by timing (all $\gamma$-ray pulsars and three X-ray binaries) have their positions assigned to be coincident with the corresponding astrophysical sources \citep[as located in other wavelengths or by LAT pulsar timing; ][]{smith08}. However, for population searches, we maintain the uncertainty of their detections around these positions.
In order to test systematics we also analyze the case in which all 1FGL  sources have their positional uncertainties enlarged by increasing  their corresponding semi-minor and semi-major axes by 20\%. This is  done consistently both for the real set of 1FGL sources, and for each of the simulated sets generated by Monte Carlo (see below), and is meant as a check of the results obtained with nominal uncertainty values.  

Let ${\cal C}(A)$ represent the number of coincidences between candidate counterparts for 
population $A$ and LAT sources. In case of finding the same {\it Fermi} detections spatially coincident with several astrophysical objects of the same class, we count all of the coincidences as one. An example of this is seen for millisecond pulsars pertaining to the same globular cluster, e.g., 47~Tuc. To determine the number of excess coincidences above the noise level, i.e., the sources that are expected to correlate by chance, we
subtract the background $b$ caused by these random coincidences, given the number and distribution of LAT sources and 
members of the testing population. 

To obtain this latter number, we produce Monte Carlo simulations shuffling the orientation of the elliptical position uncertainty (but not its size) and the centroid position of each of the actual 1FGL sources, thus generating sets of fake LAT source detections. Each simulated 1FGL
catalog is constructed to have the total number of sources (1451 detections), and both the longitude and latitude distributions of the real set.  In order to obtain the most conservative results we proceed to simulate fake LAT source catalogs by maintaining the latitude (longitude) histogram of the real set with 5, and 10 (30, 60) degrees binning, and subsequently take the largest value for the expected average number of random positional coincidences among them (and this is called $b$).

The number of excess coincidences above chance associations is then
$ {\cal E}(A)={\cal C}(A)-b(A)$. This number is used to test the null hypothesis:
{\it Population $A$ is not $\gamma$-ray emitting at flux-levels detectable by the 1st-year LAT catalog}. The predicted number of source coincidences for this hypothesis is equal to 0, and the total expected events if this hypothesis is valid is equal to $b$. The greater the excess of the real number of correlations over the corresponding $b$-value for that population, the easier it is to rule out the null hypothesis. 

The testing power of a sample of finite size is limited: if using the same set of data, claiming the discovery
of one population affects the level of confidence by which one can claim the discovery of a second. 
In order to control the reliability of our results, we require that 
{\it the combination of all of our claims} be bounded by a probability of $10^{-5}$, which then becomes the total budget ${\cal B}$. This low probability provides an overall significance of about 5$\sigma$, which implies individual claims of populations must arise with higher confidence.
The total budget can then be divided into individual ones for each population,
$P_A$, $P_B$, etc., such that
$ \sum_i P_i={\cal B}. $
Then, population $A$ will be claimed as detected with x\% CL if and only if:
\begin{itemize}
\item the Poissonian cumulative probability (CP) for obtaining the real number of spatial coincidences (or more) as a result of chance coincidence 
is less than the a priori assigned budget $P_A$ (as opposed to being less than only the larger, total budget) and

\item the number of excesses ${\cal E}(A) $ is beyond the upper limit of the corresponding confidence interval for x\% CL, from \citet{feldman1998}.
\end{itemize}

The latter value, x\%, for each population is then the confidence level obtained using the tables in 
\citet{feldman1998} for 
which the upper end of the interval equals the real number of excesses. 
(With number of events $b$, background $b$, recall that in the null hypothesis, there are 0 expected events above background.) 
The values of  $P_i$  for each population are chosen very conservatively: 0.01\% $B$ to each of the only two populations that were unambiguously identified in the EGRET catalog, pulsars and blazars; 0.1\% $B$ to millisecond pulsars and EGRET-coincident SNRs, which were hinted at in the EGRET catalog; and the rest of the budget equally distributed into those classes where no high-energy $\gamma$-ray emission was previously reported.

Finally, we provide brief notes on the populations selected for the test.  The selections were made before the start of {\it Fermi} operations, as described in the internal LAT-document AM-09067\footnote{Available from https://oraweb.slac.stanford.edu/pls/slacquery/DOCUMENTS.STARTUP?PROJECT=GLAST}, which lists each specific source selected in each class.

\begin{itemize}

\item Blazars: The selected blazars are a subset of 
the CGRABS 
catalog \citep{CGRaBS}, which is a uniform all-sky survey of 
EGRET-like blazars, selected by their Figure-of-merit \citep[FOM: see][]{CGRaBS}.  In total the CGRABS catalog includes 1625 sources. 
To assemble the blazar test list we cut the CGRABS 
catalog at the smallest FoM value for which all high-confidence EGRET blazars of the 
3EG catalog \citep{3EGcatalog}, which are also ÒMattox-blazarsÓ \citep{mattox2001}, are included \citep[see][for a comparison of 3EG and ``Mattox''-blazars]{sowards2003}.
This cut is at FoM=0.111, and results in a total of 215 sources and constitutes the blazar list used for the application of the protocol.

\item Misaligned jet sources: We start with the compilation of extragalactic jet sources \citep{liu2002} which has 661 entries 
collected from the literature (as of December 2000). This list contains radio galaxies, radio quasars, 
BL Lac objects and Seyfert galaxies. For a redshift cut at $z < 0.032$ (chosen to select a moderate number of the closest members of this class)  the list features 51 sources:  34 are clasified as radio 
galaxies (including all $\gamma$-ray radio galaxies detected with high confidence so far), 16 
as Seyfert galaxies, and one as BL Lac object (not yet detected in $\gamma$-rays and not 
included in the CGRABS list). 3C~111 and 3C~120 are added 
on the basis of evidence for their detection by EGRET.

\item Starbursts: This list includes all starburst galaxies for which in the study of \citet{torres2004} the combination of gas content, cosmic-ray density, and distance indicated a flux detectable by the LAT in one year.

\item Halo Dwarf Galaxies: The list consists of presently-known objects in this category 
up to 1000~kly. This sample of Milky Way satellite galaxies is useful for probing localized 
$\gamma$-ray excesses due to annihilation of dark matter particles. Dwarf spheroidal galaxies (dSph) may be manifestations of the largest clumps predicted by the CDM scenario. 

\item Galaxy clusters: We restrict this sample to 30 clusters with the highest values of mass-to-distance-squared, $M/d^2$, from the HIFLUCS sample of the \citet{reiprich2002}
sample, complemented by clusters that are interesting individually (Bullet cluster, RX J1347.5), as well as 
those reported to have radio haloes. Galaxy clusters with reported hard X-ray emission, and those predicted 
from large scale structure formation simulations to be detectable by the LAT in one year \citep{pfrommer2008} are implicitly included the sample.

\item Pulsars: These are selected from the ATNF catalog \citep{ATNFcatalog}, with a cut on $\dot E > 10^{34}$ erg s$^{-1}$.

\item Millisecond Pulsars: These are also selected from the ATNF catalog to have periods 
less than 10 ms, with the same cut as above in $\dot E$.

\item Magnetars:  We include all 13 known SGRs and AXPs.

\item EGRET-SNRs:  These are all SNRs that were found to be spatially coincident with an EGRET 
source, as discussed by \citet{torres03}. 

\item  TeV shell--type SNRs: these were intended to be tested separately to answer 
the question whether these objects also emit GeV $\gamma$-rays on at least the level of the LAT 1 yr sensitivity. The sample  
comprised 4 sources (RX J1713.7$-$3946, RX J0852$-$460, RCW 86 and Cas A).

\item Star-star binaries: These are from the VIIth WR catalog of \citet{vanderHucht2001}, requiring 0.001 
$E_{kin,tot} /(4\pi d_L^2)  > 10$, resulting in a sample of 41 sources. The factor 0.001 is motivated by a reasonable
   10\% acceleration efficiency, i.e., wind-energy-to-relativistic-particle-energy-conversion efficiency, and 1\% radiative efficiency
   in these environments, while the factor 10 relates to the 
   LAT sensitivity anticipated at the time of protocol application. 

\item Star-compact object binaries: 
These are all known $\gamma$-ray binaries and microquasars (MQs).

\item Binary pulsars: We include all objects known in this category.

\item Globular clusters: Globular clusters are known to contain a relatively large number of millisecond pulsars (MSPs)
  whose individual and collective emission in the X-ray and $\gamma$-ray energy bands may be detectable by the LAT. 
Here the aim is to test for the collective emission from MSPs, 
since MSPs are principally able to accelerate leptons at the shock waves originated 
in collisions of the pulsar winds and/or inside pulsar magnetospheres, and inside a globular cluster these are subsequently 
able to Comptonize stellar and microwave background radiation. The globular clusters in this list are restricted to be closer than 6~kpc. 

\end{itemize}

\subsubsection{Protocol results}

\begin{deluxetable}{lrrrllcl}
\tablecaption{Results from Application of the Population Protocol 
  \label{tab:protocol}}
\tabletypesize{\scriptsize}
\tablewidth{0pt}

\tablehead{
\colhead{Test Population} & 
\colhead{$N_p$}  & 
\colhead{$b$} &  
\colhead{${\cal C}$} & 
\colhead{CP} & 
\colhead{$P$} & 
\colhead{(CP $< P$)?} & 
\colhead{CL}
}
\startdata
\sidehead {\bf Galactic Populations}

 Pulsars & 215 & 1.440 & 30 & $5.3 \times 10^{-29}$ & $1.0 \times 10^{-9}$ & yes &  $>99.999\%$\\
 Millisecond Pulsars & 23 & 0.050 & 7 & $1.5 \times 10^{-13} $ & $1.0 \times 10^{-8}$ & yes &  $99.89\%$\\
 EGRET SNRs & 23 & 1.590 & 13 & $1.5 \times 10^{-8} $ & $1.0 \times 10^{-8}$ &   no & \nodata \\    
 TeV SNRs & 4 &  0.920  & 3 & $ 6.6\times 10^{-2} $ & $9.9 \times 10^{-7}$ & no &  \nodata \\
 Magnetars & 13 & 0.120 & 0 & \nodata & $9.9 \times 10^{-7}$ & \nodata  & \nodata \\
 WR-binaries & 41 & 0.260 & 0 & \nodata & $9.9 \times 10^{-7}$   & \nodata  & \nodata \\
 MQ/$\gamma$-ray bin.  & 17 & 0.140 & 3 &  $4.1 \times 10^{-4} $   & $9.9 \times 10^{-7}$   & no  & \nodata \\
 Binary pulsars & 10 & 0.040 & 0 & \nodata & $9.9 \times 10^{-7}$   & \nodata  & \nodata \\
 Globular clusters & 29 & 0.240 & 4 & $1.1 \times 10^{-4}$ & $9.9 \times 10^{-7}$   & no & \nodata \\

\sidehead{\bf  Extragalactic Populations}
 Blazars & 215 & 0.480 & 61 & 0.0  & $1.0 \times 10^{-9}$ & yes & $ >99.999 \%$ \\
 Misaligned jet sources  & 53 & 0.150 & 5 & $5.5 \times 10^{-7}$ & $9.9 \times 10^{-7}$     & yes & $99.25\%$\\
 Starbursts & 15 & 0.050 & 4 & $2.5 \times 10^{-7}$ & $9.9 \times 10^{-7}$   & yes &  $97.89\%$ \\
 Galaxy clusters & 48 & 0.150 & 0 &  \nodata     & $9.9 \times 10^{-7}$  & \nodata  & \nodata \\
 Dwarf spheriodals & 18 &0.070 & 0 & \nodata & $9.9 \times 10^{-7}$   & \nodata  & \nodata \\   
\enddata
    
\end{deluxetable}

Table~\ref{tab:protocol} summarizes the results of the application of the protocol for population identification to the 1FGL catalog.  The first column shows the name of the test population. 
The second column ($N_p$) shows the number of astrophysical objects included in the test of the population.
The third column shows the expected number of random coincidences (or background, $b(A)$) between the 1451 LAT sources in the 
1FGL catalog, and the $N_p$ candidates in each population as explained above.
The fourth column (${\cal C}$) gives the actual number of positional coincidences between the members of each population and the 1FGL catalog.
The fifth column shows the cumulative probability (CP) of obtaining ${\cal C}$ or more positional coincidences between 1FGL sources and the 
astrophysical candidates purely by chance. 
The sixth column shows the a priori assigned probability ($P(A)$) 
for discovery of each population, the sum of all $P$ giving the total budget for population discovery. 
The seventh column answers yes 
or no to the question of whether the actual random probability for an equal or a larger number of excesses to occur by chance (CP) is larger than the a priori
assigned budget for this population. 
For the cases in which this answer is yes, the last column gives the confidence level of the detection of the corresponding 
population; alternatively stated, this is the confidence with which we are ruling out the hypothesis that such a population is not present among the LAT detections (the null hypothesis). 

Not surprisingly, the source populations already conclusively identified in the EGRET era are found with the highest confidence in the investigated 1FGL coincidences even when very strict thresholds were chosen for associations to be claimed, justifying a posteriori the very low budget assigned to these populations. 

To study the sensitivity of the results to the sizes of the error ellipses for the sources, we also evaluated the coincidences and chance probabilities when the extents of the ellipses are increased by a (very conservative) 20\%.  The findings for misaligned jet sources and EGRET SNR populations were not robust against this change.  When the sizes of the source location regions are increased the probabilities of chance associations necessarily increase for any given source.  For the misaligned jet source population, this was enough to push the population below the a priori probability budget.  For the EGRET SNRs, enlarging the source location regions had the offsetting effect of adding one more coincidence with 1FGL sources so the net result was a chance probability below the a priori threshold.

Spatial coincidences found between LAT sources and individual millisecond PSRs (MSPs) and those tested via their globular cluster environment are related in the following sense.  Whereas for an isolated MSP there are not many alternative scenarios for producing detectable $\gamma$-ray emission, and in fact several of them have been detected and identified by their $\gamma$-ray periodicity \citep{LATMSPs}, in the case of coincidences with globular clusters \citep[e.g., 47~Tuc, ][]{LAT47Tuc}, an ambiguity exists as long as no pulsed emission from one of MSPs in the globular cluster is found. 
Thus, the aim was to search not only for the existence of a population of MSPs, but also to distinguish between environments in which MSPs are found as populations as well. The fact that the class of globular clusters is not detected as a population but the millisecond pulsars are can be interpreted as related to the size of the population of MSPs in a globular cluster. Based on the membership of the 1FGL catalog apparently only those clusters hosting a large number of MSPs were found coincident with catalog sources (e.g., Ter 5, 47~Tuc, M28), too small a number to claim them as a population of $\gamma$-ray sources in the framework of this test.

The previously-reported relation between EGRET sources and SNRs \citep[e.g., ][]{sturner1995, torres2003} cannot be confirmed with 1FGL sources using the present test.  The chance probability of the number of coincidences seen with 1FGL sources is very small, but the a priori probability budget assigned for this population was smaller still (Table~\ref{tab:protocol}).  We note also that pulsars have been detected by the LAT in some of the SNRs in our sample.

The fact that the misaligned jet sources are found as a population supports the individual 1FGL source association findings, although the test did not distinguish between radio galaxies and Seyferts, and does not show a significant correlation with ranking via radio core flux. It is to be noted that this class is found only when the uncertainties in the positions of the LAT sources are not enlarged, when the cumulative probability is low, but found to be just below the a priori assigned budget for the class.

Starburst galaxies are detected as a population in LAT data, in numbers that add up to the noted individual detections of M82 and NGC~253, and suggests the potential for future discoveries. 

Particularly interesting is the non-detection of galaxy clusters, which is not only a statistically significant result, but even a case of zero coincidences.
With the absence of even a single galaxy cluster coincidence from the tested sample, we can conclude that X-ray bright nearby galaxy clusters and those exhibiting a radio halo do not constitute a source population above the 1FGL sensitivity limit:  the models that we used to select the candidate galaxy clusters overestimate the energy conversation into particle acceleration at GeV and greater energies. 

Dwarf spheroidals are not found coincident with LAT sources either. The result is compatible with the null hypothesis that this population does not exist above the sensitivity limit of the catalog, although in some dark-matter based models of $\gamma$-ray production they would have been expected to be seen at the sensitivity level of the 1FGL catalog. The Galactic populations of magnetars, binary pulsars, and WR binaries are similarly not detected.  The latter is particularly interesting; whereas positional associations of WR stars with LAT sources can be found using the extensive list of known WR stars, those having the largest wind energies do not present any correlation and thus we disregard WRs as $\gamma$-ray emitters at the sensitivity limit of the 1FGL catalog.

Finally, 
we note that whereas we found three sources correlated with MQ-$\gamma$-ray binaries (all of them secured by timing) the population as such is not detected. We can conclude that the binaries identified in the 1FGL catalog are certainly special objects, but probably not archetypal for a population of similar objects.  Similarly, the same occurs for TeV SNRs: we find some correlated objects, but not enough to claim a population discovery due to the large expected background ($b$) for this population.

\subsection{Automated Source Associations}
\label{run:autoassoc}

Our approach for automated source association follows closely that used for the BSL, 
although we enlarged our database of catalogs of potential counterparts  and improved our 
calibration scheme to control more precisely the expected number of false associations.
The association procedure, which follows essentially the ideas developed by \cite{mattox97b} 
for the identification of EGRET sources with flat-spectrum radio sources, is described in detail
in Appendices B and C.
Here we only summarize the essential steps of the automated source association procedure.

The automated source association is based on a list of catalogs that contain potential counterparts 
of LAT sources.
This list has been compiled based either on prior knowledge about classes of high-energy 
$\gamma$-ray emitters or on theoretical expectations.
In total, 32 catalogs (some of which are subselections from 24 primary catalogs, see Table~\ref{tab:catalogs}) 
have been searched for counterparts covering
AGNs (and in particular blazars),
nearby and starburst galaxies,
pulsars and their nebulae,
massive stars and star clusters, and
X-ray binaries.  For the BSL analysis, only 14 catalogs were searched.  In addition, some of the catalogs have been enlarged somewhat (e.g., BZCAT, CRATES, and SNR total; cf. Table~\ref{tab:catalogs} with Table~1 of the BSL).
Furthermore, since millisecond pulsars have now been established as sources of $\gamma$-rays, we split the ATNF catalog into normal pulsars and millisecond pulsars. For the latter we require the pulse period $P <$ 0.1 s and the period derivative $\dot{P} < 10^{-17}$ s/s. This results in 139 objects. The remaining 1545 objects are considered as normal pulsars. We divided those into high and low $\dot{E}/d^2$ categories, the latter being defined as $\dot{E}/d^2 \le 5 \times 10^{33}$ erg~kpc$^{-2}$~s$^{-1}$.  This results in 84 high $\dot{E}/d^2$ pulsars and 1461 low $\dot{E}/d^2$ pulsars.  Note that this energy-flux selecting is unrelated to the selection on spin-down luminosity applied for the population study in \S~\ref{run:pops}.  
For SNRs, we divided the Green catalog into two lists, one containing all objects that can be considered as point-like for the LAT (157 objects) and one containing extended supernova remnants (117 objects, diameters greater than 20$\arcmin$); these subsets are denoted Small and Large, respectively. 
We search for counterparts at radio frequencies using the VLBA Calibrator Survey and at TeV energies using the TeVCat catalog.  For the latter we divide the TeVCat catalog into Small and Large angular size subsets at 40$\arcmin$. 
We also search for coincidences between 1FGL sources and BSL (0FGL), AGILE, and EGRET sources.  
The complete list of catalogs, the numbers of objects they contain, and the references
are presented in Table~\ref{tab:catalogs}. 

\begin{deluxetable}{lrrrrrr}
\tablecaption{Catalogs Used for the Automatic Source Association and Results
\label{tab:catalogs}
}
\tablewidth{0pt}
\tablehead{

\colhead{Name} & 
\colhead{Objects} & 
\colhead{$P_{\rm prior}$} & 
\colhead{\nass} & 
\colhead{\nfalse} & 
\colhead{\nfalsemc} & 
\colhead{Ref.}
}

\startdata
LAT pulsars & 56 & 0.1 & 56 & n.a. & 0.4 & 1 \\
High $\dot{E}/d^2$ pulsars & 84 & 0.024 & 24 & n.a. & 0.6 & 2 \\
Low $\dot{E}/d^2$ pulsars & 1461 & 0.011 & 1 & n.a. & 0.3 & 2 \\
Millisecond pulsars & 139 & 0.278 & 20 & n.a. & 1.0 & 2 \\
Pulsar wind nebulae & 69 & 0.049 & 27 & 0.3 & 0.9 & 1 \\
High-mass X-ray binaries & 114 & 0.010 & 3 & n.a. & 0.3 & 3 \\
Low-mass X-ray binaries & 187 & 0.050 & 8 & 0.4 & 0.5 & 4 \\
Small ($<$20$\arcmin$) SNRs & 157 & 0.021 & 11 & 0.7 & 0.7 & 5 \\
O stars & 378 & 0.015 & 1 & $<0.1$ & $<0.1$ & 6 \\
WR stars & 226 & 0.013 & 11 & 0.3 & 0.2 & 7 \\
LBV stars & 35 & 0.026 & 2 & 0.3 & 0.6 & 8 \\
Open clusters & 1689 & 0.013 & 1 & 0.1 & 0.4 & 9 \\
Globular clusters & 147 & 0.272 & 8 & $<0.1$ & 0.5 & 10 \\
Nearby galaxies & 276 & 0.066 & 5 & 0.4 & 0.4 & 11 \\
Starburst galaxies & 14 & 0.5 & 2 & $<0.1$ & $<0.1$ & 12 \\
Blazars (BZCAT) & 2837 & 0.308 & 487 & 8.9 & 6.8 & 13 \\
Blazars (CGRaBS) & 1625 & 0.238 & 282 & 4.7 & 4.1 & 14 \\
Blazars (CRATES) & 11499 & 0.333 & 490 & 17.2 & 17.8 & 15 \\
BL Lac & 1122 & 0.224 & 218 & 2.8 & 2.8 & 16 \\
AGN & 21727 & 0.021 & 11 & 0.7 & 0.8 & 16 \\
QSO & 85221 & 0.166 & 147 & 7.3 & 4.9 & 16 \\
Seyfert galaxies & 16343 & 0.041 & 24 & 2.0 & 1.6 & 16 \\
Radio-loud Seyfert galaxies & 29 & 0.1 & 4 & $<0.1$ & $<0.1$ & 1 \\
VLBA Calibrator Survey & 4558 & 0.266 & 484 & 11.5 & 10.1 & 17 \\
Small ($<$40$\arcmin$) TeV sources & 92 & 0.037 & 42 & 0.6 & 0.8 & 18 \\
Large  ($>$40$\arcmin$) TeV sources\tablenotemark{\dag} & 11 & n.a. & 13 & n.a. & 7.5 & 18 \\
Large ($>$20$\arcmin$) SNRs\tablenotemark{\dag} & 117 & n.a. & 48 & n.a. & 18.1 & 5 \\
Dwarf galaxies\tablenotemark{\dag}  & 14 & n.a. & 7 & n.a. & 2.1 & 1 \\
1st AGILE catalog\tablenotemark{\ast}  & 47 & n.a. & 52 & n.a. & 18.6 & 19 \\
3rd EGRET catalog\tablenotemark{\ast}  & 271 & n.a. & 107 & n.a. & 25.4 & 20 \\
EGR catalog\tablenotemark{\ast} & 189 & n.a. & 66 & n.a. & 9.1 & 21 \\
Bright Source List (0FGL) & 205 & n.a. & 195\tablenotemark{a} & n.a. & 3.9 & 22 \\
\enddata

\tablerefs{
1 Collaboration internal;
2 \citet{ATNFcatalog}; 
3 \citet{HMXBcatalog}; 
4 \citet{LMXBcatalog}; 
5 \citet{SNRcatalog};   
6 \citet{OStarCatalog}; 
7 \citet{vanderHucht2001}; 
8 \citet{LBVcatalog}; 
9 \citet{OpenClustersCatalog}; 
10\citet{GlobClusterCatalog}; 
11\citet{NearbyGalaxiesCatalog}; 
12 \citet{Thompson2007}; 
13 \citet{BZcatalog}; 
14\citet{CRATES}; 
15 \citet{CGRaBS}; 
16 \citet{AGNcatalog};  
17 \citet{VLBISurvey} and ref. therein, http://astrogeo.org/vlbi/solutions/2009c\_astro/; 
18 http://tevcat.uchicago.edu/ (`Default' and `Newly Announced' categories, version 3.100);
19 \citet{AGILEcatalog}; 
20 \citet{3EGcatalog}; 
21 \citet{EGRcatalog};
22 \citet{LATBSL}}

\tablenotetext{\ast}{Catalog for which the location uncertainties of the counterparts are greater than the location uncertainties for the 1FGL sources.}
\tablenotetext{\dag}{Catalog for which the counterparts are spatially extended sources.}
\tablenotetext{a}{See \S~\ref{run:0fgl}.}

\end{deluxetable}


For each catalog in the list, we make use of Bayes' theorem to compute the posterior probabilities
$P_{ik}$ that an object $i$ from the catalog is the correct association of the LAT source $k$:
\begin{equation}
P_{ik} = \left( 1 + \frac{1-P_{\rm prior}}{P_{\rm prior}} 2 \pi \rho_k a_k b_k 
e^{\Delta_k} \right)^{-1} \, .
\end{equation}
$P_{\rm prior}$ is the prior probability that counterpart $i$ is detectable by the LAT,
$a_k$ and $b_k$ are the axes of the ellipse at 1 $\sigma$, smaller than the semimajor and semiminor axes at 95\% confidence by a factor $\sqrt{-2\log(0.05)} = 2.45$,
$\rho_k$ is the local counterpart density around source $k$, and
\begin{equation}
\Delta_k = \frac{r^2}{2} \left( \frac{\cos^2(\phi-\phi_k)}{a_k^2} + \frac{\sin^2(\phi-\phi_k)}{b_k^2} \right)
\end{equation}
for a given position angle $\phi$
between LAT source $k$ and the counterpart $i$,
$\phi_k$ being the position angle of the error ellipse, and
$r$ being the angular separation between LAT source $k$ and counterpart $i$
(see Appendix B).
For each catalog, prior probabilities $P_{\rm prior}$ are assigned so that
\begin{equation}
\nfalse = \sum_{P_{ik} \ge P_{\rm thr}} (1-P_{ik})
\label{eq:nfalsecmp}
\end{equation}
gives the expected number of false associations that have posterior probabilities
above the threshold $P_{\rm thr}$ (see Appendix C).
The corresponding prior probabilities are quoted in Table~\ref{tab:catalogs} (column 3) for 
all catalogs.

For the automated association of the 1FGL catalog we set $P_{\rm thr}=0.8$, which means
that each individual association has a $\le20\%$ chance of being spurious.
This is different from the approach we took for the BSL paper where we constrained
the expected number of false associations for each catalog to $N_{\rm false} \le 1$, which
imposed a relatively tight constraint on source classes with large numbers of associations
(such as blazars and pulsars) while source classes with only few associations had a
relatively loose constraint.
Now, each individual association stands on an equal footing by having a well defined
probability for being spurious.  We note that the First LAT AGN Catalog \citep{LATAGNCatalog} applies the same association method as we use here but includes associations for AGNs down to $P_{\rm thr}=0.1$.

For a number of catalogs in our list the Bayesian method cannot be applied since either
(1) the location uncertainty of the counterpart is larger than the location uncertainty
for the 1FGL source, or
(2) the counterpart is an extended source; see notes to Table~\ref{tab:catalogs}. 
In the first case, we consider all objects $i$ as associations for which the separation
to the LAT source $k$ is less than the quadratic sum of the 95\% confidence error
radius of counterpart $i$ and the semimajor axis $\alpha_k$.  {For the bright EGRET pulsars Crab, Geminga, Vela, and PSR J1709$-$4429, we also list associations, even though in many cases the EGRET-measured locations are formally inconsistent with the positions of the pulsars; there is no doubt that EGRET detected these pulsars.}
In the second case, we assume that the counterparts have a circular extension
and consider all objects $i$ as associations for which the circular extension overlaps with 
a circle of radius $\alpha_k$ around the LAT source $k$.

\subsubsection{Automated association summary}

The results of the automated association procedure for each of the external catalogs 
are summarized
in Table~\ref{tab:catalogs}.
For each catalog, we give the number
\begin{equation}
\nass = \sum_{P_{ik} \ge P_{\rm thr}} 1
\end{equation}
of LAT sources that have been associated with objects in a given catalog (column 4).
Furthermore, we compute the expected number of false associations \nfalse\ using
Eq.~(\ref{eq:nfalsecmp}) for those catalogs which have been associated with the
Bayesian method (column 5).
We cannot give meaningful results for pulsars and high-mass X-ray binaries since,
for identified objects, the positions have been fixed in the catalog to their high-precision locations (\S~\ref{run:construction}), 
and consequently, their posterior probabilities are by definition 1.
However, we alternatively estimated the expected number of false associations
using Monte Carlo simulations of 100 realizations of fake LAT catalogs, for which
no physical associations with counterpart catalog objects are expected  (see Appendix C).
We quote the resulting estimates \nfalsemc\ in column 6 of Table~\ref{tab:catalogs}.
We find $\nfalse \approx \nfalsemc$ which confirms that the posterior probabilities
computed by the automatic association procedure are accurate
(otherwise Eq.~(\ref{eq:nfalsecmp}) would not hold).

In total we find that 821 of the 1451 sources in the 1FGL catalog (56\%) have been 
associated with a least one non-$\gamma$-ray counterpart by the automated procedure at the 80\% confidence level.
779 1FGL sources (54\%) have been associated using the Bayesian method while the 
remaining 42 sources are spatial coincidences based on overlap of the error regions or 
source extents.
From simulations we expect that 57.3 among the 821 sources (7\%) are associated 
spuriously.
Considering only the Bayesian associations, 37.5 among the 779 sources (5\%) are 
expected to be spurious.
In the following we discuss the automated association results in some detail.  Associations with TeV sources are discussed in \S~\ref{run:tev}.

\subsubsection{Blazars}

Our association procedure contains 4 catalogs to cover the blazar source class
(BZCAT, CGRaBS, CRATES, BL Lac)
and these catalogs have
 a substantial number of objects in common.
In total we find 689 1FGL sources associated with sources from at least one of the 4 
blazar catalogs.
2 of these sources (1FGL~J0047.3$-$2512 and 1FGL~J0956.5$+$6938)
are the starburst galaxies M~82 and NGC~253 (both found in the
CRATES catalog), 
and 2 sources (1FGL~J0319.7$+$4130 and 1FGL~J1325.6$-$4300) are 
the radio galaxies NGC~1275 and Cen~A.
This leaves 685 blazar candidates among the 1FGL sources.

We further note that 282 of the 1FGL sources associated with blazars also have
counterparts in the VLBA calibrator survey (VCS) which we added to our list of catalogs
following the suggestion of \citet{kovalev2009} who found 111 associations for this
catalog among the BSL sources.
For 37 of the 484 1FGL sources associated with VCS objects, the VCS association
is the only counterpart found among all catalogs.
Most of these 37 sources are located at low Galactic latitudes, a region in which our
4 blazar catalogs are incomplete.
Many of the low-latitude VCS associations thus may be related to blazars situated close
to the Galactic plane \citep{kovalev2009}.

\subsubsection{Other AGNs}

We find 24 1FGL sources that are associated with objects from the two
Seyfert galaxy catalogs.
Among those, only two sources are not also associated with blazars:
1FGL~J0840.8$+$1310 (3C~207.0) and
1FGL~J1230.8$+$1223 \citep[M~87;][]{LATM87}
3C~207.0 is a lobe-dominated quasar and and M~87 is a radio galaxy.

147 of the 1FGL sources are associated with AGNs and QSOs from the catalog of \citet{AGNcatalog},
 yet all of these are also associated with blazars.  Therefore apparently most of the 1FGL non-blazar associations are either sources that
are in fact blazars, yet are not classified as such in our catalogs, or they are
nearby radio galaxies.
In particular, we do not find convincing evidence for coincidences of 1FGL sources
with non-blazar Seyfert galaxies.

\subsubsection{Normal Galaxies}
\label{run:galaxies}
We find 2 associations with nearby starburst galaxies \citep{LATStarbursts}
1FGL~J0047.3-2512 (NGC~253), and
1FGL~J0956.5+6938 (M~82).
Both galaxies have also been detected at TeV energies \citep{HESSstarburst}
and \citep{VERITASstarburst}
and hence can be considered as high-confidence 1FGL associations.

Seven 1FGL sources are found to coincide with dwarf galaxies:
6 are associated with the Large Magellanic Cloud (LMC), 1 is associated with the Small
Magellanic Cloud (SMC), both galaxies being extended.
These sources probably correspond to local maxima of extended emission features
and probably do not represent real point sources in the field. Regarding the LMC, in Table~\ref{tab:sources} only 5 of the sources are indicated as being associated with the LMC.
The exception is
1FGL~J0600.7$-$7037 which is also associated with 
PKS 0601$-$70, a blazar that was significantly variable
during the time span of the 1FGL catalog \citep{LAT30Doradus}.

Thus NGC~253, M~82, the LMC, and the SMC are so far the only normal galaxies that
have been associated with sources in the 1FGL catalog.  We note that 1FGL J1305.4$-$4928 is associated with NGC~4945, which is a starburst galaxy that is also classified as a Seyfert II AGN.

\subsubsection{Pulsars, pulsar wind nebulae and globular clusters}
\label{run:psr}

56 1FGL sources have been identified as pulsars
through their $\gamma$-ray pulsations.  For these sources we list only the pulsar identification in Table~\ref{tab:sources}.  This is not to be taken to mean that we have necessarily ruled out contributions from known or unknown PWNs or SNRs; in fact for the sources identified with the Crab and Vela pulsars we have also identified PWNs; these sources have both class assignments

In addition to the 56 seen pulsating, we find 3 more associations with the high $\dot{E}/d^2$ subset of pulsars from the
ATNF catalog:
\begin{itemize}
\item 1FGL~J1119.4$-$6127c (PSR~J1119$-$6127)
\item 1FGL~J1410.3$-$6128c (PSR~J1410$-$6132)
\item 1FGL~J1648.4$-$4609c (PSR~J1648$-$4611)
\end{itemize} 
These 3 sources are good candidates for young energetic $\gamma$-ray pulsars,
although 1FGL~J1119.4$-$6127 and 1FGL~J1410.3$-$6128c are also associated with SNRs and/or PWNs (see Table~\ref{tab:snr_over}).  

Among the 1FGL sources that are associated with pulsar wind nebulae (PWNs), only 6 are not 
also associated with known pulsars:
\begin{itemize}
\item 1FGL~J1134.8$-$6055 (PWN~G293.8$+$0.6)
\item 1FGL~J1552.4$-$5609 (PWN~G326.3$-$1.8)
\item 1FGL~J1635.7$-$4715c (PWN~G337.2$+$0.1)
\item 1FGL~J1640.8$-$4634c (PWN~G338.3$-$0.0)
\item 1FGL~J1745.6$-$2900c, the Galactic center source (PWN~G359.95$-$0.04)
\item 1FGL~J1746.4$-$2849c (PWN~G0.13$-$0.11)
\end{itemize}
It remains to be shown whether the LAT indeed detects these PWNs, or whether
the $\gamma$-ray emission arises from the yet-unknown pulsars that power the nebulae, or potentially from an associated SNR.  Because of the ambiguity we list only two positional associations with PWNs in Table~\ref{tab:sources}, for 1FGL~J1635.7$-$4715c and 1FGL~J1746.4$-$2849c, which do not also have associations with known SNRs.  The others are included in Table~\ref{tab:snr_over}.

Among the 20 1FGL sources that are associated with millisecond pulsars, 11 are 
not associated with known $\gamma$-ray pulsars.
Among those 11, 5 are associated with globular clusters and the remaining 6 may indeed be
Galactic field $\gamma$-ray millisecond pulsars:
\begin{itemize}
\item 1FGL~J0610.7$-$2059 (PSR~J0610$-$2100)
\item 1FGL~J1024.6$-$0718 (PSR~J1024$-$0719)
\item 1FGL~J1600.7$-$3055 (PSR~J1600$-$3053)
\item 1FGL~J1713.9$+$0750 (PSR~J1713$+$0747)
\item 1FGL~J1811.3$-$1959c (PSR~J1810$-$2005)
\item 1FGL~J1959.6$+$2047 (PSR~B1957$+$20)
\end{itemize}

Finally, we find that 8 1FGL sources are associated with globular clusters.
None of those have alternative associations different from millisecond pulsars or low-mass
X-ray binaries (which both are known source populations residing in globular clusters),
which makes the reality of these associations even more plausible.  

\subsubsection{Supernova remnants}

Our automated association procedure associates 59 1FGL sources with SNRs.  Not counting associations that also include pulsars detected by the LAT, the total is 41 (Tab.~\ref{tab:snr_over}).
Of those, 5 are associated with small angular size (diameter less than 20$^\prime$) SNRs:
\begin{itemize}
\item 1FGL~J1134.8$-$6055 (SNR~G293.8$+$00.6, also associated with PWN G293.8$+$0.6)
\item 1FGL~J1213.7$-$6240 (SNR~G298.6$-$00.0)
\item 1FGL~J1617.5$-$5105 (SNR~G332.4$-$00.4)
\item 1FGL~J1640.8$-$4634 (SNR~G338.3$-$00.0, also associated with PWN G338.3$-$0.0)
\item 1FGL~J2323.4$+$5849 \citep[SNR G111.7$-$02.1, aka Cas A;][]{LATCasA}
\end{itemize}
Except for 1FGL~J2323.4$+$5849 and possibly 1FGL~J1213.7$-$6240, the presence
of alternative associations to PWNs or a low-mass X-ray binary make the
physical association of these sources to SNRs questionable.

Some of the associations with SNRs have already been suggested based on
morphology analyses of the LAT sources:
\begin{itemize}
\item 1FGL~J1856.1$+$0122 \citep[G034.7$-$00.4, aka W44;][]{LATW44}
\item 1FGL~J1922.9$+$1411 \citep[G049.2$-$00.7, aka W51C;][] {LATW51C}
\item 1FGL~J0617.2$+$2233 \citep[G189.1$+$03.0, aka IC~443;][]{LATIC443}
\end{itemize}
We consider these three SNRs to be identified sources in the 1FGL catalog; see \S~\ref{run:ids}.  Further interesting associations due to the presence of OH masers in the
SNR \citep{Hewitt2009} are
1FGL~J1805.2$-$2137c and 1FGL~J1806.8$-$2109c (both overlapping G008.7$-$00.1, also known as W30).

\subsubsection{Association of 1FGL~J1745.6$-$2900c with the Galactic center}
\label{run:gc}

With a position $(l,b)=(359.941^\circ,-0.051^\circ)$ and a 95\% confinement radius of $1.1^\prime$,
1FGL~J1745.6$-$2900c is the source closest to the Galactic center.
In this direction, many catalogs contain objects and consequently we
find a large number of formal associations to this source.
Specifically, 1FGL~J1745.6$-$2900c is formally associated with
the pulsar wind nebula G359.95$-$0.04,
the SNR G000.0$+$00.0,
the VCS object J1745$-$2900,
4 low-mass X-ray binaries, 6 LBV stars, and 10 Wolf-Rayet stars and 2 TeV sources.
We are unable to distinguish on the basis of our association scheme among
these possibilities, although some are more plausible physically.
Eventually, the spectral energy distribution of the source or any characteristic time-variability 
may help to narrow down the possibilities.

\subsubsection{X-ray binaries}

Three 1FGL sources have
been identified by their orbital modulations as high-mass X-ray binaries (HMXB):
\begin{itemize}
\item 1FGL~J0240.5$+$6113 \citep[LS~I$+61^\circ303$;][]{LATLSI+61}
\item 1FGL~J1826.2$-$1450 \citep[LS~5039;][]{LATLS5039}
\item 1FGL~J2032.4$+$4057 \citep[CygX-3;][]{LATCygX3}
\end{itemize}

Formally, we associate five 1FGL sources with low-mass X-ray binaries (LMXB).
Three of them are also associated with globular clusters, and hence their combined emission from
millisecond pulsars appears to be the more plausible counterpart of the 1FGL sources
(see \S~\ref{run:psr}) and we do not list LMXBs as a separate source class.
One association corresponds to the Galactic Center source (cf.~section \ref{run:gc}).
And the remaining association,
1FGL~J1617.5$-$5105c (1E~161348$-$5055.1) is also associated with the SNR G332.4$-$00.4.
Thus, none of the LMXB associations gives strong evidence that we indeed detect
$\gamma$-ray emission from this source class.

\subsubsection{O stars, Wolf-Rayet stars, Luminous Blue Variable stars and open clusters}

The automated association procedure finds one O star (Cyg OB2-4) associated with
1FGL~J2032.2$+$4127, yet this source is known to be a $\gamma$-ray pulsar
\citep[PSR~J2032$+$4127, ][]{LATPulsars,LATBSPs, Camilo09}.
In this case, the unusually large density of O stars in the Cyg OB2 association \citep{Knoedlseder2000}
leads to source confusion, and ignoring this association we do not find any evidence
for $\gamma$-ray emission from O stars in the 1FGL catalog.

Formally, two 1FGL sources were associated with Wolf-Rayet stars.
The first is the Galactic center source 1FGL~J1745.6$-$2900c, which, as stated above, has many
possible alternative associations, so we definitely cannot establish a physical link between
the LAT source and the Wolf-Rayet stars.
The second is 1FGL~J2032.4$+$4057 which has been identified as Cyg~X-3 \citep{LATCygX3}.
Cyg~X-3 is a compact binary system that indeed comprises a Wolf-Rayet star, hence here the
association is indeed correct.
It is unlikely that the $\gamma$-ray emission is indeed arising from the Wolf-Rayet
star \citep{LATCygX3}.  Thus we also do not find any evidence for $\gamma$-ray emission from 
Wolf-Rayet stars in the 1FGL catalog.

Two 1FGL sources were associated with Luminous Blue Variable (LBV) stars:
1FGL~J1745.6$-$2900c, the Galactic center source, which for the same reason as given above
we do not consider as a relevant association, and
1FGL~J1746.4$-$2849c, which is associated with FMM~362 and the Pistol Star.
However, the latter source is also associated with the PWN G0.13$-$0.11,
so also here it is difficult to establish a physical link.

Finally, we note that the LBV star $\eta$ Carinae was not formally associated by
our procedure to the nearby source 1FGL~J1045.2$-$5942.
The formal posterior association probability for $\eta$ Carinae is $0.76$, hence below 
our threshold (0.8) for listing an association.  
The angular separation is $1.7^\prime$ which is slightly larger than the
95\% containment radius of $1.4^\prime$.
Thus $\eta$ Carinae falls just outside the 95\% error radius of 1FGL~J1045.2$-$5942.
On the other hand, 1FGL~J1045.2$-$5942 was associated with the open cluster
Trumpler~16 (the only association to an open cluster for the 1FGL catalog), which
besides $\eta$ Carinae houses many massive stars, similar to the Cyg~OB2 association.
Recently, the young energetic pulsar PSR~J2032$+$4127 was found by the LAT in
Cyg~OB2 (see above) and it is possible that young energetic pulsars are also
hidden in Trumpler~16.

\subsubsection{Associations with Bright Source List sources}
\label{run:0fgl}
Table~\ref{tab:catalogs} reports 195 associations between the 205 BSL (0FGL) sources and those of the 1FGL catalog.  Strictly applied, the automated association procedure found 186 associations.  We have manually adopted nine more for BSL sources confidently associated with blazars or pulsars.  Regarding the latter, the 1FGL positions are formally 0, indicating that the source positions were assigned to the more-accurately known positions of the pulsars (see \S~\ref{run:construction}), affecting the statistical relationship between the 0FGL and 1FGL positions for these.

The 10 BSL sources that do not have clear counterparts in the 1FGL catalog are listed in Table~\ref{tab:orphans}.  Each is in the Galactic ridge, where for the 1FGL catalog analysis we have recognized difficulties detecting and characterizing sources (\S~\ref{run:ridge}).

\begin{deluxetable}{lrr}
\tablewidth{0pt}
\tablecaption{Bright Source List (0FGL) Sources Without 1FGL Counterparts
\label{tab:orphans}
}

\tablehead{
\colhead{Name 0FGL} & 
\colhead{$l$} & 
\colhead{$b$} 
}

\startdata
J1106.4-6055 &  290.52$\degr$ &    -0.60$\degr$   \\
J1115.8-6108 &   291.66 &    -0.38\\
J1615.6-5049 &   332.35  &   -0.01   \\
J1622.4-4945 &   333.87 &    -0.01\\
J1634.9-4737 &   336.84 &    -0.03\\
J1741.4-3046 &   357.96 &    -0.19\\
J1821.4-1444  &   16.43 &    -0.22\\
J1836.1-0727  &   24.56 &    -0.03\\
J1900.0+0356 &    37.42 &    -0.11\\
J1024.0-5754  &  284.35 &   -0.45 \\
\enddata
\end{deluxetable}

\subsubsection{Associations with EGRET and AGILE sources}

The sources in the 1FGL catalog have positional matches with 107 of the 271 3EG sources \citep{3EGcatalog} (four 3EG sources are each resolved into two 1FGL sources) and 66 of the 188 EGR sources \citep{EGRcatalog}.  A few more of the EGRET sources are close to, but not formally consistent with, 1FGL source locations; the EGRET positions for the bright pulsars were offset from the true positions, for example.  Almost all of the AGNs labeled in the 3EG catalog as good candidates are seen by LAT.  One of the exceptions is 3EG J1230$-$0247 (EGRc J1233$-$0318), which was seen only early in the CGRO mission.  

The 1FGL catalog clearly does not account for a large fraction of the sources seen by EGRET.  In light of  the high sensitivity of the LAT and the fact that the LAT sees most of the EGRET catalog AGNs, which are known to be variable, the absence of more EGRET sources from the LAT catalog cannot be attributed primarily to time variability.  A more likely explanation would seem to be that many of the EGRET sources were not discrete sources but were degree scale or larger diffuse structures not included in the model of Galactic diffuse emission used for analysis of the EGRET data.   Its improved angular resolution and high photon statistics at GeV energies make the LAT far less sensitive to such structures, and the model of Galactic interstellar diffuse emission itself has incorporated far more detail than was available in the EGRET era. See \S~\ref{run:ridge}, however, for a discussion of how even some 1FGL sources may be affected by the modeling of the diffuse $\gamma$-ray emission.  

All 47 of the sources in the first AGILE catalog \citep{AGILEcatalog} have corresponding sources in the 1FGL catalog.  A number of the 1AGL sources map to multiple 1FGL sources, and a few of the 1AGL sources are close but not formally consistent in position with the 1FGL sources.  Nevertheless, the two present high-energy $\gamma$-ray telescopes do appear to be consistent in their detections of bright sources.


\subsection{Firm Identifications}
\label{run:ids}
Firm identifications, indicated in the main table by capitals in the Class column, require more than a high-probability positional association.  The strongest test for identification is time variability, either periodicity or correlation with variability seen at another wavelength.  The 56 pulsars that have class PSR all show high-confidence (statistical probability of chance occurrence less than 10$^{-6}$) periodicity caused by the rotation of the neutron star \citep{LATPulsars, PSR0034, MOREBSP,LATPSRB1509MSH1552}.  Similar confidence levels apply to the three X-ray binary systems whose orbital periods are detected in the LAT data:  LSI +61 303 \citep{LATLSI+61}, LS5039 \citep{LATLS5039}, and Cygnus X-3 \citep{LATCygX3}. 

With the large number of blazars detected by the LAT and the significant variability seen in many of these, the search for correlated variability that can provide firm identifications is a major effort that has not yet been carried out systematically for the LAT data.  We have therefore chosen to list as firm identifications only those blazars for which publications exist showing such variability.  These are just 3C~273 \citep{LAT3C273}, 3C~279 \citep{ATEL2154}, 3C~454.3 \citep{LAT3C454.3}, PKS~1502+106 \citep{LATPKS1502} and PKS~1510$-$08 \citep[e.g., ][]{ATEL1743}.  Additional studies will undoubtedly expand this list. 

Another approach to firm identification, slightly less robust than time variability, is morphology: finding spatial extent in a $\gamma$-ray source that matches resolved emission at other wavelengths.  Some SNRs have measurable spatial extents in the LAT data and can be considered firm identifications; here we cite W44 \citep{LATW44}, W51C \citep{LATW51C}, and IC~443 \citep{LATIC443}.  The analyses for the 1FGL catalog assume point-like emission, and so the positions and fluxes are not as well characterized as they would be in analyses that take into account the finite angular extents of these sources.  A special case is 1FGL J1322.0$-$4515, which appears to be part of one of the lobes of the emission from radio galaxy Centaurus A \citep{LATCenALOBES}.  Studies of source morphology are ongoing. 

The Large Magellanic Cloud (LMC) is an extreme example \citep{LAT30Doradus} of the issue of analyzing an extended source with tools designed to find point sources. As described in \S~\ref{run:galaxies} the catalog contains 5 sources that are likely to be related to the diffuse $\gamma$-ray emission of the LMC.  The LMC can be considered a firmly identified LAT source, but it is not a point source or an ensemble of point sources.

\section{TeV Source Associations}
\label{run:tev}

1FGL sources that are positionally associated with sources seen by the ground-based TeV telescopes are of particular interest because the TeV band overlaps with the LAT energy range, suggesting the potential for such sources to be physically related.  As described in Table~\ref{tab:catalogs}, we investigated associations with the sources in the TeVCat compilation of detections from ground-based observatories.  The compilation is growing with time, and information about the sources is subject to updates and refinements, but at any given time TeVCat represents a snapshot of current knowledge of the TeV sky. 

The association analysis was made separately for extended ($>40^\prime$ diameter) TeV sources (11) and smaller sources.  The latter are typically much smaller than 40$^\prime$ and were treated like point sources for the association analysis.  The `TeV' column of Table~\ref{tab:sources} lists associations with extended sources as `E' and smaller angular-size sources as `P'.  As the table indicates, about half of the TeVCat sources have positionally-plausible associations with 1FGL sources.

The associations include the Milagro sources MGRO J1908+06, MGRO J2019+37, and MGRO J2031+41 \citep{Milagro07}, as well
as MGRO J0632+17 and MGRO J2228+61 \citep{Milagro09, Goodman09}, reported as part of a search for spatial correlations between the Milagro skymap and sources in the $Fermi$ BSL \citep{Milagro09}. A number of the Milagro detections are pulsars with PWNs. The association between GeV $\gamma$-ray PSRs and the PWNs visible in TeV $\gamma$-rays seems well established, as has already been discussed in \cite{LATPulsars}.

Other source classes are represented among the TeVCat associations.  The LAT PSR 1FGL J1023.0$-$5746 is spatially consistent with HESS J1023$-$575, itself not yet firmly identified, but noted for its possible
connection to the young stellar cluster Westerlund 2 in the star-forming region RCW49 \citep{HESS1023}.   Blazars, particularly BL Lac objects are also solidly connected between the GeV and TeV energy ranges \citep{LATTeVblazars}, as are two starburst galaxies \citep{LATStarbursts}, and the HMXB sources LS~I ~+61~303 \citep{LATLSI+61} and LS~5039 \citep{LATLS5039}.  SNRs such as W51C \citep{LATW51C} and Cas A \citep{LATCasA} also connect the two energy regimes. 

Of course sources that are positionally consistent between the LAT and TeV telescopes but have no obvious associations with objects at longer wavelengths are also of interest; Table~\ref{tab:sources} lists a few of these.  Establishing a physical connection through spectral or variability studies may help determine the nature of these sources.

As discussed in Section~\ref{run:gc}, the Galactic Center region is particularly complex.  Investigations of the associations with the TeV $\gamma$-ray sources known in this region -- HESS J1745$-$290 \citep{HESSGCMNRAS}, HESS J1745$-$303 \citep{HESS1745} and HESS J1741$-$302 \citep{HESS1741} are outside the scope of this paper and will be discussed elsewhere.

\section{Conclusion}

The 1451 sources in this First {\it Fermi}-LAT catalog (1FGL) represent the most complete understanding to date of sources in the GeV sky.  The catalog clearly contains a number of populations of $\gamma$-ray emitters.  It offers a multitude of opportunities for additional research, both on sources with likely associations or identifications and on those sources that remain without apparent counterparts. 

\acknowledgments

The {\it Fermi}-LAT Collaboration acknowledges generous ongoing support
from a number of agencies and institutes that have supported both the
development and the operation of the LAT as well as scientific data analysis.
These include the National Aeronautics and Space Administration and the
Department of Energy in the United States, the Commissariat \`a l'Energie Atomique
and the Centre National de la Recherche Scientifique / Institut National de Physique
Nucl\'eaire et de Physique des Particules in France, the Agenzia Spaziale Italiana
and the Istituto Nazionale di Fisica Nucleare in Italy, the Ministry of Education,
Culture, Sports, Science and Technology (MEXT), High Energy Accelerator Research
Organization (KEK) and Japan Aerospace Exploration Agency (JAXA) in Japan, and
the K.~A.~Wallenberg Foundation, the Swedish Research Council and the
Swedish National Space Board in Sweden.

Additional support for science analysis during the operations phase is gratefully
acknowledged from the Istituto Nazionale di Astrofisica in Italy and the Centre National d'\'Etudes Spatiales in France.

The $Fermi$-LAT collaboration acknowledges useful comments from Y. Kovalev on an early version of this work.

This work made extensive use of the ATNF pulsar  catalog\footnote{http://www.atnf.csiro.au/research/pulsar/psrcat}  \citep{ATNFcatalog}.



{\it Facilities:} \facility{Fermi LAT}.




\appendix

\section{Estimation of Detection Threshold}

An approximate, but reasonably accurate, expression for 
the detection threshold for a point source at any point in the sky can
be obtained by assuming that the diffuse background is locally uniform
and considering only one source. We start by constructing the log
Likelihood function given by a sum of the logarithms of Poisson probabilities for
detecting $n_i$ events in some bin $i$, where $i$ labels position in
the sky and energy, when the model predicts $\lambda_i$: $\log L =
\sum_i n_i \log\lambda_i - \lambda_i$. In the above approximation we
can replace the sum over $i$ by an integral over energy and the
angular separation between the source location and the event
direction, $(E,\theta)$. The expectation value for the detected
counts density is $n(E,\theta)=T_0 A_\mathrm{eff}(E) \left[S(E)
  \mathrm{PSF}(\theta,E) + B(E)\right]$ where $S(E)$ and $B(E)$
describe the spectra of the source and background events,
$\mathrm{PSF}(\theta,E)$ the spatial distribution of events from a
point source, $A_\mathrm{eff}(E)$ the effective area of the instrument and $T_0$
the observation duration. To estimate $TS$ for the detection in this case
we calculate the Likelihood under two hypotheses: the maximum
likelihood hypothesis, $L^*$, where the source term is included in the
model, and the null hypothesis, $L^0$, where the counts are presumed
to arise from the background only (also assuming the number of source
counts is small over the full ROI in comparison to the total
background counts).  From these we derive $TS=2(\log L^*-\log L^0)$.  Writing the local
source-to-background ratio as $g(\theta,E) =
S(E)\mathrm{PSF}(\theta,E) / B(E)$ we get:

\begin{eqnarray}
TS & = & 2 T_0 \int_{E_{\rm min}}^{E_{\rm max}} A_\mathrm{eff}(E) \, dE \left(
               \int_0^\pi B(E) \left[1 + g(\theta,E)\right]
               \log\left[1 + g(\theta,E)\right] \, d\Omega - S(E) \right)
\label{eq:ts} \\
 & = & T_0 \int_{\log E_{\rm min}}^{\log E_{\rm max}} W(E) \, d\log E \\
W(E) & = & 2 E A_\mathrm{eff}(E) B(E)
    \int_0^\pi \left[1 + g(\theta,E)\right] \log\left[1 + g(\theta,E)\right] - g(\theta,E) \, d\Omega
\label{eq:wts}
\end{eqnarray}

\begin{figure}
\epsscale{.80}
\plotone{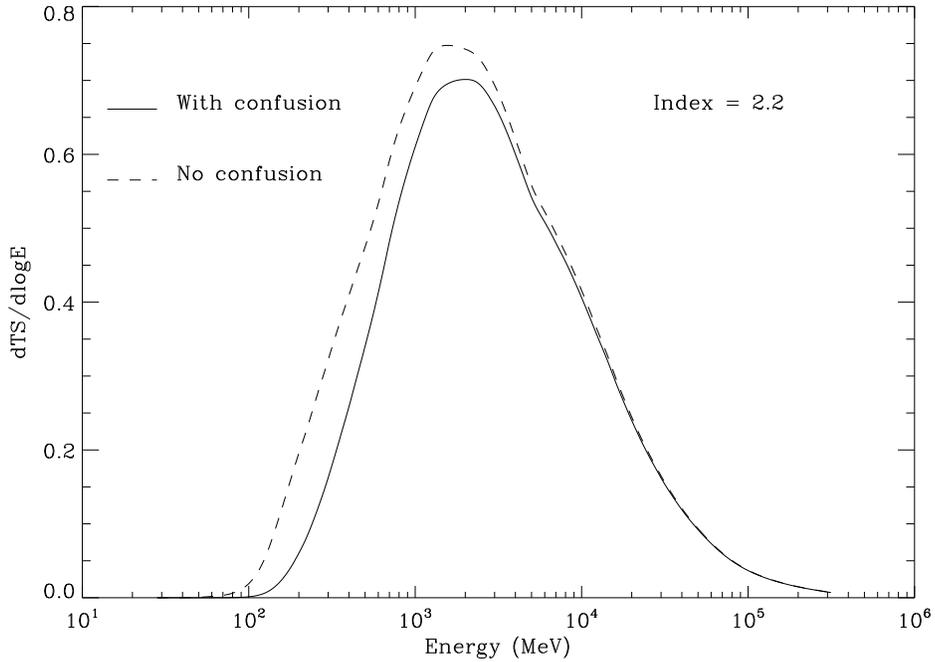}
\caption{Theoretical contribution ($W(E)$ of Eq.~\ref{eq:wts})
to Test Statistic per Ms and per log($E$)
interval as a function of energy for a power-law source over the average
background at $|b| > 10\degr$.
The assumed photon spectral index is 2.2.
The dashed line is for an isolated source.
The full line includes approximately the effect of source confusion.}
\label{fig:tsweight}
\end{figure}

Here $W(E)$ is the contribution to $TS$ per unit $\log E$.  It is
illustrated in Figure~\ref{fig:tsweight} (dashed line) for a power-law
source spectrum with index 2.2.  At low energy (below 1~GeV) faint
sources are always background limited, i.e., $g(\theta,E)$ is small
even at $\theta = 0$. In that limit, the dominant term of the
integrand in the integral over $\theta$ in Eq.~\ref{eq:wts} is
$g(\theta,E)^2/2$.  Noting that $\int_0^\pi {\rm PSF}(\theta,E)^2
d\Omega = C_{\rm sh}(E) \sigma(E)^{-2}$, in which $\sigma(E)$ is the
angular resolution and $C_{\rm sh}(E)$ is a shape factor weakly
dependent on energy, we get
\begin{equation}
\label{eq:wback}
W(E) = E C_{\rm sh}(E) \frac{A_\mathrm{eff}(E) S(E)^2}{B(E) \sigma(E)^2}
\end{equation}
This explicitly shows that the weight is proportional to the ratio
of source counts over background counts within the angular resolution.
The strong improvement of the PSF with energy (approximately as $E^{-0.8}$)
means that for a $E^{-2.2}$ source (which is not far from the background
spectral shape) $W(E) \propto A_\mathrm{eff}(E) E^{0.4}$.
Since $A_\mathrm{eff}(E)$ improves with energy up to 1~GeV, this
explains the rising part of $W(E)$.
At high energy (above several GeV) the PSF is narrow enough that even the
faint sources are limited by their own count rates ($g(0,E) > 1$)
and Eq.~\ref{eq:wback} no longer applies.
When the source density is large, a first-order way to account for confusion
is to limit the integral over angles in Eq.~\ref{eq:wts} to $\theta_{\rm max}$
such that $\pi \theta_{\rm max}^2 = \Omega_{\rm tot} / N_{\rm src}$ is
the average solid angle per source. This is shown by the solid line
on Figure~\ref{fig:tsweight}. The effect is of course larger where
the PSF is broader, at low energy.

\begin{figure}
\epsscale{.80}
\plotone{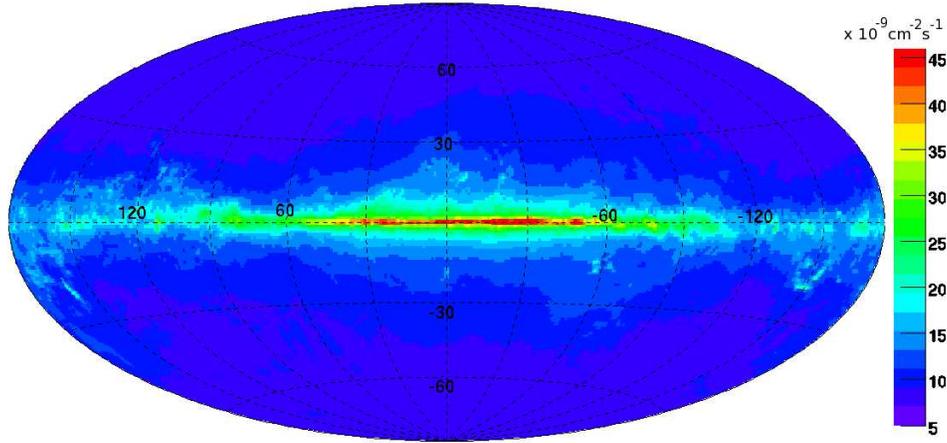}
\caption{Flux (E$>$100~MeV in ph cm$^{-2}$ s$^{-1}$)
needed to reach $TS$ = 25 in the LAT data for the 11-month time range
considered in this paper, as a function of position in Galactic coordinates.
The assumed photon spectral index is 2.2.}
\label{fig:sensmap}
\end{figure}

\begin{figure}
\epsscale{.80}
\plotone{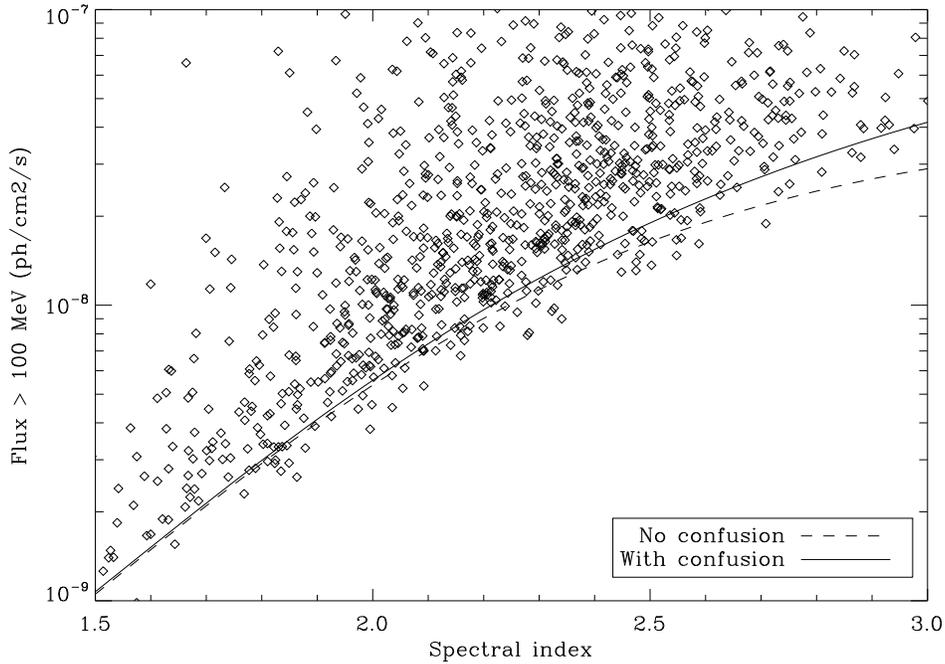}
\caption{Photon flux above 100~MeV of sources at $|b| > 10\degr$
as a function of spectral index.
The dashed line shows the theoretical detection threshold at $TS = 25$
for an isolated source over the average background at $|b| > 10\degr$.
The full line includes approximately the effect of source confusion
as on Figure~\ref{fig:tsweight}.}
\label{fig:sensspec}
\end{figure}

Setting $TS = 25$ in Eq.~\ref{eq:ts}, assuming a given source spectral shape
and solving for the normalization of the source spectrum
provides the detection threshold.
The spatial dependence of that threshold is shown in Figure~\ref{fig:sensmap}.
The Galactic diffuse and isotropic backgrounds are taken from the model
(\S~\ref{run:diffusemodel}). $T_0$ for each point is derived from 
the \textit{Fermi} pointing history during the 11 months, and 
the source spectrum is assumed to be $E^{-2.2}$,
the average spectral slope of the 1FGL sources.
Although the nonuniform exposure affects this map somewhat,
the dominant factor is the strong diffuse emission along the Galactic plane.
Because of the strong energy dependence of the PSF,
the detection threshold depends very sensitively on the spectral index as well.
Figure~\ref{fig:sensspec} illustrates this in terms of the photon flux
above 100~MeV, which ranges from $10^{-9}$ to $ 4 \times 10^{-8}$
ph cm$^{-2}$ s$^{-1}$ going from very hard ($\Gamma = 1.5$)
to soft ($\Gamma = 3$)  sources. The recipe for source confusion, given above,
is used.
We note that a few sources are below the line. This can happen for 
purely statistical
reasons, or because the background and exposure depend
a little on the direction, even after taking out the Galactic plane.

\section{Association Method}

The method implemented for automatic association of the 1FGL
sources essentially follows the ideas developed by \cite{mattox97b} for the identification
of EGRET sources with flat-spectrum radio sources.
It makes use of Bayes' theorem to compute the posterior probability \post\ that a 
counterpart $i$ from a list of potential counterparts supplied in the form of a
{\it counterpart catalog} is the correct association of a LAT source $k$:
\begin{equation}
\post = \frac{p_{ik}(\data|\id) \prior}
                     {p_{ik}(\data|\id) \prior + p_{ik}(\data|\cc) P_{i}(\cc)} .
\label{eq:bayes}
\end{equation}
\prior\ is the prior probability that counterpart $i$ is detectable by the LAT,
$P_{i}(\cc) = 1 - \prior$ is the prior probability that counterpart $i$ is not detectable by the LAT,
$p_{ik}(\data|\id)$ is the probability density for the detectable counterpart 
$i$ to have an angular separation $r$ and position angle
$\phi$ from a LAT source $k$, and
$p_{ik}(\data|\cc)$ is the probability density for source $k$ being only by chance situated
at an angular separation $r$ and position angle $\phi$ from counterpart $i$.

Under the assumption that  source location confidence region is centrally peaked, the probability density $p_{ik}(\data|\id)$ is obtained by differentiation of the
probability $P_{ik}(\data|\id)$ that the LAT source $k$ is located at an angular
separation smaller than $r$ under the position angle $\phi$:
\begin{equation}
p_{ik}(\data|\id) = \frac{d^2 P_{ik}(\data|\id)}{r dr d\phi} .
\end{equation}
By defining
\begin{equation}
\Delta_k = \ln \max(L_k) - \ln L_k(\data)
\end{equation}
as the difference between the log-likelihood maximum of the LAT source $k$ and the 
log-likelihood at position (\data) and by making use of Wilks's theorem that
$2\Delta_k$ is distributed as $\chi^2_2$ in the null hypothesis
\cite{wilks1938}
one can write
\begin{equation}
P_{ik}(\data|\id) = 1 - \int_{2 \Delta_k}^{\infty} \chi^2_2(x) dx = 1 - e^{-\Delta_k} .
\end{equation}
We approximate $\Delta_k$ as an elliptical paraboloid which is defined by the 95\% confidence
elliptical error region:
\begin{equation}
\Delta_k = \frac{r^2}{2} \left( \frac{\cos^2(\phi-\phi_k)}{a_k^2} + \frac{\sin^2(\phi-\phi_k)}{b_k^2} \right)
\end{equation}
where $a_k$ and $b_k$ are the axes of the ellipse at 1 $\sigma$, smaller than the semimajor ({\tt Conf\_95\_SemiMajor} in the FITS version of the 1FGL catalog; App. D) and semiminor ({\tt Conf\_95\_SemiMinor}) axes at 95\% confidence by a factor $\sqrt{-2\log(0.05)} = 2.45$, $\phi_k$ is the position angle of the error ellipse ({\tt Conf\_95\_PosAng}) and $\phi$ is the
 given position angle 
between LAT source $k$ and the counterpart $i$.
This can be transformed to
\begin{equation}
p_{ik}(\data|\id) = \frac{1}{2\pi a_k b_k} e^{-\Delta_k} .
\end{equation}

The chance coincidence probability density $p_{ik}(\data|\cc)$ is determined from
the local density $\rho_k$ of counterparts around LAT source $k$ as proposed
by \cite{sutherland92}:
\begin{equation}
p_{ik}(\data|\cc) = \frac{d^2 P_{ik}(\data|\cc)}{r dr d\phi} = \rho_k .
\end{equation}
To compute this density we count the number of counterparts $N_k$ in the counterpart
catalog under consideration within a radius of $r_0=4^\circ$ around the location of the
LAT source $k$ and divide by the solid angle of the search region:
\begin{equation}
\rho_k = \frac{N_k}{\pi r_0^2}
\label{eq:density}
\end{equation}
Note that the counterpart $i$ is included in $N_k$ which guarantees that $N_k \ge 1$.

As a last step, we implement the reasonable condition that a counterpart $i$ cannot be
associated with more than one LAT source.
This is done by introducing $\nlat+1$ mutually exclusive hypotheses
($\nlat=1451$ being the number of sources in the first year catalog):
\begin{itemize}
\item[] \idp: Object $i$ is a counterpart of LAT source $k$ and of none of the other LAT sources.
\item[] \ccp: Object $i$ is not a counterpart of any LAT source.
\end{itemize}
The probabilities for these new hypotheses are computed using
\begin{eqnarray}
 \tilde{P}_{ik}(\idp|\data) & = & P_{ik}(\id|\data) \prod_{k' \neq k} P_{ik'}(\cc|\data) \nonumber \\
 \tilde{P}_{i}(\ccp|\data) & = & \prod_{k'} P_{ik'}(\cc|\data) .
 \label{eq:catcorr}
\end{eqnarray}
where
\begin{equation}
P_{ik}(\cc|\data) = 1 - P_{ik}(\id|\data) .
\end{equation}
Since we dropped from the set of hypotheses all of the cases where an object $i$ is associated 
to more than a single LAT source, the sum over all probabilities
\begin{equation}
 \tilde{S}_i = \sum_k \tilde{P}_{ik}(\idp|\data) + \tilde{P}_{i}(\ccp|\data)
\end{equation}
is $\le 1$, and we thus renormalize using
\begin{equation}
P_{ik}(\idp|\data) = \frac{\tilde{P}_{ik}(\idp|\data)}{\tilde{S}_i}
\end{equation}
to obtain the posterior probability that object $i$ is a counterpart of the LAT source $k$ and of 
none of the other LAT sources.
Practically, $\tilde{S}_i<1$ only if the error ellipses of neighboring LAT sources overlap, which
is rather unlikely.
Thus, to a good approximation we have
$P_{ik}(\idp|\data) = P_{ik}(\id|\data)$
for basically all sources.

The above procedure that leads to the computation of $P_{ik}(\idp|\data)$ for a specific
catalog of counterpart candidates has been implemented in the ScienceTools executable
{\tt gtsrcid}.
We used version {\tt v2r2p3} of this tool for counterpart association for the first year
catalog.
To simplify, in what follows we will write
$P_{ik}$
instead of 
$P_{ik}(\idp|\data)$ 
for the posterior association probability of LAT source $k$ with object $i$ of a given 
counterpart catalog.

\section{Calibration of prior probabilities}

Before Eq.~(\ref{eq:bayes}) can be used for source association, 
prior probabilities \prior\ have to be specified for each counterpart $i$.
Here we make the simplifying assumption that within a given counterpart catalog the
prior probabilities for all sources $i$ are identical:
\begin{equation}
\prior = P(\id) .
\end{equation}
To assign $P(\id)$ for a given counterpart catalog, we require the relation
\begin{equation}
  \nfalse = \sum_{P_{ik} \ge P_{\rm thr}} (1-P_{ik})
  \label{eq:require}
\end{equation}
to hold, where
\nfalse\ is the number of false associations that have posterior probabilities $P_{ik}$ above 
our selected threshold $\pthres=0.8$.
For this purpose we determined by means of Monte Carlo simulations of 100 fake LAT catalogs
the expected number of false associations, \nfalsemc, as function of $P(\id)$.
For a given $P(\id)$, we obtained \nfalsemc\ by counting the number of sources in each 
fake catalog that have been associated, and by dividing this number by 100, 
i.e. by the number of fake catalogs that have been simulated.
$P(\id)$ has then been varied until $\nfalse = \nfalsemc$ was fulfilled, which then fixed
the proper prior probability $P(\id)$ for the counterpart catalog.

\begin{figure}
\plotone{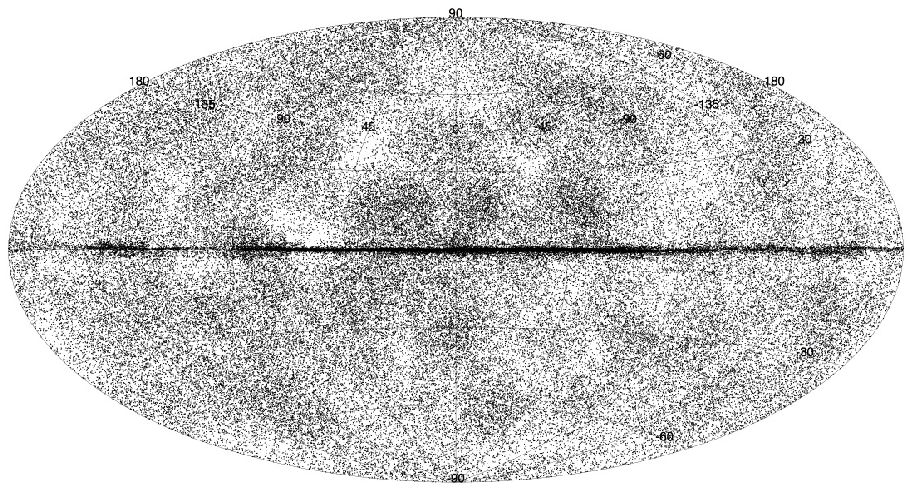}
\caption{Locations of sources in the 100 fake 1FGL catalogs used for calibration of prior 
probabilities.}
\label{fig:fake}
\end{figure}

The fake catalogs were created by randomly displacing 1FGL sources within a ring 
from $2^\circ$ to $10^\circ$ in radius around their nominal position.
Since the 1FGL catalog comprises a distinct population of Galactic sources that obey a rather
narrow latitude distribution, we limited source displacement in Galactic latitude to 
$b \pm b_{\rm max}$, where
\begin{equation}
b_{\rm max} = r_{\rm max} \left( 1.0 - \textrm{sech}^2 \left( \frac{b}{b_0} \right) \right) ,
\end{equation}
$r_{\rm max} = 10^\circ$, 
$b$ is the Galactic latitude of the 1FGL source, and
$b_0 = 5^\circ$ is the angular scale height above the Galactic plane for which the latitude
displacement is reduced.
We further required $b_{\rm max} \ge 0.2^\circ$ to allow for a non-zero latitude displacement
of sources in the Galactic plane, and required any source to be shifted by at least 
$r_{\rm min} = 2^\circ$ 
away from its original location.
For illustration, we show in Figure~\ref{fig:fake} the locations of the sources in the 100 fake catalogs
that were used for calibration.

\begin{figure}
\plottwo{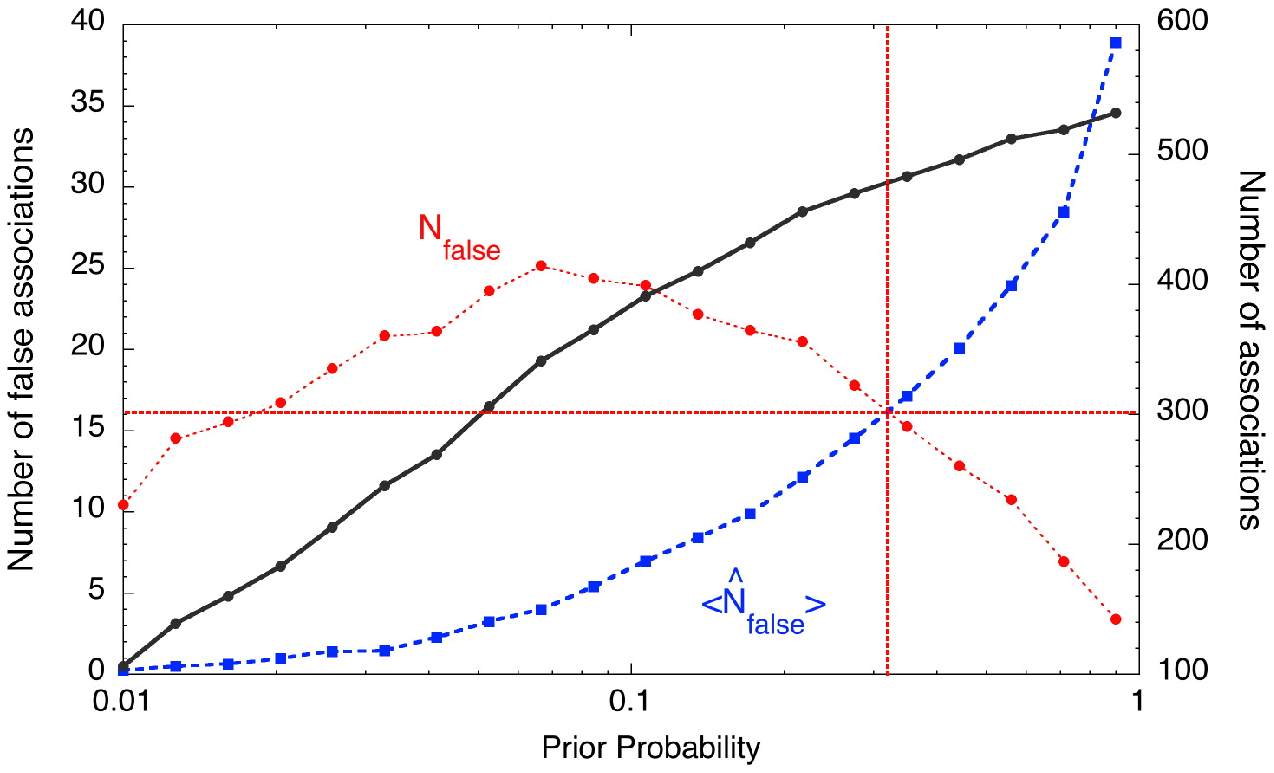}{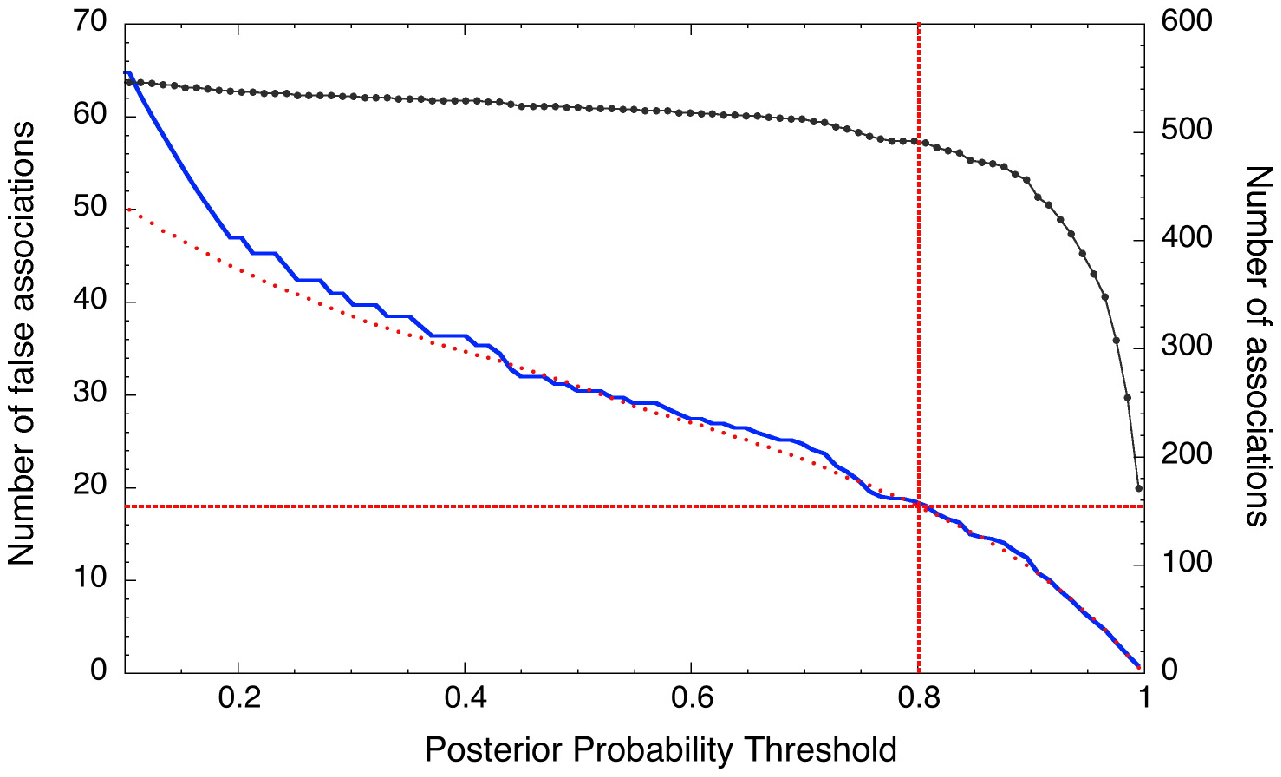}
\caption{Number of false associations as function of prior probability $P(\id)$ for
$\pthres=0.8$ (left panel), and as function of posterior probability for $P(\id)=0.33$.
Dashed red lines correspond to \nfalsemc, solid blue lines represent \nfalse\ (left axis).
The solid black lines show \nass\ (right axis).}
\label{fig:example}
\end{figure}

As example, we show 
\nfalse\ (determined from Eq.~\ref{eq:require}) and
\nfalsemc\ (determined from the simulation) as function of $P(\id)$ for the CRATES 
catalog of flat spectrum radio sources in the left panel of Figure~\ref{fig:example}.
The intersection of both curves determines the prior probability, which in this
case has been determined to $P(\id)=0.33$.
We also show in Figure~\ref{fig:example} the number of associations that is computed
using
\begin{equation}
\nass = \sum_{P_{ik} \ge P_{\rm thr}} 1 \, .
\end{equation}
as solid black lines.

Although the calibration has been performed for $\pthres=0.8$, it turns out that, once the
prior probabilities are set, Eq.~(\ref{eq:require}) is fulfilled for a large range of posterior
probability thresholds.
We illustrate this property in the right panel of Figure~\ref{fig:example}, which shows
for the CRATES catalog \nfalsemc\ and \nfalse\ as function of \pthres\ for $P(\id)=0.33$.
Obviously, both quantities are in good agreement.

\section{Description of the FITS version of the 1FGL catalog}
\label{run:fits}
The FITS format version of the 1FGL catalog\footnote{The file is available  from the $Fermi$ Science Support Center, http://fermi.gsfc.nasa.gov/ssc} has three binary table extensions.  The extension {\tt LAT\_Point\_Source\_Catalog Extension} has all of the information about the sources, including the monthly light curves (Tab.~\ref{tab:columns}).  

The extension {\tt Hist\_Start} lists the Mission Elapsed Time (seconds since 00:00 UTC on 2000 January 1) of the start of each bin of the monthly light curves.  The final entry is the ending time of the last bin.  

The extension {\tt GTI} is a standard Good-Time Interval listing the precise time intervals (start and stop in MET) included in the data analysis.  The number of intervals is fairly large because on most orbits ($\sim$95~min) $Fermi$ passes through the South Atlantic Anomaly (SAA), and science data taking is stopped during these times.  In addition, data taking is briefly interrupted on each non-SAA-crossing orbit, as $Fermi$ crosses the ascending node.  Filtering of time intervals with large rocking angles, other data gaps, or operation in non-standard configurations introduces some more entries.  The GTI is provided for reference and would be useful, e.g., for reconstructing the precise data set that was used for the 1FGL analysis.

\begin{deluxetable}{llll}
\setlength{\tabcolsep}{0.04in}
\tablewidth{0pt}
\tabletypesize{\scriptsize}
\tablecaption{LAT 1FGL FITS format:  LAT\_Point\_Source\_Catalog Extension\label{tab:columns}}
\tablehead{
\colhead{Column} &
\colhead{Format} &
\colhead{Unit} &
\colhead{Description}
}
\startdata
Source\_Name & 18A & \nodata & \nodata  \\
RA & E & deg & Right Ascension \\
DEC & E & deg & Declination \\
GLON & E & deg & Galactic Longitude \\
GLAT & E & deg & Galactic Latitude \\
Conf\_68\_SemiMajor & E & deg & Long radius of error ellipse at 68\% confidence \\
Conf\_68\_SemiMinor & E & deg & Short radius of error ellipse at 68\% confidence \\
Conf\_68\_PosAng & E & deg & Position angle of the 68\% long axis from celestial North, \\
 & \nodata & & positive toward increasing RA (eastward) \\
Conf\_95\_SemiMajor & E & deg & Long radius of error ellipse at 95\% confidence \\
Conf\_95\_SemiMinor & E & deg & Short radius of error ellipse at 95\% confidence \\
Conf\_95\_PosAng & E & deg & Position angle of the 95\% long axis from celestial North, \\
 & & & positive toward increasing RA (eastward) \\
Signif\_Avg & E & \nodata & Source significance in sigma units (derived from Test Statistic) \\
Pivot\_Energy & E & MeV & Energy at which error on differential flux is minimal \\
Flux\_Density & E & cm$^{-2}$ MeV$^{-1}$ s$^{-1}$ & Differential flux at Pivot\_Energy \\
Unc\_Flux\_Density & E & cm$^{-2}$ MeV$^{-1}$ s$^{-1}$ & 1 $\sigma$  error on differential flux at Pivot\_Energy \\
Spectral\_Index & E & \nodata & Best fit power law slope \\
Unc\_Spectral\_Index & E & \nodata & 1 $\sigma$  error on best fit power law slope \\
Flux1000 & E & cm$^{-2}$ s$^{-1}$ & Integral flux from 1 to 100 GeV \\
Unc\_Flux1000 & E & cm$^{-2}$ s$^{-1}$ & 1 $\sigma$  error on integral flux from 1 to 100 GeV \\
Energy\_Flux & E & erg cm$^{-2}$ s$^{-1}$ & Energy flux from 100 MeV to 100 GeV \\
Unc\_Energy\_Flux & E & erg cm$^{-2}$ s$^{-1}$ & 1 $\sigma$  error on energy flux from 100 MeV to 100 GeV \\
Curvature\_Index & E & \nodata & Measure of how spectrum follows power-law (currently simple $\chi^2$ \\
Flux30\_100 & E & cm$^{-2}$ s$^{-1}$ & Integral flux from 30 to 100 MeV (not filled) \\
Unc\_Flux30\_100 & E & cm$^{-2}$ s$^{-1}$ & 1 $\sigma$  error on integral flux from 30 to 100 MeV (not filled) \\
Sqrt\_TS30\_100 & E & \nodata & Square root of the Test Statistic between 30 and 100 MeV (not filled) \\
Flux100\_300 & E & cm$^{-2}$ s$^{-1}$ & Integral flux from 100 to 300 MeV \\
Unc\_Flux100\_300 & E & cm$^{-2}$ s$^{-1}$ & 1 $\sigma$  error on integral flux from 100 to 300 MeV\tablenotemark{a}\\
Sqrt\_TS100\_300 & E & \nodata & Square root of the Test Statistic between 100 and 300 MeV \\
Flux300\_1000 & E & cm$^{-2}$ s$^{-1}$ & Integral flux from 300 MeV to 1 GeV \\
Unc\_Flux300\_1000 & E & cm$^{-2}$ s$^{-1}$ & 1 $\sigma$  error on integral flux from 300 MeV to 1 GeV\tablenotemark{a} \\
Sqrt\_TS300\_1000 & E & \nodata & Square root of the Test Statistic between 300 MeV and 1 GeV \\
Flux1000\_3000 & E & cm$^{-2}$ s$^{-1}$ & Integral flux from 1 to 3 GeV \\
Unc\_Flux1000\_3000 & E & cm$^{-2}$ s$^{-1}$ & 1 $\sigma$  error on integral flux from 1 to 3 GeV\tablenotemark{a} \\
Sqrt\_TS1000\_3000 & E & \nodata & Square root of the Test Statistic between 1 and 3 GeV \\
Flux3000\_10000 & E & cm$^{-2}$ s$^{-1}$ & Integral flux from 3 to 10 GeV \\
Unc\_Flux3000\_10000 & E & cm$^{-2}$ s$^{-1}$ & 1 $\sigma$  error on integral flux from 3 to 10 GeV\tablenotemark{a} \\
Sqrt\_TS3000\_10000 & E & \nodata & Square root of the Test Statistic between 3 and 10 GeV \\
Flux10000\_100000 & E & cm$^{-2}$ s$^{-1}$ & Integral flux from 10 to 100 GeV \\
Unc\_Flux10000\_100000 & E & cm$^{-2}$ s$^{-1}$ & 1 $\sigma$ error on integral flux from 10 to 100 GeV\tablenotemark{a} \\
Sqrt\_TS10000\_100000 & E & \nodata & Square root of the Test Statistic between 10 and 100 GeV \\
Variability\_Index & E & \nodata & Measure of source variability (currently simple $\chi^2$) \\
Signif\_Peak & E & \nodata & Source significance in peak interval in $\sigma$ units \\
Flux\_Peak & E & cm$^{-2}$ s$^{-1}$ & Peak integral flux from 100 MeV to 100 GeV \\
Unc\_Flux\_Peak & E & cm$^{-2}$ s$^{-1}$ &  1 $\sigma$  error on peak integral flux \\
Time\_Peak & D & s (MET) & Time of center of interval in which peak flux was measured \\
Peak\_Interval & E & s & Length of interval in which peak flux was measured \\
Flux\_History & 11E & cm$^{-2}$ s$^{-1}$ & Integral flux from 100 MeV to 100 GeV in each interval \\
Unc\_Flux\_History & 11E & cm$^{-2}$ s$^{-1}$ &  Error on integral flux in each interval using method \\
&  & &  indicated in Unc\_Flag\_History column and added in quadrature \\
&  & &  with 3\% systematic component. \\
Unc\_Flag\_History & 11B &  &  1 if it is half of the difference between the 2$\sigma$ upper limit \\
& & &  and the maximum-likelihood value given in Flux\_History, 0 if it is the \\
& & &  1$\sigma$ uncertainty derived from a significant detection in the interval\\ 
0FGL\_Name & 18A & \nodata & Name of corresponding 0FGL source, if any \\
ASSOC\_GAM1 & 18A & \nodata & Name of likely corresponding 1AGL source \\
ASSOC\_GAM2 & 18A & \nodata & Name of likely corresponding 3EG source \\
ASSOC\_GAM3 & 18A & \nodata & Name of likely corresponding EGR source \\
TEVCAT\_FLAG & A & \nodata & P if positional association with $<$40$^\prime$ source in TeVCat \\
   & \nodata & & E if associated with a more extended source in TeVCat, N if no TeV association \\
CLASS1 & 3A & \nodata & Class designation for associated source; see Table~\ref{tab:classes} \\
CLASS2 & 3A & \nodata & Second class designation for associated source \\
ASSOC1 & 24A & \nodata & Name of identified or likely associated source \\
ASSOC2 & 24A & \nodata & Alternate name of identified or likely associated source \\
Flags & I & \nodata & Source flags (binary coding as in Table~\ref{tab:flags}) \\
\enddata
\tablenotetext{a} {The upper limit is set equal to 0 if the flux in the corresponding energy band is an upper limit ($TS < 10$ in that band).  The upper limits are 2 $\sigma$.}
\end{deluxetable}
 
\end{document}